\documentclass[12pt]{article}

\usepackage{epsfig}

\newcommand{\ltwid}{\mathrel{\raise.3ex\hbox{$<$\kern-.75em\lower1ex\hbox{$\sim$}}}}
\newcommand{\gtwid}{\mathrel{\raise.3ex\hbox{$>$\kern-.75em\lower1ex\hbox{$\sim$}}}}

\begin{document}

\begin{titlepage}

\begin{flushright}
UFIFT-QG-09-06
\end{flushright}

\vskip 1cm

\begin{center}
{\bf How Far Are We from the Quantum Theory of Gravity?}
\end{center}

\vskip 1cm

\begin{center}
R. P. Woodard$^{\dagger}$
\end{center}

\begin{center}
\it{Department of Physics, University of Florida \\
Gainesville, FL 32611, UNITED STATES}
\end{center}

\vspace{.5cm}

\begin{center}
ABSTRACT
\end{center}
I give a pedagogical explanation of what it is about quantization that
makes general relativity go from being a nearly perfect classical theory
to a very problematic quantum one. I also explain why some quantization
of gravity is unavoidable, why quantum field theories have divergences,
why the divergences of quantum general relativity are worse than those
of the other forces, what physicists think this means and what they
might do with a consistent theory of quantum gravity if they had one.
Finally, I discuss the quantum gravitational data that have recently
become available from cosmology.

\begin{flushleft}
\end{flushleft}

\begin{flushleft}
$^{\dagger}$ e-mail: woodard@phys.ufl.edu
\end{flushleft}

\end{titlepage}

\section{Introduction}

Gravity was the first of the fundamental forces to be recognized and it
will be the last to be understood. Its early recognition came because
it is both long range and universal; gravity acts over macroscopic
distances, and no one has ever found a way to screen it. Mankind's
problems comprehending quantum gravity are the subject of this article,
but they derive from two basic facts:
\begin{enumerate}
\item{Humans are not good at guessing fundamental principles without
guidance from nature; and}
\item{Gravity is such a weak interaction that nature doesn't provide much
guidance in the quantum regime of microscopic sources.}
\end{enumerate}
We know there is something wrong with perturbative quantum general
relativity because one cannot consistently absorb the divergences it
produces without introducing new degrees of freedom that would make the
universe blow up \cite{Bryce,HV,DvN,DTvN,KS,GS,Ven}. Obtaining that result
was the work of great physicists over many decades, but their achievement
will remain incomplete until we can identify the problem and fix it.

I wish I could present a solution to quantum gravity, or at least a
program that could reasonably be expected to provide a solution. Because
that is not possible I shall endeavor to instead give a clear explanation
of the problem. This entails answering a number of questions:
\begin{itemize}
\item{What is the distinction between classical physics and quantum
physics that makes general relativity give such a wonderful classical
theory of gravity and such a problematic quantum one?}
\item{Why do we have to quantize gravity?}
\item{Why do quantum field theories have divergences?}
\item{Why are the divergences of quantum general relativity worse than those
of the other forces?}
\item{How bad is the problem?}
\item{What are the main schools of thought about quantizing gravity and
why do they disagree?}
\item{What would we do with the theory of quantum gravity if we had it?}
\end{itemize}
I will also comment on the quantum gravitational data that has recently
become available.

The tale I have to tell is of necessity a complex one, requiring many
digressive explanations. However, there is no need for the exposition to
transcend the knowledge one expects of any physics graduate student.
Because every one of the basic issues behind the problem of quantum
gravity has a counterpart in either electrodynamics or introductory
quantum mechanics, I shall use those subjects as paradigms. This is not
condescension; even experts can benefit from occasionally viewing a
tough problem in a general way, without becoming lost in technical
details.

I shall work in four spacetime dimensions, with coordinate points
labeled thus, $x^{\mu} = (ct,\vec{x})$. A symbol with an arrow over it
denotes a 3-vector, and I employ the usual notations for scalar and vector
products. Differentiation with respect to time is denoted with a
dot; the gradient operator is $\vec{\nabla}$. To be clear, and for
future reference, Maxwell's equations in MKS units are,
\begin{eqnarray}
\epsilon_0 \vec{\nabla} \cdot \vec{E} = \rho \qquad & , & \qquad
\vec{\nabla} \cdot \vec{B} = 0 \; , \label{Max1} \\
\frac1{\mu_0} \vec{\nabla} \times \vec{B} - \epsilon_0 \dot{\vec{E}}
= \vec{J}  \qquad & , & \qquad \vec{\nabla} \times \vec{E} + \dot{\vec{B}}
= 0 \; . \label{Max2}
\end{eqnarray}
Here the various fields are: electric, $\vec{E}(t,\vec{x})$; magnetic,
$\vec{B}(t,\vec{x})$; the charge density, $\rho(t,\vec{x})$; and the
current density, $\vec{J}(t,\vec{x})$. The two constants are the electric
permittivity of free space, $\epsilon_0$, and the magnetic permeability of
free space, $\mu_0$. I denote spatial Fourier transforms with a tilde,
\begin{equation}
\widetilde{f}(t,\vec{k}) = \int \!\! d^3x \, e^{-i \vec{k} \cdot \vec{x}}
f(t,\vec{x}) \qquad \Longleftrightarrow \qquad f(t,\vec{x}) = \int \!\!
\frac{d^3k}{(2\pi)^3} \, e^{i \vec{k} \cdot \vec{x}} \widetilde{f}(t,\vec{k})
\; .
\end{equation}
Finally, I use the term ``classical'' to mean ``not quantum,'' irrespective
of relativity. So it is perfectly valid to speak of ``classical general
relativity.'' The adjective for suspending relativity is ``nonrelativistic.''

\section{Perturbative Quantum General Relativity}

The central problem of quantum gravity is that the computational
techniques used with great success for the other forces do not give
consistent results when applied to quantum general relativity. To
understand the problem I will have to explain a little bit about
general relativity, what it means to quantize a theory, and the
only technique we so far have for computing things in quantum field
theory. Then I describe renormalization from the perspective of
polarization in electrodynamics, and I explain why the only way of
consistently renormalizing perturbative quantum general relativity
would make the universe virulently unstable. The section closes with
a brief discussion of fixes that don't work.

\subsection{General Relativity}

One defines a physical theory by specifying three things:
\begin{itemize}
\item{The dynamical variable;}
\item{How the dynamical variable affects the rest of physics; and}
\item{How the rest of physics affects the dynamical variable.}
\end{itemize}
For example, the fundamental dynamical variables of electromagnetism
are the scalar potential $\Phi(t,\vec{x})$ and the vector potential
$\vec{A}(t,\vec{x})$, whose derivatives give the electric and magnetic
fields,
\begin{equation}
\vec{E} = -\vec{\nabla} \Phi - \dot{\vec{A}} \qquad {\rm and} \qquad
\vec{B} = \vec{\nabla} \times \vec{A} \; . \label{EandB}
\end{equation}
They affect a particle of charge $q$ which has position $\vec{x}$ and
velocity $\vec{v}$ through the Lorentz force,
\begin{equation}
\vec{F} = q \vec{E}(t,\vec{x}) + q \vec{v} \times \vec{B}(t,\vec{x}) \; .
\label{Lorentz}
\end{equation}
And the various charges and currents affect them through the Maxwell
equations (\ref{Max1}-\ref{Max2}), which have the general structure,
\begin{equation}
\partial^2 \Bigl(\Phi,\vec{A}\Bigr) = \Bigl(\rho,\vec{J}\Bigr) \; .
\label{generic}
\end{equation}
My notation is that $\partial^2 (\Phi,\vec{A})$ stands for the derivative
with respect to any two coordinates of any combination of $\Phi$ and
$\vec{A}$. For the purposes of this article no greater specificity is
required than to note:
\begin{enumerate}
\item{The electrodynamic field equations are linear in the potentials;}
\item{The electrodynamic field equations involve second derivatives of
the potentials; and}
\item{The electrodynamic field equations are sourced by the charge density
$\rho(t,\vec{x})$ and the current density $\vec{J}(t,\vec{x})$.}
\end{enumerate}

In these terms, the dynamical variable of general relativity is known
as the {\it metric field}, $g_{\mu\nu}(t,\vec{x})$. It is a $4 \times 4$,
symmetric matrix whose components are functions of space and time. It
affects the rest of physics by controlling the physical distances and
times between points. Recall that in special relativity the frame independent
concept of the distance between nearby points $x^{\mu}$ and $x^{\mu} +
dx^{\mu}$ is given by the ``invariant interval'',
\begin{equation}
\Bigl(ds^2\Bigr)_{\rm special \atop relativity} = -c^2 dt^2 + d\vec{x} \cdot
d\vec{x} \; .
\end{equation}
The invariant interval of general relativity is,
\begin{equation}
\Bigl(ds^2\Bigr)_{\rm general \atop relativity} = \sum_{\mu=0}^3 \sum_{\nu=0}^3
g_{\mu\nu}(t,\vec{x}) dx^{\mu} dx^{\nu} \; .
\end{equation}
The fact that the metric defines true distances and times affects how
derivatives and integrals are constructed in a way that is analogous to
coupling rule of electrodynamics. The details need not concern us.

The rest of physics affects the metric through the Einstein equations
which have the general form,
\begin{equation}
\Biggl[ A(g) \partial^2 g + B(g) \partial g \partial g + \Lambda g
\Biggr]_{\mu\nu} = \frac{8 \pi G}{c^4} \, T_{\mu\nu} \; . \label{Einstein}
\end{equation}
My notation is that $\partial^2 g$ stands for the derivative of any
component of the metric tensor with respect to any two coordinates.
In the same sense, $\partial g \partial g$ stands for any product of first
derivatives of the metric. The parameter $\Lambda$ is known as the {\it
cosmological constant}. It is often set to zero in general relativity but
retaining it will facilitate our eventual discussion of renormalization,
and nature seems to have chosen a small nonzero value for it \cite{SNIA,Yun}.
At the level I propose to work one need only note:
\begin{enumerate}
\item{The gravitational field equations are not linear in the metric;}
\item{The gravitational field equations involve up to second derivatives
of the metric; and}
\item{The gravitational field equations are sourced by the stress-energy
tensor $T_{\mu\nu}(t,\vec{x})$.}
\end{enumerate}

The stress-energy tensor $T_{\mu\nu}$ has the units of energy per volume
(which is the same as force per area, or stress) and its various
components have the meanings,
\begin{eqnarray}
T_{00} & : & {\rm energy\ density} \; , \\
T_{i0} & : & {\rm momentum\ density\ in\ the\ i\!-\!th\ direction} \; , \\
T_{0j} & : & {\rm energy\ flux\ in\ the\ j\!-\!th\ direction} \; , \\
T_{ij} & : & {\rm i\!-\!th\ component\ of\ momentum\ flux\ in\ the\ j\!-\!th\
direction.} \qquad
\end{eqnarray}
(The ``flux'' of any quantity in the j-th direction is the amount of
that quantity which passes through a unit area perpendicular to the
j-th direction, per unit time.) The stress-energy tensor is composed
of the other fields in physics, and it also depends on the metric.
For example, if we set the metric equal to the value it has in
special relativity (that is, $g_{00} = -1$, $g_{0i} = 0$ and $g_{ij}
= \delta_{ij}$), the energy density contributed by electromagnetism
is,
\begin{equation}
\Bigl[T_{00}(t,\vec{x})\Bigr]_{\rm flat \atop space} = \frac12 \epsilon_0
\Bigl[ \vec{E} \cdot \vec{E} + c^2 \vec{B} \cdot \vec{B}\Bigr] \; .
\label{EMrho}
\end{equation}
Of course gravity cannot be absent if there are electrodynamic (or
any other) fields present so the actual expression for $T_{00}$
depends upon the metric in a manner that need not concern us. It is
worth commenting that the nonlinearity of the gravitational field
equations, and the fact that their sources depend upon the metric
they help to determine, means that no general solution of the
Einstein equations is known. This is why physicists resort to the
perturbative approximation technique I will describe in subsection
2.3.

It turns out that all fundamental force fields have a triune nature: one
part is completely arbitrary, another part is totally determined by the
sources of the force, and the final part consists of independent degrees
of freedom. For example, the completely arbitrary part of electromagnetism
is the ability to change the scalar potential and vector potentials by a
gauge transformation,
\begin{eqnarray}
\Phi(t,\vec{x}) & \longrightarrow & \Phi(t,\vec{x}) + \dot{\theta}(t,\vec{x})
\; , \label{dPhi} \\
\vec{A}(t,\vec{x}) & \longrightarrow & \vec{A}(t,\vec{x}) - \vec{\nabla}
\theta(t,\vec{x}) \; , \label{dA}
\end{eqnarray}
where $\theta(t,\vec{x})$ is an arbitrary function of space and time. One
can easily check that the transformation (\ref{dPhi}-\ref{dA}) makes no
change in the electric and magnetic fields (\ref{EandB}). One can also show
that it makes Maxwell's equations (\ref{Max1}-\ref{Max2}) consistent with
current conservation.

The other two parts of any force field are illustrated by solving
(\ref{Max1}-\ref{Max2}) for the magnetic field, assuming the current
density is some known function,
\begin{eqnarray}
\lefteqn{\vec{B}(t,\vec{x}) = \int \!\! \frac{d^3k}{(2\pi)^3} \, e^{i\vec{k}
\cdot \vec{x}} \int_0^t \!\!\ dt' \, \frac{\sin[ck (t \!-\! t')]}{c k
\epsilon_0} \, i \vec{k} \times \widetilde{\vec{J}}(t',\vec{k}) } \nonumber \\
& & \hspace{2cm} + \int \!\! \frac{d^3k}{(2\pi)^3} \, e^{i\vec{k} \cdot\vec{x}}
\Biggl\{ \widetilde{\vec{B}}_0(\vec{k}) \cos(c k t) - \frac{i}{c k} \,
\vec{k} \!\times\! \widetilde{\vec{E}}_0(\vec{k}) \sin(c k t) \Biggr\} .
\qquad \label{genB}
\end{eqnarray}
The first line of (\ref{genB}) give the part of the magnetic field which
depends upon its source, which is the current density. This part of the
field contains important physics, but it isn't an independent degree of
freedom because knowing the current density $\vec{J}(t,\vec{x})$ fixes it
completely. The technical name for such a field is ``constrained''. The
``dynamical'' part of $\vec{B}(t,\vec{x})$ comes on the final line of
(\ref{genB}) which contains the independent, purely electromagnetic degrees
of freedom that would be present even if the current density was zero for
all time. These extra degrees of freedom consist of pure electromagnetic
radiation that corresponds to the ``photons'' of quantum electrodynamics.

Gravity has the same general structure as electromagnetism and all
the other fundamental force fields: part of the metric can be
changed arbitrarily by a symmetry transformation known as {\it
general coordinate invariance}; another part of the metric is
determined by the stress-energy tensor; and a third part contains
independent degrees of freedom, the pure gravitational radiation
which comprises the gravitons of quantum gravity. Because the
Einstein equations (\ref{Einstein}) are not linear, the breakup of
$g_{\mu\nu}(t,\vec{x})$ into these three constituents is vastly more
complex than for electromagnetism, but that is another detail which
need not concern us here, although it is a major headache for people
who work in gravity. A fact of great importance for the discussion
of renormalization is that general coordinate invariance implies
{\it the left hand side of the Einstein equation (\ref{Einstein}) is
the unique combination of the metric and no more than two of its
derivatives which is consistent with the conservation of
stress-energy.}

To understand quantum field theory it is important to be clear about the
distinction between the constrained and dynamical parts of a force field. It
is the constrained part of electrodynamics which holds the Hydrogen atom
together. For example, if $\vec{q}(t)$ and $\vec{p}(t)$ are the position and
momentum of the electron then the Hamiltonian for nonrelativistic Hydrogen is,
\begin{equation}
H = \frac{\Vert \vec{p} \Vert^2}{2 m} - e \Phi(\vec{q}) \qquad {\rm where}
\qquad \Phi(\vec{q}) = \frac{e}{4 \pi \epsilon_0 \Vert \vec{q}\Vert} \; .
\label{nonrel}
\end{equation}
Note that the scalar potential $\Phi(\vec{q})$ is a completely determined
function of the electron's position $\vec{q}(t)$. If we were doing quantum
mechanics it would be a quantum operator, but only because the electron's
position is a quantum operator; it would not possess any independent quantum
degrees of freedom of its own. Of course the nonrelativistic Hamiltonian
(\ref{nonrel}) is only an approximation and the full, relativistic system
does incorporate quantized photon degrees of freedom. These degrees of
freedom are needed to explain the Lamb shift of about one part in $10^6$ in
the frequencies of light emitted from decays of the $\mbox{}^2 S_{\frac12}$ and
$\mbox{}^2 P_{\frac12}$ states \cite{Lamb}. However, it is worth emphasizing
that the basic structure of the atom is determined by the constrained part of
the electrodynamic potentials and one has to work quite hard to see even the
first evidence for quantized photons from it. For example, the Lamb shift was
only detected by stimulating a transition {\it between} the two levels.

Now consider the vastly weaker gravitational force. We have so far not been
able to directly detect gravitational radiation, much less the gravitational
radiation from a quantum transition, or the even subtler shift due to
quantized gravitons. The gravitational effects which hold the solar system
together derive from the constrained part of metric. There is only indirect
evidence that gravitational radiation exists \cite{Will}, and there is no
evidence at all for its quantization. Which brings me to one of the major
points of this article: {\it some quantization of gravity is inevitable
because part of the metric depends upon the other fields whose quantum nature
has been well established.} It turns out that the first divergences of
quantum gravity are due to quantum effects from these other fields
\cite{HV,DvN,DTvN}. The quantum effects of gravitons --- if there is
gravitational radiation, and if it is quantized --- do cause problems
\cite{GS,Ven}, but these difficulties occur at higher order in the
approximation scheme described in section 2.3. {\it So the central problem of
quantum general relativity has nothing to do with gravitons.}

\subsection{Quantum Mechanics}

It will be noted that I have written field equations such as
(\ref{Max1}-\ref{Max2}), and even solved them (\ref{genB}), without
stating whether the system is classical or quantum. There is a good
reason for this: it doesn't matter! {\it The operator equations of
motion in the Heisenberg Picture of quantum mechanics are the same as
those for the corresponding classical theory.} Further, ``solving''
these equations means precisely the same thing: one expresses the
dynamical variables at any time in terms of the initial values of the
dynamical variables. Those initial values are the fundamental degrees of
freedom of the system, and the only difference between classical physics
and quantum physics is what they represent. In classical physics the
initial values are just numbers and each of them can take any value,
whereas in quantum physics they are non-commuting operators which must
obey the Uncertainty Principle.

An example which has great significance for our discussion is the simple
harmonic oscillator. The dynamical variable is the position as a function
of time, $q(t)$. For the moment we imagine this system to exist in isolation,
so it has no effect on the rest of the universe. Nor does the rest of the
universe affect it; its dynamics are controlled by its Lagrangian and
Euler-Lagrange equations,
\begin{equation}
L = \frac12 m \dot{q}^2 - \frac12 m \omega^2 q^2 \quad \Longrightarrow
\quad \frac{\partial L}{\partial q} - \frac{d}{dt} \frac{\partial L}{\partial
\dot{q}} = -m \Bigl[ \ddot{q}(t) + \omega^2 q(t) \Bigr] = 0 \; . \label{SHO}
\end{equation}
The general solution in the sense of fundamental theory is,
\begin{equation}
q(t) = q_0 \cos(\omega t) + \frac{\dot{q}_0}{\omega} \, \sin(\omega t) \; ,
\end{equation}
where $q_0 = q(0)$ and $\dot{q}_0 = \dot{q}(0)$.

Now break up the trigonometric functions using the Euler relation,
\begin{equation}
\cos(\omega t) = \frac12 \Bigl[ e^{i\omega t} + e^{-i\omega t}\Bigr] \qquad ,
\qquad \sin(\omega t) = \frac{i}2 \Bigl[- e^{i\omega t} + e^{-i\omega t}
\Bigr] \; .
\end{equation}
The solution can be expressed as a sum of positive and negative frequencies,
\begin{equation}
q(t) = \frac12 \, e^{-i \omega t} \Bigl[q_0 + \frac{i \dot{q}_0}{\omega}\Bigr]
+ \frac12 \, e^{i \omega t} \Bigl[q_0 - \frac{i \dot{q}_0}{\omega}\Bigr] \; .
\end{equation}
As noted above, this same solution applies both for the classical theory
and for the quantum one. In the latter we can recognize that the operator
coefficient of $e^{-i\omega t}$ must lower the energy by $\hbar \omega$,
whereas the operator coefficient of $e^{i\omega t}$ must raise it by the
same amount. The canonical momentum tells us how $q_0$ and $\dot{q}_0$
commute,
\begin{equation}
p = \frac{\partial L}{\partial \dot{q}} = m \dot{q} \qquad \Longrightarrow
\qquad \Bigl[q_0 ,\dot{q}_0\Bigr] = \frac{i \hbar}{m} \; .
\end{equation}
We can use this to canonically normalize the raising and lowering
operators,
\begin{equation}
a \equiv \sqrt{\frac{m \omega}{2 \hbar}} \Bigl(q_0 + \frac{i \dot{q}_0}{
\omega}\Bigr) \qquad \Longrightarrow \qquad \Bigl[ a, a^{\dagger}\Bigr]
= 1 \; .
\end{equation}
And the final result for the position operator takes the form,
\begin{equation}
q(t) = a \, \varepsilon(t) + a^{\dagger} \, \varepsilon^*(t) \; , \label{HOq}
\end{equation}
where the ``mode coordinates'' are,
\begin{equation}
a \equiv \sqrt{\frac{m \omega}{2 \hbar}} \Bigl(q_0 + \frac{i \dot{q}_0}{
\omega}\Bigr) \qquad {\rm and} \qquad a^{\dagger} \equiv \sqrt{\frac{m
\omega}{2 \hbar}} \Bigl(q_0 - \frac{i \dot{q}_0}{\omega}\Bigr) \; ,
\end{equation}
and the ``mode functions'' are,
\begin{equation}
\varepsilon(t) \equiv \sqrt{\frac{\hbar}{2 m \omega}} \, e^{-i \omega t}
\qquad {\rm and} \qquad \varepsilon^*(t) \equiv \sqrt{\frac{\hbar}{2 m \omega}}
\, e^{i \omega t} \; . \label{mf1}
\end{equation}

The harmonic oscillator is so important because of an amazing fact: {\it the
spatial Fourier components of every quantum field degenerate to independent
harmonic oscillators in the limit that interactions vanish.} For example,
if we turn off the current density in (\ref{genB}) then the spatial Fourier
transform of the magnetic field vector $\widetilde{\vec{B}}(t,\vec{k})$
behaves as a pair of independent harmonic oscillators for each wave vector
$\vec{k}$. The resulting free field mode sum can be rendered thus,
\begin{equation}
\vec{B}(t,\vec{x}) = \int \!\! \frac{d^3k}{(2\pi)^3} \sum_{\lambda = \pm}
\Biggl\{ i\vec{k} \times \vec{\varepsilon}\Bigl(t,\vec{x};\vec{k},\lambda\Bigr)
\, a(\vec{k},\lambda) - i\vec{k} \times \vec{\varepsilon}^{~*}\!\Bigl(t,\vec{x};
\vec{k},\lambda\Bigr) \, a^{\dagger}(\vec{k},\lambda)\Biggr\} . \label{Bmodesum}
\end{equation}
In this expression the canonically normalized mode coordinates are,
\begin{eqnarray}
a(\vec{k},\lambda) & \equiv & i \sqrt{\frac{\epsilon_0}{4 \hbar c k}} \,
\Bigl[\widehat{\theta} \!-\! i \lambda \widehat{\phi}\Bigr] \cdot \Bigl[c
\widehat{r} \!\times\! \widetilde{\vec{B}}_0(\vec{k}) - \widetilde{\vec{E}}_0(
\vec{k})\Bigr] \; , \\
a^{\dagger}(\vec{k},\lambda) & \equiv & -i \sqrt{\frac{\epsilon_0}{4\hbar c k}}
\, \Bigl[\widehat{\theta} \!+\! i \lambda \widehat{\phi}\Bigr] \cdot \Bigl[c
\widehat{r} \!\times\! \widetilde{\vec{B}}_0(-\vec{k}) - \widetilde{\vec{E}}_0(
-\vec{k})\Bigr] \; .
\end{eqnarray}
The associated mode functions are,
\begin{equation}
\vec{\varepsilon}\Bigl(t,\vec{x};\vec{k},\lambda\Bigr) \equiv
\sqrt{\frac{\hbar}{4 \epsilon_0 c k}} \, \Bigl[ \widehat{\theta} + i \lambda
\widehat{\phi}\Bigr] e^{-i k c t + i \vec{k} \cdot \vec{x}} \; . \label{mf2}
\end{equation}
And it should be noted that I have expressed the wave number
$\vec{k}$ in spherical coordinates with the usual spherical unit
vectors,
\begin{eqnarray}
\widehat{r} & \equiv & \Bigl(\sin(\theta) \cos(\phi), \sin(\theta) \sin(\phi),
\cos(\theta)\Bigr) \; , \\
\widehat{\theta} & \equiv & \Bigl(\cos(\theta) \cos(\phi), \cos(\theta)
\sin(\phi), -\sin(\theta)\Bigr) \; , \\
\widehat{\phi} & \equiv & \Bigl(-\sin(\phi), \cos(\phi),0\Bigr) \; .
\end{eqnarray}

In quantum electrodynamics acting $a^{\dagger}(\vec{k},\lambda)$ on a state
adds a photon with energy $E = \hbar k c$, 3-momentum $\vec{p} = \hbar
\vec{k}$ and polarization $\lambda$. ($\lambda = +1$ stands for left-handed,
circular polarization and $\lambda = -1$ is right-handed.) Acting
$a(\vec{k},\lambda)$ would remove a photon with the same quantum numbers.
It should be noted that the ``particles'' of fundamental theory are always
the Fourier modes of quantum fields. For example, the electrons and positrons
of quantum electrodynamics are represented by terms in the free field mode sum
for the electron field,
\begin{equation}
\Psi_i(t,\vec{x}) = \int \!\! \frac{d^3k}{(2\pi)^3} \sum_{s = \pm \frac12}
\Biggl\{ \varepsilon_i\Bigl(t,\vec{x};\vec{k},s\Bigr) \, b(\vec{k},s) +
\overline{\varepsilon}_i\Bigl(t,\vec{x};\vec{k},s\Bigr) \,
c^{\dagger}(\vec{k},s)\Biggr\} . \label{Psimodesum}
\end{equation}
Acting $c^{\dagger}(\vec{k},s)$ on a state adds a positron of energy $E =
\sqrt{(m c^2)^2\!+\!(\hbar c k)^2}$, 3-momentum $\vec{p} = \hbar \vec{k}$ and
$z$-component spin (in its rest frame) $s = \pm \frac12$. Acting $b(\vec{k},s)$
on a state removes an electron with the same 4-momentum and spin. As always,
the mode coordinates are simply linear combinations of the initial values
of the dynamical variable, which are the true degrees of freedom of system,
\begin{equation}
b(\vec{k},s) \equiv \sqrt{\frac{c}{2 E}} \sum_{i=1}^4 u^*_i(\vec{k},s)
\widetilde{\Psi}_i(0,\vec{k}) \quad , \quad c^{\dagger}(\vec{k},s)
\equiv \sqrt{\frac{c}{2 E}} \sum_{i=1}^4 v^*_i(\vec{k},s)
\widetilde{\Psi}_i(0,\vec{k}) \; .
\end{equation}
Here the spinor wave functions are,
\begin{eqnarray}
u(\vec{k},s) & \equiv & \sqrt{\frac{\hbar c}{2 (E \!+\! m c^2)}}
\left(\matrix{ \Bigl[\frac{E + m c^2}{\hbar c} \, I \!-\! \vec{k} \cdot
\vec{\sigma}\Bigr] \xi(s) \cr \Bigl[\frac{E + m c^2}{\hbar c} \, I \!+\!
\vec{k} \cdot \vec{\sigma}\Bigr] \xi(s)}\right) \; , \label{uspin} \\
v(\vec{k},s) & \equiv & \sqrt{\frac{\hbar c}{2 (E \!+\! m c^2)}}
\left(\matrix{ \Bigl[\frac{E + m c^2}{\hbar c} \, I \!-\! \vec{k} \cdot
\vec{\sigma}\Bigr] \eta(s) \cr -\Bigl[\frac{E + m c^2}{\hbar c} \, I \!+\!
\vec{k} \cdot \vec{\sigma}\Bigr] \eta(s)}\right) \; ,
\end{eqnarray}
and the various 2-component quantities are familiar from nonrelativistic
quantum mechanics,
\begin{eqnarray}
I \equiv \left(\matrix{1 & 0 \cr 0 & 1}\right) \; , \;
\sigma^1 = \left(\matrix{0 & 1 \cr 1 & 0}\right) \; , \;
\sigma^2 = \left(\matrix{0 & -i \cr i & 0}\right) \; , \;
\sigma^3 = \left(\matrix{1 & 0 \cr 0 & -1}\right) \; , \\
\xi\Bigl(+\frac12\Bigr) \equiv \left(\matrix{1 \cr 0}\right) \; , \;
\xi\Bigl(-\frac12\Bigr) \equiv \left(\matrix{0 \cr 1}\right) \; , \;
\eta\Bigl(+\frac12\Bigr) \equiv \left(\matrix{0 \cr 1}\right) \; , \;
\eta\Bigl(-\frac12\Bigr) \equiv \left(\matrix{-1 \cr 0}\right) \; .
\end{eqnarray}
The mode functions in (\ref{Psimodesum}) are,
\begin{equation}
\varepsilon_i\Bigl(t,\vec{x};\vec{k},s\Bigr) \equiv \sqrt{\frac{\hbar^2 c}{2E}}
\, e^{-iE t/\hbar +i \vec{k} \cdot \vec{x}} \, u_i(\vec{k},s) \quad , \quad
\overline{\varepsilon}_i \equiv \sqrt{\frac{\hbar^2 c}{2E}} \, e^{iE t/\hbar -
i\vec{k} \cdot \vec{x}} \, v_i(\vec{k},s) \; . \label{Psimodefunc}
\end{equation}

At this point I must discuss a little about quantum states. Operators such
as $q_0$ and $\dot{q}_0$ have the potential for being anything; it is the
state wave function which describes how they are distributed. These wave
functions are time independent in the Heisenberg Picture of quantum mechanics;
it is the operators which evolve in time. In the position representation
we can write the wave function as $\psi(x)$ and the two fundamental operators
act by multiplication and differentiation,
\begin{eqnarray}
q_0 \, \psi(x) & = & x \, \psi(x) \; , \\
\dot{q}_0 \, \psi(x) & = & -\frac{i \hbar}{m} \, \psi'(x) \; .
\end{eqnarray}
The inner product between any two states is defined by integration,
\begin{equation}
\Bigl\langle \phi \Bigl\vert \psi \Bigr\rangle \equiv \int_{-\infty}^{\infty}
\!\!\! dx \, \phi^*(x) \psi(x) \; .
\end{equation}
And a very important property of any state is {\it normalization},
\begin{equation}
\Bigl\langle \psi \Bigl\vert \psi \Bigr\rangle \equiv \int_{-\infty}^{\infty}
\!\!\! dx \, \psi^*(x) \psi(x) = 1 \; . \label{normalize}
\end{equation}
(This is what puts the ``quantum'' in quantum mechanics.) The expectation
value of any operator $\mathcal{O}(q_0,\dot{q}_0)$ in the state $\vert
\psi\rangle$ is,
\begin{equation}
\Bigl\langle \psi \Bigl\vert \mathcal{O}\Bigl(q_0,\dot{q}_0\Bigr) \Bigr\vert
\psi \Bigr\rangle = \int_{-\infty}^{\infty} \!\!\! dx \, \psi^*(x) \,
\mathcal{O}\Bigl(x,-\frac{i \hbar}{m} \frac{\partial}{\partial x}\Bigr) \,
\psi(x) \; . \label{VEV}
\end{equation}
Of course expression (\ref{VEV}) suffices for time dependent operators
such as $q(t)$ because they can be expressed in terms of $q_0$ and
$\dot{q}_0$.

The notions of state wave functions, and inner products involving them, all
have straightforward generalizations to quantum field theory (if one is good
with functional calculus!) however, they are very seldom used. The reason for
this is that quantum field theories possess an infinite number of mode
coordinates, one or more for every wave vector $\vec{k}$. Only a finite
number of these modes can be excited because it costs energy to excite a
mode and there is only a limited amount of free energy available. Hence there
are so many more ground state modes than excited ones that most quantum field
theoretic effects derive from the vast number of modes in their ground states.

The Uncertainty Principle provides a powerful, intuitive way of using
classical physics to understand the effects of modes which are in their
ground states. When the expectation values of $q_0$ and $\dot{q}_0$ are
zero we can express the Uncertainty Principle as follows,
\begin{equation}
\langle \psi \vert q_0^2 \vert \psi \rangle \cdot
\langle \psi \vert \dot{q}_0^2 \vert \psi \rangle
\geq \frac{\hbar^2}{4 m^2} \; .
\end{equation}
Equality is achieved for a {\it Minimum Uncertainty} state. For such a
state we can think of the Hamiltonian as a function of $q_0$ alone,
\begin{equation}
H \equiv \frac12 m \dot{q}_0^2 + \frac12 m \omega^2 q_0^2 \longrightarrow
\frac{\hbar^2}{8 m q^2_0} + \frac12 m \omega^2 q^2_0 \equiv E(q_0) \; .
\end{equation}
The term $\hbar^2/8m q_0^2$ in $E(q_0)$ is known as {\it Uncertainty
Pressure}. It reflects the physical import of the Uncertainty Principle,
which is that concentrating $q_0$ more tightly about its mean value of zero
makes $\dot{q}_0$ proportionately less concentrated. With Uncertainty
Pressure we can understand many quantum effects classically. For example,
the minimum energy is,
\begin{equation}
\frac{\partial E}{\partial q_0} = -\frac{\hbar^2}{4 m q_0^3} + m \omega^2
q_0 \qquad \Longrightarrow \qquad q_{\rm min} = \sqrt{\frac{\hbar}{2 m \omega}}
\;\; {\rm and} \;\; E_{\rm min} = \frac12 \hbar \omega \; .
\end{equation}
One doesn't always get the factors of two right with this level of analysis
but it is a powerful technique for understanding complex things in a simple
way, and I will apply it to explain where the divergences of quantum gravity
arise and why they are worse than those associated with the other forces.

\subsection{Perturbation Theory}

The alert reader will have noted that the exact solutions of the previous
subsections were all obtained for noninteracting theories. There is a good
reason for that: {\it not a single interacting field theory has been solved
in four spacetime dimensions.} Note the essential distinction between
``exactly solving a theory'', which means expressing the dynamical variable
at any time in term of arbitrary initial value data, and obtaining an ``exact
solution to the field equations,'' which means solving the equations of
motion for one particular choice of initial values. The various ``exact
solutions'' of Einstein's equations involve setting almost all the degrees
of freedom to zero. This is fine for classical physics, but it is not
permitted in quantum mechanics. For example, in free quantum electrodynamics
the mode coordinates $a(\vec{k},\lambda)$ and $a^{\dagger}(\vec{k},\lambda)$
of expression (\ref{Bmodesum}) are not commuting variables,
\begin{equation}
\Bigl[a(\vec{k},\lambda) , a^{\dagger}(\vec{k}' , \lambda')\Bigr] =
\delta_{\lambda \lambda'} (2\pi)^3 \delta^3( \vec{k} \!-\! \vec{k}') \; .
\end{equation}
A classical picture is possible, but one must imagine that each mode
experiences the 0-point motion we found above for the harmonic oscillator.

Because there is no hope of exactly solving the field equations for arbitrary
initial value data, computing anything in quantum field theory requires the
use of approximation techniques. The standard one is known as {\it
perturbation theory} and a good way of motivating it is by observing that,
even though our expression (\ref{genB}) is correct, it doesn't give the
magnetic field because the current density $\vec{J}(t,\vec{x})$ is affected
by the electrodynamic potentials. One can see this in quantum electrodynamics
for which the charge density and the current density are formed from the
electron field $\Psi_i(t,\vec{x})$,
\begin{equation}
\rho(t,\vec{x}) = \frac{e}{\hbar} \Psi^{\dagger}(t,\vec{x}) \Psi(t,\vec{x})
\quad , \quad \vec{J}(t,\vec{x}) = \frac{e c}{\hbar} \Psi^{\dagger}(t,\vec{x})
\left(\matrix{-\vec{\sigma} & 0 \cr 0 & \vec{\sigma}}\right) \Psi(t,\vec{x})
\; .
\end{equation}
And the Dirac equation for $\Psi$ involves the potentials $\Phi$ and $\vec{A}$,
\begin{equation}
\Bigl[ \frac{\partial}{\partial ct} \!-\! \frac{i e \Phi}{\hbar c} \Bigr] \Psi
+ \left(\matrix{-\vec{\sigma} & 0 \cr 0 & \vec{\sigma}}\right) \cdot
\Bigl(\vec{\nabla} \!+\! \frac{ie \vec{A}}{\hbar}\Bigr) \Psi + \frac{i mc}{
\hbar} \left(\matrix{0 & I \cr I & 0}\right) \Psi = 0 \; . \label{Dirac}
\end{equation}
One can see what must happen, even without working out the details:
\begin{itemize}
\item{0-point motions of $\Psi$ modes engender $\Phi$ and $\vec{A}$;}
\item{These electromagnetic fields change $\Psi$;}
\item{Which changes $\Phi$ and $\vec{A}$;}
\item{Which changes $\Psi$, and so on.}
\end{itemize}
The progression is endless, and we could just have easily begun it with
0-point motions of the electromagnetic fields affecting $\Psi$! The process
could be made to terminate if an external force fixed $\Psi$, or if we made
some very special choice of initial values. But there are no external forces
in fundamental theory, and quantum mechanics demands that we consider generic
initial value data.

The cycle of action and reaction I have sketched precludes obtaining the
exact solution, but a good approximate solution can be found if each cycle of
action and reaction induces a smaller subsequent cycle. So we start with
noninteracting, charged matter and electrodynamics, which defines the 0th
order. In quantum electrodynamics the 0th order would be the free field mode
sums (\ref{Psimodesum}) and the analogue of (\ref{Bmodesum}) for the vector
and scalar potentials. The 1st order perturbation for the electrodynamic
potentials comes from solving Maxwell equations (\ref{Max1}-\ref{Max2})
with the charge density and current density formed from the fixed, 0th order
electron field (\ref{Psimodesum}). The 1st order perturbation for the
electron field $\Psi^1$ comes solving the Dirac equation (\ref{Dirac}) with
the terms involving the electron charge $e$ evaluated at 0th order,
\begin{equation}
\frac{\partial \Psi^1}{\partial ct} + \left(\matrix{-\vec{\sigma} & 0 \cr 0
& \vec{\sigma}}\right) \cdot \vec{\nabla} \Psi^1 + \frac{i mc}{\hbar}
\left(\matrix{0 & I \cr I & 0}\right) \Psi^1 =
\frac{i e \Phi^0 \Psi^0}{\hbar c} - \left(\matrix{-\vec{\sigma} & 0 \cr
0 & \vec{\sigma}}\right) \cdot \frac{ie \vec{A}^0}{\hbar} \Psi^0 \; .
\end{equation}
The second order perturbation is obtained by computing what the first order
fields do, and so on. If the effect of each cycle is reduced by a small enough
factor then we do not need to carry the process out for many cycles before the
theoretical prediction is as accurate as any experiment that can be performed.
This is how perturbation theory works.

It should be clear from the sketch I have given that each order of quantum
electrodynamic perturbation theory comes with an extra power of the electron
charge $e$. It should also be clear that the effects of different lower order
modes add in perturbation theory. In a fixed volume $V$ one sums over the
Fourier wave vectors $\vec{k}$ using the famous density of states formula,
\begin{equation}
\sum_{\vec{k}} = V \int \!\! \frac{d^3k}{(2\pi)^3} \; . \label{dofs}
\end{equation}
Because we typically want densities, the factor of $V$ drops out. Because
there are as many positron modes as electron modes, with the same kinematic
properties and opposite charge, it turns out that the odd powers of $e$
cancel out and the net effect comes from even powers. It also happens that
one always gets a certain number of $2\pi$'s from (\ref{dofs}), and some
factors of $\epsilon_0$, $\hbar$ and $c$ in the dimensionless combination,
\begin{equation}
\frac{e^2}{4\pi^2 \epsilon_0 \hbar c} \equiv \frac{\alpha}{\pi} \simeq
\frac1{430} \; . \label{alpha}
\end{equation}
This is the expansion parameter of quantum electrodynamics and its smallness
is what makes perturbation theory effective.

I have glossed over quite a number of details in describing how perturbation
theory works for quantum electrodynamics, and I am not going to be any more
thorough for quantum general relativity. However, the coupling constant
is important because it controls how reliable perturbation theory ought to
be. One can see from the Einstein equation (\ref{Einstein}) that the source
terms involve $G/c^4$, so one more power of this must obviously appear at
each successive order in perturbation theory. Each higher order in
perturbation theory also contributes a factor of $1/\hbar c$, just as in
quantum electrodynamics (and all other quantum field theories). The resulting
combination of fundamental constants has the dimension of an inverse energy
squared,
\begin{equation}
\frac{G}{\hbar c^5} \simeq \Bigl(\frac1{2.0 \times 10^9~{\rm J}}\Bigr)^2
\simeq \Bigl(\frac1{1.2 \times 10^{19}~{\rm GeV}}\Bigr)^2 \; .
\end{equation}
Therefore quantum gravitational perturbation theory must amount to an
expansion in the dimensionless parameter $G E^2/\hbar c^5$, where $E$ is
some energy in whatever process is under study.

An issue which sometimes confuses people is that the series approximation
perturbation theory generates is only asymptotic, rather than convergent.
Suppose we have function $S(x)$ and we obtain an $N$-term series approximation
$S_N(x)$ in terms of some standard set of functions $f_n(x)$ which are
organized so that, for the $x$-range of interest,
\begin{equation}
f_0(x) > f_1(x) > f_2(x) > f_3(x) > \ldots \; .
\end{equation}
The typical case is that $f_n(x) = x^n$ but I want to allow for fractional
powers or powers times logarithms. The series approximation $S_N(x)$ takes
the form,
\begin{equation}
S_N(x) = \sum_{n=0}^{N-1} s_n f_n(x) \; ,
\end{equation}
where the $s_n$ are numbers. We say that the series is {\it convergent} if
taking $N$ to infinity recovers the original function $S(x)$,
\begin{equation}
{\rm Convergent:} \qquad \Longrightarrow \qquad \lim_{N \rightarrow \infty}
\Bigl[S(x) \!-\! S_N(x) \Bigr] = 0 \; .
\end{equation}
We say the series is {\it asymptotic} at $x = x_0$ if the difference between
$S(x)$ and $S_N(x)$ goes to zero faster than $f_{N-1}(x)$ as $x$ approaches
$x_0$, for any fixed $N$,
\begin{equation}
{\rm Asymptotic:} \qquad \Longrightarrow \qquad \lim_{x \rightarrow x_0}
\Biggl[\frac{S(x) \!-\! S_N(x)}{f_{N-1}(x)} \Biggr] = 0 \; .
\end{equation}
It is entirely possible for a series to possess both properties but it is a
fact that perturbative expansions of quantum field theory are only asymptotic
(for zero coupling constant), not convergent.

The difference between a convergent series and an asymptotic series is
beautifully illustrated by a special function known as the ``Exponential
Integral'' \cite{AS},
\begin{equation}
{\rm E}_1(x) \equiv \int_{x}^{\infty} \!\!\!\! dt \, \frac{e^{-t}}{t} \; .
\label{E1}
\end{equation}
A convergent series for small $x$ can be obtained by extracting the integral
down to $x=1$ as a constant, and then adding zero in the form of $-\ln(x) +
\ln(x)$,
\begin{equation}
{\rm E}_1(x) = \int_1^{\infty} \!\!\! dt \, \frac{e^{-t}}{t} + \int_0^1 \!\!\!
dt \, \Bigl[ \frac{e^{-t} \!-\! 1}{t}\Bigr] - \ln(x) - \int_0^{x} \!\!\! dt \,
\Bigl[\frac{e^{-t} \!-\! 1}{t}\Bigr] \; . \label{rearrange}
\end{equation}
The first two integrals of (\ref{rearrange}) just give minus the
Euler-Mascheroni constant $\gamma \approx .577$. The final integral of
(\ref{rearrange}) can be evaluated by expanding $(e^{-t} \!-\! 1)/t$ in powers
of $t$ and then integrating termwise. The resulting expansion is,
\begin{equation}
{\rm E}_1(x) = - \gamma - \ln(x) - \sum_{n=1}^{\infty} \frac{(-1)^n x^n}{n
\cdot n!} \; . \label{smallx}
\end{equation}
Of course this series is also asymptotic for $x \rightarrow 0$.

One can get a asymptotic expansion of ${\rm E}_1(x)$ for large $x$ by
successive partial integrations,
\begin{equation}
{\rm E}_1(x) = \sum_{n=0}^{N-1} \frac{(-1)^n n! \, e^{-x}}{x^{n+1}} + (-1)^N
N! \int_{x}^{\infty} \!\!\! dt \frac{e^{-t}}{t^{N+1}} \; . \label{E1exact}
\end{equation}
The asymptotic series (for $x \rightarrow \infty$) is the first term of
(\ref{E1exact}), the second term is the remainder. Its magnitude can be
bounded by replacing the $1/t^{N+1}$ with $1/x^{N+1}$,
\begin{equation}
R_N \equiv N! \int_{x}^{\infty} \!\!\! dt \frac{e^{-t}}{t^{N+1}} <
\frac{N!}{x^{N+1}} \int_{x}^{\infty} \!\!\! dt \, e^{-t} = \frac{N! \,
e^{-x}}{x^{N+1}} \; .
\end{equation}
Hence the magnitude of the difference between ${\rm E}_1(x)$ and $S_N(x)$ is,
\begin{equation}
\Biggl\vert {\rm E}_1(x) \!-\! \sum_{n=0}^{N-1} \frac{(-1)^n n! \,
e^{-x}}{x^{n+1}} \Biggr\vert < \frac{N! e^{-x}}{x^{N+1}} \; .
\end{equation}
This proves the series is asymptotic as $x \rightarrow \infty$. To see that
it is not convergent, note that the sum of $R_N$ and $R_{N+1}$ can be
evaluated exactly,
\begin{equation}
R_N + R_{N+1} = -N! \int_{x}^{\infty} \!\!\! dt \, \frac{\partial}{\partial t}
\Bigl[ \frac{e^{-t}}{t^{N+1}} \Bigr] = \frac{N! \, e^{-x}}{x^{N+1}} \; .
\label{bound}
\end{equation}
Increasing $N$ by one changes the right hand side of (\ref{bound}) by a factor
of $(N\!+\!1)/x$. As long as this is less than one, accuracy is increased, but
for any fixed value of $x$ there must eventually be an $N$ past which
accuracy deteriorates. The best asymptotic series approximation for
${\rm E}_1(x)$ is therefore obtained by carrying the expansion out to about
$N \sim 1/x$ and no further.

The fact that nonconvergent asymptotic series expansions cannot be made
arbitrarily accurate bothers the mathematically inclined. It does preclude
defining a quantum field theory by its perturbative expansion, however,
whether or not there is any practical problem depends upon the value of the
coupling constant and the state of the experimental art. If the coupling
constant is so small that no experiment can measure the deviation between
reality and the best asymptotic series result then there is no operational
problem. That is the case for quantum electrodynamics at conventional energy
scales. A process which involves $2N$ vertices will acquire a factor of
$(\alpha/2\pi)^N$ from its coupling constants and the factors of $2$ and $\pi$
from momentum integrations. There will also be a multiplicity factor of about
$(2N \!-\! 1)!!$, so one expects the series to begin diverging at about $N
\sim \pi/\alpha \sim 430$. At that point the accuracy is so great one can
only estimate it using Stirling's approximation,
\begin{equation}
\Bigl(\frac{\alpha}{2\pi}\Bigr)^N \times (2N \!-\! 1)!! \longrightarrow
\sqrt{2} \Bigl(\frac{\alpha N}{e \pi}\Bigr)^{N} \sim 10^{-187} \; .
\end{equation}
No experiment can approach that accuracy; the current state of the art is
sensitive to about the {\it fourth} order in $\alpha$, not the {\it four
hundredth}!

On the other hand, the coupling constant of the strong interaction is large
enough, at low energies, that the asymptotic expansion is practically
worthless. So how about quantum general relativity? We have seen that
perturbation theory generates an expansion in powers of $G E^2/\hbar c^5$.
For the highest proton energy we shall be able to reach at the LHC this number
would be about,
\begin{equation}
\frac{G E^2}{\hbar c^5} \sim \Bigl(\frac{7 \times 10^3~{\rm GeV}}{1.2 \times
10^{19}~{\rm GeV}}\Bigr)^2 \sim 3.4 \times 10^{-31} \; . \label{LHC}
\end{equation}
This means the perturbative series for quantum general relativity
should be wonderfully more accurate than that of quantum
electrodynamics, for which analogous factor is $\alpha/2\pi \sim 1.2
\times 10^{-3}$. In fact it should be so good that some special
circumstance would be needed to make quantum gravitational effects
observable at all. I will have more to say about this later but let
us for now note that the asymptotic nature of perturbative results
is not high on the list of problems for quantum gravity.

\subsection{Renormalization}

It turns out that the perturbative corrections from any 4-dimensional
quantum field theory diverge when they are expressed in terms of the
parameters such as $e$ and $m$ which appear in the field equations. The
reason for this divergence is very simple: all the modes contribute a
little bit, and there are so many modes that one gets a divergence from
the sum (\ref{dofs}) over them. For all the other quantum field theories
these divergences can be absorbed by regarding the parameters of the field
equations to depend in a divergent way upon physically measured quantities
such as the electron charge and mass. When that is done correctly the
perturbative corrections really are small, if the coupling constant is,
and they agree wonderfully with experiment.

The procedure for using the parameters of the field equations to absorb
divergences is known as {\it renormalization}. I will describe how it
works by showing that the vacuum polarization of quantum electrodynamics
is completely analogous to the phenomenon of classical polarization in a
medium. Then I will use the fact that gravity couples to stress-energy,
rather than charge, to show that renormalizing the divergences of quantum
gravity involves adding new sorts of terms to the gravitational field
equations. These new terms would remove all the divergences \cite{KS},
but we will see in the next subsection that they would also make the
universe blow up. That is the central problem of perturbative quantum
general relativity.

Let us consider the phenomenon of polarization in a static, classical
medium. The actual charge density of the medium consists of an enormous
number of positive and negative point charges $q_{\alpha}$ located at
equilibrium positions $\vec{X}_{\alpha}$,
\begin{equation}
\rho(\vec{x}) \Bigl\vert_{\rm undisturbed \atop medium} = \sum_{\alpha}
q_{\alpha} \delta^3(\vec{x} \!-\! \vec{X}_{\alpha}) \; .
\end{equation}
The medium as a whole is electrically neutral; it is only
microscopically that one can see its vast collection of positive and
negative charges. If we apply a static electric field
$\vec{E}(\vec{x})$, the charges shift to new equilibrium positions
$\vec{X}_{\alpha} \rightarrow \vec{X}_{\alpha} + \Delta
\vec{x}_{\alpha}$. We can expand the density of each charge around
its equilibrium value,
\begin{equation}
q_{\alpha} \delta^3\Bigl(\vec{x} \!-\! \vec{X}_{\alpha} \!-\!
\Delta \vec{x}_{\alpha}\Bigr) = q_{\alpha} \delta^3(\vec{x} \!-\!
\vec{X}_{\alpha}) - \vec{\nabla} \cdot \Bigl[ q_{\alpha} \Delta
\vec{x}_{\alpha} \delta^3(\vec{x} \!-\! \vec{X}_{\alpha}) \Bigr] + \ldots
\end{equation}
Hence we can write the charge density of the disturbed medium as its
undisturbed value plus a series of corrections,
\begin{equation}
\rho(\vec{x}) \Bigl\vert_{\rm disturbed \atop medium} =
\rho(\vec{x}) \Bigl\vert_{\rm undisturbed \atop medium} - \vec{\nabla} \cdot
\Biggl[\sum_{\alpha} q_{\alpha} \Delta \vec{x}_{\alpha} \delta^3(\vec{x} \!-\!
\vec{X}_{\alpha}) \Biggr] + \ldots \label{exprho}
\end{equation}

We have already noted that the undisturbed charge density appears to be
zero on macroscopic scales because the positive and negative charges cancel
one another. However, the sum in the square bracketed term on the right hand
side of (\ref{exprho}) tends to give a coherent effect because the positive
charges move with the applied electric field and the negative charges move
the other way. The higher terms in the expansion tend to be small because the
charges don't move much, so the result is,
\begin{equation}
\rho(\vec{x}) \Bigl\vert_{\rm disturbed \atop medium} \simeq
- \vec{\nabla} \cdot \vec{P}(\vec{x}) \; ,
\end{equation}
where the {\it polarization} is,
\begin{equation}
\vec{P}(\vec{x}) \equiv \sum_{\alpha} q_{\alpha} \Delta \vec{x}_{\alpha}
\delta^3(\vec{x} \!-\! \vec{X}_{\alpha}) \; .
\end{equation}

Now suppose we add a small number of ``free'' charges to the vast collection
of ``bound'' ones in the medium. The Gauss law equation reads,
\begin{equation}
\epsilon_0 \vec{\nabla} \cdot \vec{E} \simeq \rho_{\rm free} - \vec{\nabla}
\cdot \vec{P} \; .
\end{equation}
The smart way to solve this equation is to combine $\vec{P}$ with
$\epsilon_0 \vec{E}$ to form the electric displacement $\vec{D}$,
\begin{equation}
\vec{\nabla} \cdot \Bigl( \epsilon_0 \vec{E} + \vec{P} \Bigr) \equiv
\vec{\nabla} \cdot \vec{D} = \rho_{\rm free} \; .
\end{equation}
For the case of a linear, isotropic medium the polarization is proportional
to the electric field,
\begin{equation}
\vec{P}(\vec{x}) = \Delta \epsilon \times \vec{E}(\vec{x}) \; . \label{liniso}
\end{equation}
In that case we can subsume the effect of the medium into a change of the
electric permittivity,
\begin{equation}
\epsilon \vec{\nabla} \cdot \vec{E} = \rho_{\rm free} \qquad {\rm where}
\qquad \epsilon \equiv \epsilon_0 + \Delta \epsilon \; . \label{rhofree}
\end{equation}

Let us now switch from classical physics with a medium to quantum
electrodynamics in empty space. Of course the space can never really be
empty because it is pervaded by the electron field $\Psi_i(t,\vec{x})$,
which gives the charge density of quantum electrodynamics,
\begin{equation}
\rho(t,\vec{x}) = \frac{e}{\hbar} \, \Psi^*_i(t,\vec{x}) \Psi_i(t,\vec{x})
\; .
\end{equation}
As we saw in expression (\ref{Psimodesum}), the electron field consists
of a an infinite collection of the operators which create and destroy
charged particles. We also saw that it is possible to think about this sum
of operators classically provided one imagines each mode to be executing
0-point motion. When an electric field is applied these 0-point motions
change, and that produces an observable, coherent effect, just as it does
for the classical medium and for the same reasons.

From the free field expansion (\ref{Psimodesum}) one can see that the
spacetime dependence of 0-point motion is characterized by mode functions
(\ref{Psimodefunc}). The part of interest to us is the oscillatory factor
on the electron creation operator $b^{\dagger}(\vec{k},s)$ in
$\Psi^*(t,\vec{x})$,
\begin{equation}
e^{i Et/\hbar - i \vec{k} \cdot \vec{x}} \qquad {\rm where} \qquad
E = \sqrt{ (m c^2)^2 + (\hbar c k)^2} \; . \label{Ener1}
\end{equation}
We might cancel the spatial phase by combining this with a positron
creation operator of opposite momentum and spin in $\Psi(t,\vec{x})$,
\begin{equation}
e^{iE t/\hbar - i \vec{k} \cdot \vec{x}} b^{\dagger}(\vec{k},s) \times
e^{iE t/\hbar + i \vec{k} \cdot \vec{x}} c^{\dagger}(-\vec{k},-s) \; ,
\label{Ener2}
\end{equation}
but nothing can be done about the temporal phase factor. This temporal
phase factor means that effects from this ``virtual'' electron-positron
pair cannot remain coherent longer than about $\Delta t \sim \hbar/E$.
This is a very short time; the longest lived mode is the one with $\vec{k}
= 0$,
\begin{equation}
\Bigl( \Delta t\Bigr)_{\vec{k} = 0} \sim \frac{\hbar}{m c^2} \sim
\frac{10^{-34}~{\rm J\!-\!s}}{(9 \!\times\! 10^{-31}~{\rm kg})
(3 \!\times\! 10^8~{\rm m/s})^2} \sim 10^{-22}~{\rm s} \; . \label{short}
\end{equation}

Let us compute the polarization induced by a virtual pair of wave number
$\vec{k}$. As we have seen, quantum physics tells us they effectively exist
for a time $\Delta t \sim \hbar/E$, but the rest of the analysis is completely
classical. The equation of motion for a charge $e$ acted upon by an electric
field $\vec{E}$ is,
\begin{equation}
\frac{d}{d t} \Biggl[\frac{m \vec{v}}{\sqrt{1 \!-\! v^2/c^2}} \Biggr]
= e \vec{E} \; .
\end{equation}
Over an interval as short as (\ref{short}) we can regard the energy of the
charge as constant,
\begin{equation}
\frac{m}{\sqrt{1 \!-\! v^2/c^2}} \simeq \frac{E}{c^2} \; .
\end{equation}
We can also forget about the variation in $\vec{E}(\vec{x})$ as the particle
moves, so the induced deviation in time $\Delta t = \hbar/E$ is,
\begin{equation}
\Delta \vec{x} \simeq \frac{e c^2 \Delta t^2}{2 E} \, \vec{E} =
\frac{e \hbar^2 c^2}{2 E^3} \, \vec{E} \; . \label{Delx}
\end{equation}

It remains only to add the electron and positron contributions, and then
sum over modes to find the total induced polarization,
\begin{equation}
\vec{P}(\vec{x}) = \int \!\! \frac{d^3k}{(2\pi)^3} \Biggl[ e \times
\Bigl(\frac{e \hbar^2 c^2}{2 E^3}\Bigr) \times \vec{E}(\vec{x})- e \times
\Bigl(\frac{-e \hbar^2 c^2}{2 E^3}\Bigr) \times \vec{E}(\vec{x}) \Biggr]
\; . \label{vacP}
\end{equation}
Expression (\ref{vacP}) is proportional to the electric field, just like a
linear, isotropic medium, so we can identify the change in the
permittivity as,
\begin{eqnarray}
\Delta \epsilon & = & e^2 \hbar^2 c^2 \int \!\! \frac{d^3k}{(2\pi)^3}
\frac1{[(m c^2)^2 \!+\!  (\hbar c k)^2]^{\frac32}} \; , \\
& = & \frac{e^2}{2\pi^2 \hbar c} \int_0^{\infty} \!\!\! d(\hbar c k)
\frac{(\hbar c k )^2}{[(m c^2)^2 \!+\! (\hbar c k)^2]^{\frac32}} \; .
\label{Deps}
\end{eqnarray}
Our expression for $\Delta \epsilon$ diverges logarithmically! On the other
hand, this effect is ubiquitous because it comes from vacuum fluctuations,
without any medium being present. That means we will never observe $\Delta
\epsilon$ independently of $\epsilon_0$, only their sum,\footnote{Readers
familiar with quantum field theory will recognize that $\epsilon$ is
proportional to $Z_2$, the fermion field strength renormalization.}
\begin{equation}
\epsilon \equiv \epsilon_0 + \Delta \epsilon \; .
\end{equation}
It is this sum which must be finite, not $\Delta \epsilon$ or $\epsilon_0$
separately. So what particle theorists do is to adjust the parameter in the
field equation $\epsilon_0$ to be conventional value of about $8.85 \times
10^{-12}~{\rm F/m}$ {\it minus} $\Delta \epsilon$. That is how renormalization
works.

Some people find it disconcerting to have a parameter from the field equations
changed when it appears in physical predictions. However, much of this sense
of wrongness derives from insufficient experience with nonlinear systems. In a
linear system, such as electrodynamics becomes when its sources are held fixed,
there is indeed a simple relation between parameters in the equations and
physical predictions. For example, a stationary point charge $q$ induces a
Coulomb field,
\begin{equation}
\rho(t,\vec{x}) = q \delta^3(\vec{x}) \qquad \Longrightarrow \qquad
\vec{E}(t,\vec{x}) = \frac{q \vec{x}}{4\pi \epsilon_0 \Vert \vec{x}\Vert^3}
\; .
\end{equation}
The parameters $q$ and $\epsilon_0$ that enter the classical field equations
are the same ones which appear in the observed long range field. But even
electrodynamics becomes nonlinear when one permits its sources to respond to
electromagnetic fields, and this generally causes the observed quantities to
differ from their cognates in the equations. For example, the conduction
electrons in a metal behave, for many purposes, as if they are free to move
inside the metal, but with a charge and mass different from their values in
empty space.

Renormalization is the hallmark of nonlinear systems, even classical ones.
We encountered it in quantum electrodynamics because that theory is nonlinear,
not because it is quantum mechanical. Similar effects occur in nonlinear
classical systems. For example, the combined mass of the Earth-Moon system
is a little less than the sum of their masses in isolation, owing to their
gravitational interaction energy,
\begin{equation}
-\frac{G M_E M_M}{R_{EM} c^2} \sim -\frac{(7 \times 10^{-11}~\frac{\rm N}{{\rm
m}^2}) (6 \times 10^{24}~{\rm kg}) (7 \times 10^{22}~{\rm kg})}{(4 \times
10^8~{\rm m}) (3 \times 10^8~\frac{\rm m}{\rm s})^2} \sim -8 \times
10^9~{\rm kg} \; .
\end{equation}
Nor is the Earth's mass equal to the sum of the masses of its constituents.
If we imagine it to be a uniform sphere the actual mass is less by about,
\begin{equation}
-\frac{3 G M_E^2}{5 R_E c^2} \sim -\frac{3 (7 \times 10^{-11}~\frac{\rm N}{{
\rm m}^2}) (6 \times 10^{24}~{\rm kg})^2}{5 (6 \times 10^6~{\rm m}) (3 \times
10^8~\frac{\rm m}{\rm s})^2} \sim -3 \times 10^{15}~{\rm kg} \; .
\end{equation}
This ``gravitational renormalization'' effect obviously becomes stronger the
more compact the mass is. Arnowitt, Deser and Misner have shown the
renormalization is actually infinite for a point mass \cite{ADM}.

Some people are resigned to renormalization in principle, but disturbed by
the fact that quantum field theoretic renormalizations involves divergent
quantities. This bothered even the physicists who devised renormalization! 
They eventually accepted it for two reasons:
\begin{itemize}
\item{As I have emphasized, renormalization is inevitable in nonlinear
systems, so we would need to choose $\epsilon_0$ to make the observed quantity
$\epsilon_0 + \Delta \epsilon$ agree with experiment, even if $\Delta
\epsilon$ had been finite; and}
\item{Once this is done, along with the analogous things for the electron
mass and charge, all other quantum electrodynamic corrections are quite
small and in wonderful agreement with experiment.}
\end{itemize}

To see this last point let us return to vacuum polarization and do a better
job of accounting for the spatial variation while still (incorrectly)
ignoring temporal variation. The fundamental field equation is Gauss's law,
\begin{equation}
\epsilon_0 \vec{\nabla} \cdot \vec{E}(t,\vec{x}) = \frac{e}{\hbar} \,
\Psi^*_i(t,\vec{x}) \Psi_i(t,\vec{x}) \; .
\end{equation}
If the electric field is not constant in space we need to compute the
vacuum polarization of each wave vector $\vec{p}$. Taking the spatial Fourier
transform of the source and using Parseval's theorem gives,
\begin{equation}
\int \!\! d^3x \, e^{-i\vec{p} \cdot \vec{x}} \frac{e}{\hbar} \, \Psi^*_i(t,
\vec{x}) \Psi_i(t,\vec{x}) = \frac{e}{\hbar} \int \!\! \frac{d^3k}{(2\pi)^3} \,
\widetilde{\Psi}^*_i(t,\vec{k}) \widetilde{\Psi}(t,\vec{p} \!-\! \vec{k}) \; .
\end{equation}
From the free field mode sum (\ref{Psimodesum}) it is apparent that there are
two energies involved, not the single one of expression (\ref{Ener1}),
\begin{equation}
E(\vec{k}) \equiv \sqrt{(m c^2)^2 + (\hbar c)^2 \Vert \vec{k}\Vert^2}
\quad {\rm and} \quad
E(\vec{p} \!-\! \vec{k}) \equiv \sqrt{(m c^2)^2 + (\hbar c)^2 \Vert \vec{p}
\!-\! \vec{k}\Vert^2} \; .
\end{equation}
In place of expression (\ref{Ener2}) we should expect the two operators to
contribute as follows,
\begin{eqnarray}
\widetilde{\Psi}^*(t,\vec{k}) & \longrightarrow & \exp\Bigl[\frac{i E(\vec{k})
\, t}{\hbar}\Bigr] \times b^{\dagger}(\vec{k},s) \; , \\
\widetilde{\Psi}(t,\vec{p} \!-\! \vec{k}) & \longrightarrow &
\exp\Bigl[\frac{i E(\vec{p} \!-\! \vec{k}) \, t}{\hbar}\Bigr] \times
c^{\dagger}(\vec{p} \!-\! \vec{k},-s) \; .
\end{eqnarray}

Now recall that the induced displacement we computed in (\ref{Delx})
involved the inverse third power of ``the'' energy. As we saw, there are
really two energies and it turns out that it is their sum which appears
in the full quantum field theoretic result. There is also a factor of
$\frac83$, so we can write the induced polarization as,
\begin{equation}
\widetilde{\vec{P}}(\vec{p}) = \frac83 e^2 \hbar^2 c^2
\int \!\! \frac{d^3k}{(2\pi)^3} \, \frac1{[E(\vec{k}) \!+\! E(\vec{p}
\!-\! \vec{k})]^3} \times \widetilde{\vec{E}}(\vec{p}) \; .
\end{equation}
We can write the position space result in terms of a momentum dependent
shift in the permittivity,
\begin{equation}
\vec{P}(\vec{x}) = \int \!\! \frac{d^3p}{(2 \pi)^3} \, e^{i\vec{p} \cdot
\vec{x}} \Delta \epsilon(\vec{p}) \!\! \int \!\! d^3x' \, e^{-i \vec{p} \cdot
\vec{x}'} \vec{E}(\vec{x}') \; , \label{trueP}
\end{equation}
where the momentum dependent permittivity shift is,
\begin{equation}
\Delta \epsilon(\vec{p}) = \frac{8 e^2}{3\hbar c} \int\!\!
\frac{d^3k}{(2\pi)^3} \, \frac{\hbar^3 c^3}{[E(\vec{k}) \!+\! E(\vec{p}
\!-\! \vec{k})]^3} \; . \label{kint}
\end{equation}

One way of understanding (\ref{kint}) is by expanding the energy denominator
around $\vec{p} = 0$,
\begin{equation}
E(\vec{k}) \!+\! E(\vec{p} \!-\! \vec{k}) = 2 E(\vec{k}) \Biggl[1 -
\frac{\hbar^2 c^2 \vec{k} \!\cdot\! \vec{p}}{2 E^2(\vec{k})} +
\frac{\hbar^2 c^2 p^2}{4 E^2(\vec{k})} - \frac{\hbar^4 c^4 (\vec{k} \!\cdot\!
\vec{p})^2}{4 E^4(\vec{k})} + \ldots \Biggr] \label{pexp}
\end{equation}
Substituting just the first term of (\ref{pexp}) into (\ref{kint}) gives
($\frac13 \times$) the logarithmically divergent expression (\ref{Deps}) we
called $\Delta \epsilon$. The contributions from all the higher terms of the
expansion (\ref{pexp}) are finite. Using some mathematical methods that were
developed for quantum field theory they can be evaluated to give,
\begin{equation}
\Delta \epsilon_{\rm fin}(\vec{p}) = -\frac{e^2}{2 \pi^2 \hbar c} \int_0^1
\!\!\! d\tau \, \tau (1\!-\!\tau) \ln\Bigl[1 \!+\! \tau (1 \!-\!\tau)
\frac{\hbar^2 p^2}{m^2 c^2} \Bigr] \; . \label{epsfin}
\end{equation}
We can therefore think of the total permittivity as,
\begin{equation}
\epsilon(\vec{p}) = \epsilon_0 + \Delta \epsilon(0) +
\Delta \epsilon_{\rm fin}(\vec{p}) = \epsilon_{\rm meas} + \Delta
\epsilon_{\rm fin}(\vec{p}) \; , \label{finperm}
\end{equation}
where $\epsilon_{\rm meas} \simeq 8.85 \times 10^{-12}~{\rm F/m}$ is the
measured value of the electric permittivity.

Expressions (\ref{epsfin}-\ref{finperm}) illustrate why particle
theorists are so pleased with renormalization. First, all the divergences
are gone, as promised. Second, the quantum corrections from $\Delta
\epsilon_{\rm fin}(\vec{p})$ are small, both because of the initial factor
of $e^2$ and also because the ratio $\hbar^2 p^2/m^2 c^2$ is minuscule for
most applications. If we express the wave number in terms of a wave length,
$p = 2\pi/\lambda$ then the ratio is,
\begin{equation}
\Bigl(\frac{\hbar p}{m c}\Bigr)^2 \simeq \Bigl(\frac{2 \!\times\!
10^{-12}~{\rm m}}{\lambda}\Bigr)^2 \; . \label{ratio}
\end{equation}
Even at the Bohr radius of about $\lambda \sim 5 \times 10^{-11}~{\rm m}$
the ratio (\ref{ratio}) would be only about $10^{-3}$. Finally, and most
important of all, the small effects from $\Delta \epsilon_{\rm fin}(\vec{p})$
have been verified in many experiments.

One of the strangest effects from $\Delta \epsilon_{\rm fin}(\vec{p})$ is
that the electrodynamic force grows stronger at short distances. During the
1990's electrons and positrons were brought to within about $\lambda \sim
10^{-18}~{\rm m}$ at colliders such as the SLC at Stanford University and
LEP at the European Nuclear Research Center (CERN). At these separations
the fractional change in permittivity is,
\begin{equation}
\frac{\Delta \epsilon_{\rm fin}(\vec{p})}{\epsilon_{\rm meas}} \simeq
-\frac{2\alpha}{3 \pi} \times \ln\Bigl(\frac{2 \!\times\! 10^{-12}~{\rm m}}{
10^{-18}~{\rm m}}\Bigr) \simeq -.02 \; .
\end{equation}
Because the electric force goes like $1/\epsilon$, this 2\% reduction in the
permittivity implies a 2\% increase in the force, and that is just what was
seen. The phenomenon is known as {\it running of the coupling constants.}
We can understand it very simply from the fact that increasing the wave
number $p$ increases the energy of the virtual electron-positron pair,
which decreases the time they can exist and hence the degree to which they
can polarize the vacuum. So the charge screening at high $p$ must be less
than at low $p$.

Because the force fields of the strong interaction attract one another, it
turns out that the strong interaction becomes weaker at large $p$. Discovering
this in 1973 was what won the 2004 Nobel Prize for Politzer \cite{Pol},
Wilczek and Gross \cite{WG}. Our belief that we know the correct theory of
the strong interaction is mostly based upon pushing to very small separations,
at which point perturbative predictions from this theory become reliable.
The curious fact that strong interactions are much stronger than
electromagnetism at low energies, and grow weaker at high energies, whereas
electromagnetism gets stronger, is one thing which makes particle physicists
suspect both interactions are part of a Grand Unified Theory whose unity
becomes manifest at very high energy.

We are finally ready to consider how quantum matter affects the Einstein
equation (\ref{Einstein}). For definiteness let us focus on the 0th order
electromagnetic contribution to the energy density (\ref{EMrho}),
\begin{equation}
\frac{8\pi G}{c^4} \times \frac12 \epsilon_0 \Bigl[\vec{E}(t,\vec{x}) \cdot
\vec{E}(t,\vec{x}) + c^2 \vec{B}(t,\vec{x}) \cdot \vec{B}(t,\vec{x})\Bigr] \; .
\end{equation}
At lowest order this is a product of two free field mode sums, just like the
charge density of quantum electrodynamics,
\begin{equation}
\frac{e}{\hbar} \, \Psi^*(t,\vec{x}) \Psi(t,\vec{x}) \; .
\end{equation}
As might be expected, the terms they induce on the right hand side of their
respective field equations are very similar. For quantum electrodynamics in
the static limit we got,
\begin{equation}
\vec{\nabla} \cdot \vec{P}(\vec{x}) = \int \!\! \frac{d^3p}{(2\pi)^3} \,
e^{i \vec{p} \cdot \vec{x}} \!\! \int \!\! \frac{d^3k}{(2\pi)^3} \,
\frac{\frac83 e^2 (\hbar c \Vert \vec{p} \Vert)^2}{[ E(\vec{k}) \!+\!
E(\vec{p} \!-\! \vec{k})]^3} \times \widetilde{\Phi}(\vec{p}) \; . \label{QED1}
\end{equation}
For quantum gravity the analogous result takes the form,
\begin{equation}
\mathcal{G}_{\mu\nu}(\vec{x}) = \int \!\! \frac{d^3p}{(2\pi)^3} \,
e^{i \vec{p} \cdot \vec{x}} \!\! \int \!\! \frac{d^3k}{(2\pi)^3} \,
\frac{\frac{8\pi G}{c^4} \; \mathcal{E}^4}{[E(\vec{k}) \!+\! E(\vec{p} \!-\!
\vec{k})]^3} \times \widetilde{h}(\vec{p}) \; , \label{QG1}
\end{equation}
where $h_{\mu\nu}(t,\vec{x})$ is the deviation of the metric from its
quiescent value and I have suppressed its indices in (\ref{QG1}) because
more than one component can contribute to any one of the ten Einstein
equations. The symbol $\mathcal{E}^4$ stands for any combination of four
quantities built from $\vec{k}$ and $\vec{p} \!-\! \vec{k}$ and having the
dimensions of energy$^4$. Possible values for $\mathcal{E}^4$ include,
\begin{equation}
(\hbar c \Vert \vec{p}\Vert)^4 \;\; , \;\;
[\hbar^2 c^2 \vec{k} \cdot (\vec{p} \!-\! \vec{k})]^2 \;\; , \;\;
E^2(\vec{k}) E^2(\vec{p} \!-\! \vec{k}) \;\; {\rm and} \;\;
[E(\vec{k}) \!+\! E(\vec{p} \!-\! \vec{k})]^4 \; . \label{possibilities}
\end{equation}
If (\ref{QG1}) represents a contribution from electrodynamics then the
energy of a photon with wave vector $\vec{k}$ is $E(\vec{k}) = \hbar c \Vert
\vec{k} \Vert$.

There are two outstanding differences between the quantum electrodynamic
result (\ref{QED1}) and its quantum gravitational analogue (\ref{QG1}):
\begin{itemize}
\item{The quantum gravitational integrand (\ref{QG1}) contains four factors
of energy in the numerator as opposed to only two in the quantum electrodynamic
numerator (\ref{QED1}); and}
\item{All four of the energy factors $\mathcal{E}^4$ in the quantum
gravitational integrand (\ref{QG1}) can involve $\vec{k}$ whereas the two
factors of $\hbar c \Vert \vec{p} \Vert$ in the quantum electrodynamic
integrand (\ref{QED1}) do not.}
\end{itemize}
The first difference derives from the fact that stress-energy is the
source for gravity in general relativity, whereas the source for
electromagnetism is charge. The second difference derives from the fact that
the 0-point energy of a bosonic mode such as a photon is always positive
(the 0-point energy of a fermionic meode such as an electron is always
negative) whereas the same wave vector $\vec{k}$ contributes both positive
and negative charges. That is why one gets no coherent effect from the
undisturbed charges of a neutral medium; the lowest coherent effect in
quantum electrodynamics comes from the {\it difference} of positive and
negative charges subjected to an electric field, whereas quantum general
relativity receives a coherent effect from the 0-point energy of each mode.

The two differences I have noted explain why the divergences of quantum
general relativity are so much worse than those of quantum electrodynamics.
To be quantitative, let us evaluate the $\vec{k}$ mode sum of (\ref{QG1})
with the second possibility from the list (\ref{possibilities}) for the
numerator factors $\mathcal{E}^4$. After some tedious but standard reductions
we reach the form,
\begin{eqnarray}
\lefteqn{\frac{8\pi G}{c^4} \!\! \int \!\! \frac{d^3k}{(2\pi)^3} \,
\frac{[\hbar^2 c^2 \vec{k} \cdot (\vec{p} \!-\! \vec{k})]^2}{[\hbar c
\Vert\vec{k}\Vert \!+\!  \hbar c \Vert\vec{p} \!-\! \vec{k}\Vert]^3} }
\nonumber \\
& & \hspace{-.3cm} = \frac{3 G\hbar}{\pi c^3} \!\! \int_0^1 \!\! dx \, x
(1\!-\!x) \!\! \int_0^{\infty} \!\!\! dk \, k^2 \, \frac{[k^4 \!+\! (\frac13
\!-\! \frac{10}3 x \!+\! \frac{10}3 x^2) k^2 p^2 \!+\! x^2 (1 \!-\! x)^2 p^4]
}{[k^2 \!+\! x (1 \!-\! x) p^2]^{\frac32}} \; . \qquad \label{quartic}
\end{eqnarray}
Expression (\ref{quartic}) is a quarticly divergent integral, which is
typical of the first order corrections in quantum general relativity. To
make it well defined I will cut off the upper limit at $k = K$. This
procedure is an example of a technique known as {\it regularization} which
is employed to make mathematical sense of the divergences of quantum field
theory before they are removed by renormalization. I should really have done
it at the beginning of the computation, with a better technique, and used it
consistently throughout (which is, rest assured, the standard practice) but
this level of rigor is superfluous if one just wants an explanation of the
problem without precise numbers.

Once regulated, expression (\ref{quartic}) becomes well defined. We can
evaluate it exactly but it suffices to give the expansion for large $K$,
\begin{eqnarray}
\lefteqn{\frac{3 G\hbar}{\pi c^3} \!\! \int_0^1 \!\! dx \, x (1\!-\!x) \!\!
\int_0^{K} \!\!\! dk \, k^2 \, \frac{[k^4 \!+\! (\frac13 \!-\! \frac{10}3 x
\!+\! \frac{10}3 x^2) k^2 p^2 \!+\! x^2 (1 \!-\! x)^2 p^4]}{[k^2 \!+\! x
(1 \!-\! x) p^2]^{\frac32}} } \nonumber \\
& & = \frac{3G\hbar}{\pi c^3} \Biggl\{ \frac{K^4}{4} \!-\! \frac{K^2 p^2}{24}
\!+\! \frac{61}{2^4 \!\cdot\! 7!!} \, p^4 \ln\Bigl(\frac{2 K}{p}\Bigr) \!-\!
\frac{190921}{2^7 \!\cdot\! (7!!)^2} \, p^4 + O\Bigl(\frac{p^6}{K^2}\Bigr)
\Biggr\} . \qquad
\end{eqnarray}
All contributions to (\ref{QG1}) have this same form. It is useful to break
them up into a part that just has positive powers of $p^2$ but diverges when
$K$ goes to infinity and another part which can have logarithms of $p$ or
even inverse powers, but remains finite. For the logarithm term this breakup
requires the introduction of a fixed length scale $L$,
\begin{equation}
\ln\Bigl(\frac{2K}{p}\Bigr) = \ln(2 L K) - \ln(L p) = \frac12 \ln(L^2 K^2)
-\frac12 \ln(L^2 p^2) + \ln(2) \; .
\end{equation}
We can therefore express the first order electromagnetic correction to the
Einstein equations as the sum of divergent and finite parts having the form,
\begin{eqnarray}
\mathcal{G}^{\rm div}_{\mu\nu}(\vec{x}) &\!=\!& \frac{G\hbar}{c^3} \int \!\!
\frac{d^3p}{(2\pi)^3} \, e^{i \vec{p} \cdot \vec{x}} \Biggl\{A K^4 + B K^2 p^2
+ C \ln(L^2 K^2) \, p^4\Biggr\} \times \widetilde{h}(\vec{p}) \; , \qquad \\
& \!=\! & \frac{G\hbar}{c^3} \Biggl\{ A K^4 h(\vec{x}) - B K^2 \nabla^2
h(\vec{x}) + C \ln(L^2 K^2) \nabla^4 h(\vec{x}) \Biggr\} \; , \qquad
\label{locdiv} \\
\mathcal{G}^{\rm fin}_{\mu\nu}(\vec{x}) &\!=\!& \frac{G\hbar}{c^3} \int \!\!
\frac{d^3p}{(2\pi)^3} \, e^{i \vec{p} \cdot \vec{x}} \Biggl\{-C \ln(L^2 p^2)
\, p^4 + D p^4\Biggr\} \times \widetilde{h}(\vec{p}) \; . \label{nonloc}
\end{eqnarray}
Here $A$, $B$, $C$ and $D$ are pure numbers of order one and I am still
suppressing the indices of the field $h_{\mu\nu}$. Had I considered 0-point
contributions from a massive field, such as the electron, the resulting
$\mathcal{G}_{\mu\nu}^{\rm fin}$ would have had a more complicated structure
involving the mass, but the divergent part would have had the same form as
(\ref{locdiv}).

One might think it is the most highly divergent parts of (\ref{locdiv}) which
cause problems for quantum general relativity but that is not so. We can see
this by taking the static, linearized limit of the Einstein equations
(\ref{Einstein}), and including the effect we have just discussed from
quantum 0-point motions,
\begin{equation}
\Bigl[\nabla^2 h + \Lambda h \Bigr]_{\mu\nu} = \mathcal{G}_{\mu\nu}
+ \frac{8\pi G}{c^4} \, \Bigl(T_{\mu\nu}\Bigr)_{\rm linear} . \label{newEin}
\end{equation}
(Because $\mathcal{G}_{\mu\nu}$ already includes the effect of matter 0-point
motions one should think of the linearized stress-energy tensor as a
classical source, just like $\rho_{\rm free}$ in our discussion
(\ref{rhofree}) of electromagnetic polarization.) Now multiply (\ref{newEin})
by $c^4/8\pi G$ and bring the $\mathcal{G}_{\mu\nu}$ terms to the left hand
side, as we do for electromagnetic polarization,
\begin{eqnarray}
\lefteqn{\Biggl\{ -\frac{C \hbar c \ln(L^2 K^2)}{8\pi} \, \nabla^4 h +
\Biggl[\frac{c^4}{8\pi G} \!+\! \frac{B \hbar c K^2}{8 \pi}\Biggr]
\nabla^2 h} \nonumber \\
& & \hspace{3cm} + \Biggl[ \frac{c^4 \Lambda}{8\pi G} \!-\! \frac{A \hbar
c K^4}{8\pi}\Biggr] h \Biggr\}_{\mu\nu} + \frac{c^4}{8\pi G} \,
\mathcal{G}_{\mu\nu}^{\rm fin} = \Bigl( T_{\mu\nu}\Bigr)_{\rm linear} . \qquad
\end{eqnarray}
We see that the quadratic and quartic divergences can be absorbed into
renormalizations of Newton's constant and the cosmological constant,
\begin{eqnarray}
\frac{c^4}{8\pi G} + \frac{B \hbar c K^2}{8\pi} & \equiv &
\Bigl(\frac{c^4}{8\pi G}\Bigr)_{\rm meas} \; , \label{QGren1} \\
\frac{c^4 \Lambda}{8\pi G} \!-\! \frac{A \hbar c K^4}{8\pi} & \equiv &
\Bigl(\frac{c^4 \Lambda}{8\pi G}\Bigr)_{\rm meas} \; . \label{QGren2}
\end{eqnarray}
Just as neither $\epsilon_0$ nor $\Delta \epsilon(0)$ is separately
observable in electrodynamics, so it is only the combinations
(\ref{QGren1}-\ref{QGren2}) which are observable in gravity.

Alas, there is no parameter in general relativity with which to absorb the
logarithmic divergence! If only there were then the remaining, finite
quantum gravitational effects from $\mathcal{G}_{\mu\nu}^{\rm fin}$ would be
unobservably small. For example, the fractional change in Earth's surface
gravity due to quantum gravitational effects would be about,
\begin{equation}
\frac{G \hbar}{c^3 R^2_E} \, \ln\Bigl(\frac{L^2}{R_E^2}\Bigr) \sim 10^{-83}
\times \ln\Bigl(\frac{L^2}{R_E^2}\Bigr) \; ,
\end{equation}
which is negligible even if $L$ is chosen to be the Planck length of about
$10^{-35}~{\rm m}$. However, infinity is not small, and that is what one
gets from the logarithmic divergence. One can only absorb it if new, 4th
derivative terms are added to the gravitational field equations, but I will
show in the next subsection that doing so would make the universe blow up.
That is the fundamental obstacle to making sense of perturbative quantum
general relativity.

Before concluding the discussion of renormalization I need to comment on three
more issues. The first is that Einstein's equations are not linear in the
field $h_{\mu\nu}$ and I have only considered quantum corrections which are
linear in this field. Might there be additional divergences on nonlinear
terms? There are indeed such divergences but the conservation of stress-energy
prescribes how the various powers of $h_{\mu\nu}$ can enter the field
equations. With either zero derivatives or two derivatives of the full metric
the results are unique,
\begin{eqnarray}
\partial^0 & \Longrightarrow & \Lambda g_{\mu\nu} \; , \label{0ds} \\
\partial^2 & \Longrightarrow & \Bigl[ A(g) \partial^2 g + B(g) \partial g
\partial g\Bigr]_{\mu\nu} \; . \label{2ds}
\end{eqnarray}
These correspond to the two terms in the Einstein equations (\ref{Einstein}),
and knowing their linear parts in $h_{\mu\nu}$ dictates all higher (and
lower) powers as well.

The second point is that I have so far worked in the static limit. That is
not even correct for classical electrodynamics! Real media cannot polarize
infinitely rapidly, so dielectric response is frequency dependent. To compute
the actual polarization one must first Fourier transform the electric field
in time, as well as space, then multiply by the frequency and wave vector
dependent permittivity $\epsilon(\omega,\vec{k})$, and only then transform
back,
\begin{equation}
\vec{P}(t,\vec{x}) = \int \! \frac{d\omega}{2\pi} \, e^{-i\omega t} \!\!
\int \!\! \frac{d^3k}{(2\pi)^3} \, e^{i\vec{k} \cdot \vec{x}} \times
\epsilon(\omega,\vec{k}) \times \int \!\! dt' \, e^{i \omega t'} \!\!
\int \!\! d^3x' \, e^{-i \vec{k} \cdot \vec{x}'} \vec{E}(t,\vec{x}') \; .
\end{equation}
This obviously raises issues about causality! Those issues can all
be resolved but doing so in quantum field theory involves a
technique known as the ``Schwinger-Keldysh formalism'' which even
many particle theorists don't understand.\footnote{An ignorant (and
rather pompous) reviewer for the Department of Energy once
recommended terminating my funding because the Schwinger-Keldysh
field equations I had derived must be wrong on account of lacking
some contributions with which he was familiar from the usual
formalism!} (The original literature is \cite{JS,MB,LK}; for some
nice reviews see \cite{SKrev}.) They have evolved a series of tricks
to avoid having to think about it and, although I do understand the
technique, I employed such a trick to avoid a lengthy (and probably
not very illuminating) digression to explain it. The trick was to
study the static limit of no time dependence and then appeal to the
fact that the Einstein equation is the unique combination of the
metric and no more than two derivatives which is consistent with
stress-energy conservation. So the renormalizations
(\ref{QGren1}-\ref{QGren2}) which were found in the static limit of
only spatial derivatives must apply as well for space and time
derivatives in the full theory.

Finally, I should comment that stress-energy conservation allows two linearly
independent combinations of four derivatives of the metric. Each of these
terms has general form,
\begin{equation}
\Biggl[A(g) \partial^4 g + B(g) \partial g \partial^3 g + C(g) \partial^2 g
\partial^2 g + D(g) \partial g \partial g \partial^2 g + E(g) \partial g
\partial g \partial g \partial g \Biggr]_{\mu\nu} \; . \label{4ds}
\end{equation}
They are called, the ``$R^2$  counterterm'' and the ``$C^2$ counterterm''
after the curvature scalars that comprise the Lagrangian densities from which
they descend. The difference between them has to do with how the indices are
arranged, which I have suppressed. As with (\ref{0ds}) and (\ref{2ds}),
knowing just the linearized parts $\partial^4 h$ fixes all other powers.

\subsection{The Problem with Higher Derivatives}

We have seen that the renormalization of perturbative quantum general
relativity requires that the equations of motion be changed to include
terms with up to four derivatives. Stelle has shown that if such terms are
added to gravity (which we cannot any longer call general relativity)
then the resulting quantum theory is perturbatively renormalizable \cite{KS}.
However, it is also subject to a virulent instability that is totally
inconsistent with the observed reality of a universe which is 13.7 billion
years old.

This result is very old, and not specific to gravity; it was
obtained in 1850 by the great Russian physicist Ostrogradsky
\cite{MO}. Ostrogradsky's theorem is that there is a linear
instability in the Hamiltonians associated with Lagrangians which
depend upon more than one time derivative in such a way that the
dependence cannot be eliminated by partial integration \cite{MO}.
The result is so general that I can simplify the discussion by
presenting it in the context of a single, one dimensional point
particle whose position as a function of time is $q(t)$.

In the usual case of $L = L(q,\dot{q})$, the Euler-Lagrange equation is,
\begin{equation}
\frac{\partial L}{\partial q} - \frac{d}{dt} \frac{\partial L}{\partial
\dot{q}} = 0 \; . \label{ELE1}
\end{equation}
The assumption that $\frac{\partial L}{\partial \dot{q}}$ depends upon
$\dot{q}$ is known as {\it nondegeneracy}. If the Lagrangian is nondegenerate
we can write (\ref{ELE1}) in the form Newton originally laid down for the laws
of physics,
\begin{equation}
\ddot{q} = \mathcal{F}(q,\dot{q}) \qquad \Longrightarrow \qquad q(t) =
\mathcal{Q}(t,q_0,\dot{q}_0) \; . \label{newt}
\end{equation}
From this form it is apparent that solutions depend upon two pieces of initial
value data: $q_0 = q(0)$ and $\dot{q}_0 = \dot{q}(0)$.

The fact that solutions require two pieces of initial value data means that
there must be two canonical coordinates, $Q$ and $P$. They are traditionally
taken to be,
\begin{equation}
Q \equiv q \qquad {\rm and} \qquad P \equiv \frac{\partial L}{\partial \dot{q}}
\; . \label{ctrans}
\end{equation}
The assumption of nondegeneracy is that we can invert the phase space
transformation (\ref{ctrans}) to solve for $\dot{q}$ in terms of $Q$ and $P$.
That is, there exists a function $v(Q,P)$ such that,
\begin{equation}
\frac{\partial L}{\partial \dot{q}} \Biggl\vert_{q = Q \atop \dot{q} = v}
= P \; . \label{invct}
\end{equation}
For example, one finds $v(Q,P) = P/m$ for the harmonic oscillator (\ref{SHO}).

The canonical Hamiltonian is obtained by Legendre transforming on $\dot{q}$,
\begin{eqnarray}
H(Q,P) & \equiv & P \dot{q} - L \; , \\
& = & P v(Q,P) - L\Bigl(Q,v(Q,P)\Bigr) \; .
\end{eqnarray}
It is easy to check that the canonical evolution equations reproduce the
inverse phase space transformation (\ref{invct}) and the Euler-Lagrange
equation (\ref{ELE1}),
\begin{eqnarray}
\dot{Q} & \equiv & \frac{\partial H}{\partial P} = v + P \frac{\partial v}{
\partial P} - \frac{\partial L}{\partial \dot{q}} \frac{\partial v}{\partial P}
= v \; , \\
\dot{P} & \equiv & -\frac{\partial H}{\partial Q} = -P \frac{\partial v}{
\partial Q} + \frac{\partial L}{\partial q} + \frac{\partial L}{\partial
\dot{q}} \frac{\partial v}{\partial Q} = \frac{\partial L}{\partial q} \; .
\end{eqnarray}
This is what we mean by the statement, ``the Hamiltonian generates time
evolution.'' When the Lagrangian has no explicit time dependence, $H$ is also
the associated conserved quantity. Hence it is ``the'' energy by any usual
standard, of course up to canonical transformation.

Now consider a system whose Lagrangian $L(q,\dot{q},\ddot{q})$ depends
nonde\-gen\-er\-ate\-ly upon $\ddot{q}$. The Euler-Lagrange equation is,
\begin{equation}
\frac{\partial L}{\partial q} - \frac{d}{dt} \frac{\partial L}{\partial
\dot{q}} + \frac{d^2}{dt^2} \frac{\partial L}{\partial \ddot{q}} = 0 \; .
\label{ELE2}
\end{equation}
Non-degeneracy implies that $\frac{\partial L}{\partial \ddot{q}}$ depends
upon $\ddot{q}$, in which case we can cast (\ref{ELE2}) in a form radically
different from Newton's,
\begin{equation}
q^{(4)} = \mathcal{F}(q,\dot{q},\ddot{q},q^{(3)}) \qquad \Longrightarrow \qquad
q(t) = \mathcal{Q}(t,q_0,\dot{q}_0,\ddot{q}_0,q^{(3)}_0) \; .
\end{equation}

Because solutions now depend upon four pieces of initial value data there must
be four canonical coordinates. Ostrogradsky's choices for these are,
\begin{eqnarray}
Q_1 \equiv q \qquad & , & \qquad P_1 \equiv \frac{\partial L}{\partial \dot{q}}
- \frac{d}{dt} \frac{\partial L}{\partial \ddot{q}} \; , \label{ct1} \\
Q_2 \equiv \dot{q} \qquad & , & \qquad P_2 \equiv \frac{\partial L}{\partial
\ddot{q}} \; . \label{ct2}
\end{eqnarray}
The assumption of nondegeneracy is that we can invert the phase space
transformation (\ref{ct1}-\ref{ct2}) to solve for $\ddot{q}$ in terms of
$Q_1$, $Q_2$ and $P_2$. That is, there exists a function $a(Q_1,Q_2,P_2)$
such that,
\begin{equation}
\frac{\partial L}{\partial \ddot{q}} \Biggl\vert_{{q = Q_1 \atop \dot{q} =
Q_2} \atop \ddot{q} = a} = P_2 \; . \label{invct2}
\end{equation}
Note that one only needs the function $a(Q_1,Q_2,P_2)$ to depend upon {\it
three} canonical coordinates --- and not all four --- because
$L(q,\dot{q},\ddot{q})$ only depends upon three configuration space
coordinates. This simple fact has great consequence.

Ostrogradsky's Hamiltonian is obtained by Legendre transforming, just as in the
first derivative case, but now on $\dot{q} = q^{(1)}$ and $\ddot{q} = q^{(2)}$,
\begin{eqnarray}
\lefteqn{H(Q_1,Q_2,P_1,P_2) \equiv \sum_{i=1}^2 P_i q^{(i)} - L \; , } \\
& & = P_1 Q_2 + P_2 a(Q_1,Q_2,P_2) - L\Bigl(Q_1,Q_2,a(Q_1,Q_2,P_2)\Bigr) \; .
\label{Host}
\end{eqnarray}
The time evolution equations are just those suggested by the notation,
\begin{equation}
\dot{Q_i} \equiv \frac{\partial H}{\partial P_i} \qquad {\rm and} \qquad
\dot{P}_i \equiv - \frac{\partial H}{\partial Q_i} \; .
\end{equation}
Let's check that they generate time evolution. The evolution equation for
$Q_1$,
\begin{equation}
\dot{Q}_1 = \frac{\partial H}{\partial P_1} = Q_2 \; ,
\end{equation}
reproduces the phase space transformation $\dot{q} = Q_2$ in (\ref{ct2}).
The evolution equation for $Q_2$,
\begin{equation}
\dot{Q}_2 = \frac{\partial H}{\partial P_2} = a + P_2 \frac{\partial a}{
\partial P_2} - \frac{\partial L}{\partial \ddot{q}} \frac{\partial a}{\partial
P_2} = a \; ,
\end{equation}
reproduces (\ref{invct2}). The evolution equation for $P_2$,
\begin{equation}
\dot{P}_2 = -\frac{\partial H}{\partial Q_2} = -P_1 - P_2 \frac{\partial a}{
\partial Q_2} + \frac{\partial L}{\partial \dot{q}} + \frac{\partial L}{
\partial \ddot{q}} \frac{\partial a}{\partial Q_2} = -P_1 + \frac{\partial L}{
\partial \dot{q}} \; ,
\end{equation}
reproduces the phase space transformation $P_1 = \frac{\partial L}{\partial
\dot{q}} - \frac{d}{dt} \frac{\partial L}{\partial \ddot{q}}$ (\ref{ct1}). And
the evolution equation for $P_1$,
\begin{equation}
\dot{P}_1 = -\frac{\partial H}{\partial Q_1} = -P_2 \frac{\partial a}{\partial
Q_1} + \frac{\partial L}{\partial q} + \frac{\partial L}{\partial \ddot{q}}
\frac{\partial a}{\partial Q_1} = \frac{\partial L}{\partial q} \; ,
\end{equation}
reproduces the Euler-Lagrange equation (\ref{ELE2}). So Ostrogradsky's system
really does generate time evolution. When the Lagrangian contains no explicit
dependence upon time it is also the conserved Noether current. By any standard
definition, it is therefore ``the'' energy, again up to canonical
transformation.

There is one, overwhelmingly bad thing about Ostrogradsky's Hamiltonian
(\ref{Host}): {\it it is linear in the canonical momentum} $P_1$. Note the
power and generality of the result. It applies to {\it every} Lagrangian
$L(q,\dot{q},\ddot{q})$ which depends nondegenerately upon $\ddot{q}$,
independent of the details. The only assumption is nondegeneracy, and that
simply means one cannot eliminate $\ddot{q}$ by partial integration. This
is why Newton was right to assume the laws of physics take the form
(\ref{newt}) when expressed in terms of fundamental dynamical variables.

The Ostrogradskian instability is not a problem with the potential energy,
in which the dynamical variable can reach arbitrarily negative energies by
approaching a special value. It is instead an instability of the kinetic
energy in which arbitrarily negative energies are associated with a special
sort of time dependence. The problematic term in Ostrogradsky's Hamiltonian
(\ref{Host}) is $P_1 Q_2$. One makes its large by adjusting the third time
derivatives in $P_1$, which can be done while the dynamical variable $q(t)$
is still quite small.

At this point I need to debunk the misconception that unstable systems decay
``because the system wants to lower its energy.'' Total energy is conserved
in fundamental theory, so the decay of an excited atomic state into the
ground state atom plus some photons leads to no change of energy. What drives
the decay is entropy: quantum systems explore the space of classical
configurations which have the same energy, and there are many more ways for
an atom to decay as opposed to only one for it not to decay. For example, the
decay photons can go off in any direction.

This insight about what drives decays means that the Ostrogradskian
instability is {\it instantly} fatal for an interacting field theory. Recall
that each Fourier mode of such a theory contributes its own, independent
kinetic energy, and there are an infinite number of such Fourier modes. 
One's usual intuition about modes with large $\Vert \vec{k}\Vert$ is
completely wrong for theories which possess the Ostrogradskian instability.
We are used to thinking that these modes cannot be excited because doing so
costs a large energy and there is only a finite free energy available. But
that is only true when all modes carry positive energy. Exciting a negative
energy mode {\it frees up energy}, which can then be used to excite
positive energy modes. And exciting a negative energy mode with higher
$\Vert \vec{k} \Vert$ frees up even more energy. Now use expression
(\ref{dofs}) to count up the number of modes per unit volume which have
$\Vert \vec{k}\Vert < K$,
\begin{equation}
\int \!\! \frac{d^3k}{(2\pi)^3} \, \theta\Bigl(K \!-\! \Vert \vec{k}
\Vert\Bigr) = \frac{K^3}{6 \pi^2} \; .
\end{equation}
Of course there is no limit on $K$ because the higher modes actually
participate more strongly when the Ostrogradskian instability is present.
So one can see that an interacting field theory with the Ostrogradskian
instability decays instantly, no matter how weak the interaction is.

A final point is that the problematic term $P_1 Q_2$ of the Ostrogradskian
Hamiltonian (\ref{Host}) can have either sign. We have been concerned with
the fact that it can be arbitrarily negative, but it can also be arbitrarily
positive. This means that two things happen when you add a higher derivative
term to a lower derivative theory:
\begin{itemize}
\item{There can be changes in the original, lower derivative degrees freedom;
and}
\item{The higher derivative term introduces new degrees of freedom which
carry the opposite kinetic energy to the changed, lower derivative degrees
of freedom.}
\end{itemize}
The usual case is that the lower derivative theory has positive energy,
so adding a higher derivative induces negative energy degrees of freedom.
That is what happens with the $C^2$ counterterm; it adds a negative energy,
spin two graviton which would make the universe decay instantly. However,
it turns out that the $R^2$ counterterm adds a positive energy, spin zero
particle which is harmless. This represents no violation of Ostrogradsky's
theorem because the spin zero part of the metric in general relativity
carries negative energy. It is better known as the Newtonian potential and
its negative energy poses no problem for stability because this part of the
metric is completely determined by the stress-energy tensor. The new spin
zero degree of freedom induced by the $R^2$ counterterm is an independent,
purely gravitational degree of freedom, just like the gravitons.

To summarize, although adding the $R^2$ counterterm to general relativity
would be no problem, adding the $C^2$ counterterm would make the universe
blow up. If only we didn't need the $C^2$ counterterm! But careful analyses
show that we do need it for scalar particles like the Higgs \cite{HV}, for
electromagnetism \cite{DvN} and for the particles which carry the weak and
strong interactions \cite{DTvN}. We do not need it for pure gravity at first
order in perturbation theory \cite{HV}, which is why I have emphasized that
the basic problem of quantum gravity concerns quantum matter effects which
must be present, whether or not there is gravitational radiation or it is
quantized. That is not to say gravitons pose no problems. A heroic second
order computation by Goroff and Sagnotti \cite{GS}, verified by van de Ven
\cite{Ven}, demonstrates that they induce a higher derivative counterterm
as unacceptable as the $C^2$ counterterm.

\subsection{The Impact of Primordial Inflation on Two Fixes}

Inflation is defined as a period of accelerated cosmological expansion.
We know this can happen because it's taking place right now \cite{SNIA,Yun}.
Guth has proposed that a very early phase of {\it primordial inflation}
would explain why the current universe is so nearly homogeneous and
isotropic on the largest scales, why it is so nearly spatially flat, and
why it contains no exotic relics such as magnetic monopoles (which
typically occur when all the forces are unified) and primordial black
holes \cite{Guth}. There  is a lot of evidence in favor of this idea and
none against it, although physicists are not yet ready to regard it as
proven. I shall have a lot more to say about primordial inflation in
section 5 but let me for now explore the consequences for quantum gravity
of three tenets of primordial inflation:
\begin{enumerate}
\item{Quantum gravitational fluctuations in the metric at the beginning of
inflation were no larger than about one part in $10^6$;}
\item{The universe has expanded by a factor of at least $10^{51}$; and}
\item{The structures of today's universe derived from 13.7 billion years
of gravitational collapse into the tiny (one part in about $10^5$)
inhomogeneities provided by quantum fluctuations of the stress-energy tensor
near the end of inflation.}
\end{enumerate}

Assumptions 1-3 can be used to rule out two proposals which are sometimes
advanced for resolving the apparent inconsistency of perturbative quantum
general relativity:
\begin{itemize}
\item{Regard {\it all} components of the metric as classical and change the
source of gravity from the quantum stress-energy tensor to its expectation
value in some state; or}
\item{Regard spacetime as discrete at some very small length scale.}
\end{itemize}
Of course assumption \#3 immediately falsifies the first proposal. If inflation
is correct then the expectation value of the stress-energy tensor at the
end of inflation cannot retain inhomogeneities of more than about one part in
$10^{78}$, otherwise they would have been so big at the beginning of inflation
that gravitational collapse would have ensued. But the measured strength of
primordial perturbations in the cosmic microwave background and in the matter
density is about one part in $10^{5}$ \cite{WMAP,SDSS}. I remark in passing
that, if inflation is correct, {\it primordial perturbations are the first
ever data from quantum gravity.} One can see how having this data has shifted
debates over quantum gravity from philosophy and aesthetics to the
interpretation of hard evidence. Quantum gravity is coming of age.

One has to work a little harder to debunk discretization. We only have
direct evidence for the continuum nature of spacetime down to about
$10^{-18}~{\rm m}$. If space and time were discrete at some smaller scale
$\Delta L$ the strength of quantum gravitational corrections would be
roughly what one gets from taking the spatial momentum cutoff $K$ to be
$1/\Delta L$. That would remove the divergences but one also has to keep
quantum gravitational corrections sufficiently small and, it turns out that
discretization cannot accomplish this if one accepts primordial inflation.

First note that the scale of discreteness $\Delta L$ cannot be much smaller
than about the Planck length of $[\hbar G/c^3]^{\frac12} \sim 1.6 \times
10^{-35}~{\rm m}$. This might seem surprising in view of the first order
results (\ref{locdiv}),
\begin{equation}
\mathcal{G}_{\mu\nu}^{\rm div} =  \frac{G\hbar}{c^3} \Biggl\{ A K^4 h(\vec{x})
- B K^2 \nabla^2 h(\vec{x}) + C \ln(L^2 K^2) \nabla^4 h(\vec{x}) \Biggr\} \; .
\end{equation}
Just because $K \sim 1/\Delta L$ will stay finite does not preclude
renormalization, so we can still absorb the $K^4$ contribution into a
shift of the cosmological constant, and the $K^2$ contribution into a
change of the Newton constant. That leaves only the $\ln(L^2 K^2)$
correction, which would be minuscule at low scales, even if the argument
of the logarithm is enormous.

However, one must consider higher order corrections. If one {\it doesn't}
add 4th derivative terms to the classical field equations then the
divergences of quantum gravity grow worse as the order of perturbation
theory increases. For example, at second order the divergences would take
the form,
\begin{equation}
\mathcal{G}_{\mu\nu}^{\rm div} = \Bigl(\frac{G}{\hbar c^5}\Bigr)^2
\Biggl\{ A K^6 + B K^4 \nabla^2 + C K^2 \nabla^4 + D \ln(L^2 K^2)
\nabla^6 \Biggl\} h(\vec{x}) \; ,
\end{equation}
where $A$, $B$, $C$ and $D$ are numbers of order one. One can absorb
the $K^6$ divergence in the cosmological constant and the $K^4$
divergence in the Newton constant, but the other two divergences
must be regulated by the cutoff $K \sim 1/\Delta L$. The term
proportional to $\nabla^4 h(\vec{x})$ could be written as the first
order term times an extra factor of $G K^2/\hbar c^5$,
\begin{equation}
\Bigl(\frac{G}{\hbar c^5}\Bigr)^2 \times C K^2 \nabla^4 h(\vec{x})
= \frac{G K^2}{\hbar c^5} \times \Biggl\{ \frac{G}{\hbar c^5} \times
C \nabla^4 h(\vec{x})\Biggr\} \; .
\end{equation}
The term in brackets is roughly the same strength as the $\nabla^4 h$ term
we got at first order, and very small under normal circumstances, but the
initial factor of $G K^2/\hbar c^5$ would be huge if scale of discretization
drops much below the Planck length. The third order correction would contain
two factors of this huge number, et cetera. The only way to avoid eventually
getting an unacceptably large quantum correction is to prevent the cutoff
scale $\Delta L$ from dropping below the Planck Length.

Now consider Assumption \#1 from primordial inflation, that quantum
gravitational effects were small at the beginning of inflation. This 
means that, {\it in terms of the physical length measured at the 
beginning of inflation}, the scale of discretization cannot have been 
below about $10^{-35}~{\rm m}$. But Assumption \#2 says that the universe 
has expanded by a factor of at least $10^{51}$ since the beginning of 
inflation. That means the physical length between discrete points (the 
number of which cannot change with time) ought to be about $10^{16}~{\rm m}$ 
today! That is roughly the distance between stars in our part of the 
Milky Way galaxy, and of course utterly inconsistent with current, direct 
checks of continuum spacetime to about 34 orders of magnitude smaller. 
Turning the argument around, for the current physical length of 
discreteness to be less than $10^{-18}~{\rm m}$, its value at the 
beginning of primordial inflation must have been $10^{-69}~{\rm m}$, 
which is 34 orders of magnitude too small to explain why quantum 
gravitational effects are small during inflation.

Note that the same argument can be invoked for {\it any} early event during
which we have reason to believe quantum gravitational fluctuations were
small. Some events and the associated cosmological expansion factors are:
\begin{eqnarray}
{\rm Recombination} & \Longrightarrow & 10^3 \; , \\
{\rm Nucleosynthesis} & \Longrightarrow & 10^9 \; , \\
{\rm Quark\ Gluon\ Plasma} & \Longrightarrow & 10^{12} \; , \\
{\rm Electroweak\ Symmetry\ Breaking} & \Longrightarrow & 10^{15} \; .
\end{eqnarray}
Primordial inflation provides a vastly stronger bound because it came much
earlier, but the expansion since electroweak symmetry breaking (which may
well have seen the formation of the asymmetry between matter and anti-matter)
lacks only two orders of magnitude to connect the Planck length to the current
experimental bound on discreteness. We therefore conclude that, while
spacetime may well be discrete at some, very small scale, this cannot explain
what is suppressing quantum gravitational effects.

\section{General Reactions to the Problem}

We have seen that the problem of quantum gravity arises from a conflict
between four physics principles:
\begin{itemize}
\item{{\it Continuum Field Theories} possess an infinite number of modes;}
\item{{\it Quantum Mechanics} requires each mode to have a minimum amount of
energy;}
\item{{\it General Relativity} stipulates that stress-energy is the source
of gravitation; and}
\item{{\it Perturbation Theory} simply adds up the contribution from each
mode at lowest order.}
\end{itemize}

When a problem can be shown to derive from a well-defined set of propositions
then one or more of these propositions must be wrong. In the previous section
I argued that it cannot be the first two. Although spacetime may well be
discrete at some level, the expansion of the universe implies that this
discreteness must be at too small a scale to be useful for making sense of
quantum gravity. And the lowest order divergences of quantum gravity derive
from the quantum properties of matter, which have been too thoroughly checked
to abandon. It follows that the problem must lie either with general
relativity or with the use of perturbation theory. The great fault line which
divides fundamental theorists is which of these two is held suspect.

\subsection{Particle Theorists versus Relativists}

The two major schools of thought on quantum gravity consist of those who
approach the subject from the perspective of particle theory and those who
approach it from the perspective of classical general relativity.\footnote{
There are of course exceptions: string theorists who were trained as
relativists and loop quantum gravity researchers who were trained in
particle theory. I hope they will not take offense at the names I have
chosen to characterize their disciplines.} Particle theorists are much 
attached to perturbation theory, so they are willing to alter general 
relativity. This is what led to supergravity and superstring theory, and 
to study of on-shell finiteness and asymptotic safety. Relativists are 
equally attached to general relativity, so they are willing to ignore 
perturbative problems. This is what has led to loop quantum gravity.

Both views reflect the body of experience of those who hold them.
The history of particle physics has involved aggressively using
perturbation theory to derive predictions for proposed models of
fundamental interactions, and then ruthlessly discarding any model
which failed to agree with observation and experiment. Among particle
theorists the ``crackpots'' were those who became too attached to a
particular model and either failed to check it using perturbation
theory or else refused to abandon the model when perturbative checks
indicated a problem.

It should also be mentioned that particle theorists are much attracted
to the idea of unifying gravity with the other forces. This led to
progress, first with Maxwell's unified theory of electricity and
magnetism and, a century later, with the unification of the weak and
electromagnetic forces for which Weinberg, Salam and Glashow shared
the 1979 Nobel Prize. And the fact that the electroweak and strong
coupling constants become equal at about $10^{15}~{\rm GeV}$, and
that this energy is close to the Planck energy, seems (to particle
theorists) to point to a fully unified theory at very high energies.

Relativists come to quantum gravity with a completely different
historical perspective. For them general relativity is a model which
has stood the test of time. Whenever people thought there was a
problem which necessitated changing general relativity it turned out,
upon closer examination of either the theory or the data, that general
relativity was right and the proposed changes were wrong. Among relativists
the people who made mistakes were those who tinkered with the model on
the basis of incomplete data or anything less than a rigorous theoretical
analysis. The first example was none other than Albert Einstein, who in
1917 introduced the cosmological constant $\Lambda$ into his gravitational
field equations (\ref{Einstein}) in order to prevent the universe from
expanding in the simplest cosmological realization of general relativity.
Of course Hubble actually quantified this expansion in 1929 \cite{Hubble},
and Einstein could have predicted it had he just stuck with his original
formulation of general relativity. He called this the greatest blunder of
his life.

Relativists are also familiar with a vast collection of ``paradoxes''
which purport to show that either special or general relativity is wrong,
and which can only be debunked by carefully identifying false assumptions.
So it seems very natural to a relativist to reject the result of an
asymptotic series expansion, especially when divergences are present.
They distrust the idea of considering something as ``small'' when it is
actually divergent, and they won't be satisfied that there is anything
wrong with quantum general relativity until a rigorous proof is supplied
which is not based on perturbation theory.

I am myself a particle theorist but I have friends in both camps and it
is sometimes difficult to make them see any worth in the other side's
views. Although I share the fondness of my particle colleagues for
perturbation theory, let me reply to an objection they sometimes raise
about doubting the validity of perturbative results for quantum general
relativity. The objection takes the form of an exasperated question:
{\it Perturbation theory is supposed to be valid when the corrections it
generates are small and, whatever is the right fundamental theory, quantum
gravitational corrections must be small at low energies because we have
never observed a single one! How then can you people refuse to accept a
perturbative treatment of quantum general relativity?}

My answer is twofold:
\begin{enumerate}
\item{There might be low energy quantum gravitational effects which
masquerade as something else; and}
\item{Perturbation theory might not be generating the correct asymptotic
series expansion for quantum general relativity.}
\end{enumerate}
I will have more to say about the first possibility in the next subsection.
Concerning the second, suppose the actual series expansion consists not
of just powers of $G E^2/\hbar c^5$ but also logarithms,
\begin{equation}
\sum_{\ell = 0}^{\infty} \Bigl(\frac{ G E^2}{\hbar c^5}\Bigr)^{\ell}
\sum_{k = 0}^{\ell} a_{k\ell} \ln\Bigl(\frac{G E^2}{\hbar c^5}\Bigr)
= a_{00} + a_{11} \Bigl(\frac{G E^2}{\hbar c^5}\Bigr)
\ln\Bigl(\frac{G E^2}{\hbar c^5}\Bigr)
+ a_{01} \ln\Bigl(\frac{G E^2}{\hbar c^5}\Bigr) + \ldots
\end{equation}
Note first that trying to beat this into the form of an expansion in
powers of $G$ would result in uncontrollable logarithmic
divergences, which is what we see in quantum general relativity.
Note also that low energy quantum gravitational corrections would
still be unobservably small, just not quite as small as without the
logarithms. For example, at LHC energies the suppression factor
would not be the figure of $\sim 3.4 \times 10^{-31}$ we got in
expression (\ref{LHC}). We would instead get this number times its
logarithm,
\begin{equation}
\Bigl(3.4 \times 10^{-31}\Bigr) \cdot \ln\Bigl(3.4 \times 10^{-31}\Bigr) \sim
-2.4 \times 10^{-29} \; .
\end{equation}
The extra factor of about 100 represents an enormous enhancement, but the
effect is still many orders of magnitude below observability.

Exotic terms occur in many familiar asymptotic expansions, and
divergences are the typical signature of their appearance. Consider
the logarithm of the grand canonical partition function for
non-interacting, non-relativistic bosons of mass $m$ in a three
dimensional volume $V$:
\begin{equation}
\ln\Bigl(\Xi\Bigr) = V n_Q \sum_{k=1}^{\infty} k^{-\frac52} \, \exp(k
\beta \mu) \; .
\end{equation}
Here $n_Q \equiv (m k_B T/2\pi \hbar^2)^{\frac32}$ is the quantum
concentration, $\mu < 0$ is the chemical potential, and $\beta =
(k_B T)^{-1}$. Near Bose-Einstein condensation one has $0 < -\beta
\mu \ll 1$ so it should make sense to expand $\ln(\Xi)$ for small
$\beta \mu$. Straightforward perturbation theory corresponds to the
following expansion:
\begin{equation}
\ln\Bigl(\Xi\Bigr) = V n_Q \sum_{k=1}^{\infty}
k^{-\frac52} \, \sum_{\ell=0}^{\infty} \; \frac{(k \beta \mu)^{\ell}}{\ell!}
\longrightarrow V n_Q \sum_{\ell=0}^{\infty} \frac{(\beta \mu)^{\ell}}{\ell!}
\sum_{k=1}^{\infty} k^{\ell-\frac52} \; . \label{oscdiv}
\end{equation}
Although the $\ell=0$ and $\ell=1$ terms are finite, the sum over $k$ diverges
for $\ell \geq 2$.

The divergences we have encountered do not mean that higher corrections
are large, just that they are not as small as $(\beta \mu)^2$. One sees
this by expanding the second derivative around its integral approximation:
\begin{eqnarray}
\lefteqn{\frac1{V n_Q} \, {\partial^2 \ln\Bigl(\Xi\Bigr) \over \partial
(\beta \mu)^2} = \sum_{k=1}^{\infty} k^{-\frac12} \, \exp(k \beta \mu) \; ,} \\
& & = \int_0^{\infty} \!\!\! dy \; y^{-\frac12} \, \exp(y \beta \mu)
+ \sum_{k=1}^{\infty} \Bigl[ k^{-\frac12} \,
\exp(k \beta \mu) - \int_{k-1}^k \!\!\!\!\! dy \, y^{-\frac12} \,
\exp(y \beta \mu) \Bigr] , \qquad \\
& & = \Bigl({-\pi \over \beta \mu}\Bigr)^{
\frac12} + \sum_{k=1}^{\infty} \Bigl[k^{-\frac12} - 2 k^{\frac12} + 2
(k\!-\!1)^{\frac12}~\Bigr] + O\Bigl(\beta \mu\Bigr) . \qquad
\end{eqnarray}
Integration reveals the true asymptotic expansion:
\begin{equation}
\ln\Bigl(\Xi\Bigr) = V n_Q \Biggl\{\zeta\Bigl(\frac52\Bigr) +
\zeta\Bigl(\frac32\Bigr) \, \beta \mu + \frac43 \sqrt{\pi} (-\beta \mu)^{
\frac32} + O\Bigl(\beta^2 \mu^2\Bigr)\Biggr\} \; . \label{Xiexp}
\end{equation}
The oscillating series of ever-increasing divergences in the perturbative
expansion (\ref{oscdiv}) has resolved itself into a perfectly finite,
fractional power. If only we had the analytical power to check for such
behavior in quantum general relativity!

\subsection{How We'd Use Quantum Gravity if We Had It}

The unsatisfactory state of affairs in quantum general relativity is
that the only computational tool we currently possess offers the choice
between two hopelessly incorrect predictions:
\begin{itemize}
\item{Either quantum gravitational effects are infinitely strong;}
\item{Or else the universe blows up instantly.}
\end{itemize}
But there is obviously some quantum theory of gravity, because quantum
matter gravitates, and one can discuss what it would tell us if we found
it. To be specific, suppose we discovered a consistent and plausible
quantum theory of gravity whose dimensionless coupling strength is that
of quantum general relativity, $G E^2/\hbar c^5 \sim (E/10^{19}~{\rm GeV})^2$.
An irony of this subject is that, as soon as we manage to avoid predicting
infinitely strong effects, we are almost inevitably left with no observable
predictions at all because $G E^2/\hbar c^5$ is so small for any process
which can be contrived in the laboratory!

One might think observable effects could be obtained by resorting to large
masses, such as that of the Earth. This indeed gives measurable gravitational
effects from the incoherent sum over many sources, but {\it quantum}
gravitational effects derive from the energy of only a single mode. So the
energy $E$ appropriate for quantum gravitational corrections to the Earth's
potential is not the Earth's enormous rest mass energy of $M_E c^2$ but
rather the minuscule energy $E = \hbar c/R$ of the mode whose wave length
is the radius $R$ at which the potential is measured. Quantum gravitational
corrections to the Earth's classical potential of $-G M_E/R$ take the
form \cite{gravpot},
\begin{equation}
\Phi = -\frac{G M_E}{R} \Bigl\{ 1 + {\rm const} \times \frac{G \hbar}{R^2 c^3}
+ \dots \Bigr\} \; . \label{dEarth}
\end{equation}
At the surface of the Earth the fractional change would be about,
\begin{equation}
\frac{G \hbar}{R_E^2 c^3} \sim 10^{-84} \; .
\end{equation}
Even if we could measure gravity that accurately (and we cannot), this
change is vastly smaller than the classical effect from the mass of the
human taking the data!

How to observe a very weak interaction is not an unprecedented problem
in the long history of physics. There are two general approaches:
\begin{itemize}
\item{Find a regime in which the interaction is not so weak; or}
\item{Exploit some unique property of the interaction that gives rise to
effects for which there is no background from other interactions.}
\end{itemize}
I will comment on both approaches.

The obvious way of overcoming suppression by the factor $G E^2/\hbar c^5$
is to scale up the energy $E$. We cannot build accelerators which approach
interesting energy ranges but nature does this for us in four cases:
\begin{itemize}
\item{The initial singularity which must occur, on very general grounds,
either without primordial inflation \cite{HE} or with it \cite{sing};}
\item{The final stages of black hole collapse;}
\item{The final stages of black hole evaporation; and}
\item{The phase of primordial inflation.}
\end{itemize}
The first three processes can access modes of arbitrarily high energy. The
final one might reach as high as $E \sim 10^{13}~{\rm GeV}$, at which quantum
gravitational effects would be small but not negligible. So there are good
reasons to expect significant quantum gravitational effects in all four
cases; the issue is finding some signature of them that reaches us here and
now. Ideas about how to do this for the initial singularity and for black
hole collapse are still quite speculative, and we have not discovered any
black holes near the end of their existence. However, there is by now a
well-developed formalism for tracing quantum gravitational effects from
primordial inflation to the current epoch. The simplest interpretation of
the data from anisotropies in the cosmic microwave background \cite{WMAP}
and from large scale structure surveys \cite{SDSS} is that the primordial
perturbations in the gravitational potential of our universe arose from
quantum matter fluctuations near the end of inflation. I will review the
argumentation in section 5.

So much for the first approach; the other technique is to identify processes
driven by some special feature of gravity that no other force possesses. So
if we see the effect at all, no matter how weak it is, it must be from
gravity. For example, particle physicists did not discern the weak nuclear
force in the background of vastly stronger quantum electrodynamic processes
but rather because it alone mediates decays such as $\mu^- \rightarrow e^-
\nu_{\mu} \overline{\nu}_e$.

There are four special features of gravity which deserve comment:
\begin{itemize}
\item{One of the gravitational parameters is the cosmological constant
$\Lambda$;}
\item{It determines the maximum speed at which signals can propagate;}
\item{Gravitons have zero mass without being driven to zero amplitude by
the expansion of the universe; and}
\item{The gravitational interaction energy is negative.}
\end{itemize}
The third point is what my own current research concerns and it will make
more sense if I postpone it to section 5. I will discuss the first two
points briefly and the last one at greater length.

The cosmological constant $\Lambda$ multiplies a term in the Einstein
equations (\ref{Einstein}) without any derivatives. It influences the
rate at which the overall expansion of the universe is accelerating;
positive $\Lambda$ tends to make the universe accelerate whereas negative
$\Lambda$ tends to make it decelerate. (The spatially homogeneous parts
of certain matter fields also play a role.) The current universe is
accelerating \cite{SNIA,Yun}, which is consistent with $\Lambda$ having
a small, positive value,
\begin{equation}
\Bigl(\Lambda\Bigr)_{\rm meas} \sim + 10^{-52}~{\rm m}^{-2} \; . \label{Lamval}
\end{equation}

The measured value (\ref{Lamval}) of the cosmological constant is
outlandish! From equation (\ref{QGren2}) one can see that first
order quantum gravitational correction has the form $\Delta \Lambda
\sim G\hbar K^4/c^3$, where $K$ is the cutoff wave number. Of course
one must cancel the divergence when $K$ goes to infinity, but there
will obviously be a finite remainder which takes the same form with
some finite $K$. (As particle theorists used to remark during the
period renormalization was being worked out, ``Just because
something is infinite does not mean it is zero'' \cite{Weinberg}.)
The trouble is that the other scales in physics give values for
$\Delta \Lambda$ which are vastly larger than (\ref{Lamval}),
\begin{eqnarray}
{\rm Planck\ Scale} \; \Bigl(K^2 = \frac{c^3}{G \hbar}\Bigr) & \Longrightarrow
& \Delta \Lambda = \frac{c^3}{G \hbar} \sim 10^{121} \times 10^{-52}~{m}^{-2}
\; , \qquad \\
Z \; {\rm Boson\ Mass} \; \Bigl(K = \frac{m_Z c}{\hbar}\Bigr) & \Longrightarrow
& \Delta \Lambda = \frac{G c m_Z^4}{\hbar^3} \sim 10^{53} \times
10^{-52}~{m}^{-2} \; , \qquad \\
{\rm Electron\ Mass} \; \Bigl(K = \frac{m_e c}{\hbar}\Bigr) & \Longrightarrow &
\Delta \Lambda = \frac{G c m_e^4}{\hbar^3} \sim 10^{32} \times
10^{-52}~{m}^{-2} \; . \qquad
\end{eqnarray}
Particle theorists refer to a mismatch of this type as a {\it hierarchy
problem}, and the one associated with the cosmological constant is the
worst in all of fundamental theory \cite{Carroll}. Of course $\Lambda$ is
free parameter in the Einstein equations (\ref{Einstein}) and it has to
take some value, so why not precisely the number which gives (\ref{Lamval})?
That could be, but many people suspect we are missing a key principle
from quantum gravity \cite{DETF}.

Because the metric field $g_{\mu\nu}(t,\vec{x})$ determines physical
lengths and times, it controls the maximum rate at which signals can
propagate. A tiny quantum fluctuation in the metric could allow some
photons of light from a distant galaxy to reach us a little sooner
than others. Because we don't know cosmic distances very well the
potentially observable effects would be a blurring of images,
fluctuations in luminosity and a broadening of spectral lines
\cite{Ford1}. Since the earliest days of quantum gravity such
effects have been termed {\it smearing of the light-cone}
\cite{smear}. Surprisingly, the lowest order contributions can be
computed using first order perturbation theory and they give finite
results \cite{Ford2}. These results are still unobservably small,
but not by much, and they may well be detectable in future laser
interferometers \cite{Ford3}. It could be that history repeats
itself because the first quantum electrodynamic correction to the
electron magnetic moment is finite, and was derived by Schwinger in
1948 \cite{JS2} while physicists were still puzzling out how to
fully absorb all the divergences.

The final special feature of gravity is its negative interaction energy.
That raises a fascinating possibility in the context of computing
the contribution to a particle's measured mass from the interaction with
its own force fields. Every beginning physics student is taught that the
electric potential from a collection of charges $q_i$ at fixed positions
$\vec{x}_i$ is,
\begin{equation}
\Phi(\vec{x}) = \sum_i \frac{q_i}{4\pi \epsilon_0 \Vert \vec{x} \!-\!
\vec{x}_i\Vert } \; . \label{Estatic}
\end{equation}
They are also taught to compute the electrostatic interaction energy
between these charges by summing $\frac12 q_i \Phi(\vec{x}_i)$, where
$\Phi(x_{i})$ {\it is the potential due to all the other charges},
\begin{equation}
E_{\rm EM} = \sum_{i} \sum_{j\neq i} \frac{q_i q_j}{8 \pi \epsilon_0 \Vert
\vec{x}_i \!-\! \vec{x}_j\Vert } \; .
\end{equation}
The curious rule about omitting the charge's interaction with its
own field (which is the strongest contribution!) derives from the
fact that setting $\vec{x} = \vec{x}_i$ in (\ref{Estatic}) produces
a divergent potential. More advanced students are instructed to
regulate this divergence and then absorb it into the unobservable
bare mass of the particle in such a way as to make the total
self-energy of the particle agree with its measured mass. (Cf.
chapter 16 of the text by Jackson \cite{Jackson}.)

Arnowitt, Deser and Misner have worked out how this procedure changes,
on the classical level, when the gravitational interaction is included
\cite{ADM}. It is simplest to model the particle as a stationary spherical
shell of radius $R$ (that is the regularization) charge $e$ and bare mass
$m_0$. In Newtonian gravity the shell's energy would be,
\begin{equation}
E_R = m_0 c^2 + \frac{e^2}{8 \pi \epsilon_0 R} - \frac{G m_0^2}{2 R} \; .
\end{equation}
It turns out that all the effects of general relativity are
accounted for by replacing $E_R/c^2$ and $m_0$ with the full mass
$m_R$,
\begin{equation}
m_{R} c^2 = m_0 c^2 + \frac{e^2}{8 \pi \epsilon_0 R} - \frac{G m^2_{R}}{2 R}
= \frac{R c^4}{G} \Biggl[-1 + \sqrt{1 + \frac{2 G}{R c^4} \Bigl(m_0 c^2 +
\frac{e^2}{8 \pi \epsilon_0 R}\Bigr)}~\Biggr] \; .
\end{equation}
It should be noted that Arnowitt, Deser and Misner rigorously solved
the constraint equations of general relativity and electrodynamics,
and then used the asymptotic metric to compute the ADM mass. They
also developed the simple model I am presenting \cite{ADM}.

The perturbative result is obtained by expanding the square root,
\begin{equation}
m_{\rm pert} c^2 = m_0 c^2 + \frac{e^2}{8 \pi \epsilon_0 R} + \sum_{n=2}^{
\infty} \frac{(2n-3)!!}{n!} \; \Biggl(-\frac{G}{R c^4}\Biggr)^{n-1}
\Biggl(m_0 c^2 + \frac{e^2}{8 \pi \epsilon_0 R}\Biggr)^n \; , \label{pertexp}
\end{equation}
and shows an oscillating series of increasingly singular terms. The
alternating signs derive from the fact that gravity is attractive.
The positive divergence of order $e^2/R$ evokes a negative
divergence or order $G e^4/R^3$, which results in a positive
divergence of order $G^2 e^6/R^5$, and so on. The reason these terms
are increasingly singular is that the gravitational response to an
effect at one order is delayed to a higher order in perturbation
theory.

The correct result is obtained by taking $R$ to zero before expanding in the
coupling constants $e^2$ and $G$,
\begin{equation}
\lim_{R \rightarrow 0} m_{R} c^2 = \Biggl(\frac{e^2}{4 \pi \epsilon_0 G}
\Biggr)^{\frac12} c^2 = \sqrt{\alpha} M_{\rm Planck} c^2 \sim
10^{18}~{\rm GeV} \; . \label{exact}
\end{equation}
Like the expansion of $\ln(\Xi)$ in expression (\ref{Xiexp}) it is finite
but not analytic in the coupling constants $e^2$ and $G$. Unlike the expansion
of $\ln(\Xi)$, it diverges for small $G$. This is because gravity has
regulated the linear self-energy divergence which results for a
non-gravitating charged particle.

One can understand the process from the fact that gravity has a
built-in tendency to oppose divergences. A charge shell does not
want to contract in pure electromagnetism; the act of compressing it
calls forth a huge energy density concentrated in the nearby
electric field. Gravity, on the other hand, tends to make things
collapse, especially large concentrations of energy density. The
dynamical signature of this tendency is the large negative energy
density concentrated in the Newtonian gravitational potential. In
the limit of $R \rightarrow 0$ the two effects balance and a finite
total mass results.

Expressed this way, there seems to be no reason why gravitational
interactions should not cancel divergences in quantum field theory
the same way they do in classical field theory. It is significant that
the divergences of some quantum field theories --- such as quantum
electrodynamics --- are weaker than the linear ones which ADM have
shown that classical gravity controls \cite{Weisskopf}. So less
cancellation is necessary and one might expect a smaller final mass,
closer to the values of known charged particles. The frustrating
thing is that the only computational tool we possess for quantum
field theory is perturbation theory, and one cannot hope to see the
cancellation perturbatively. In perturbation theory the
gravitational response to an effect at any order must be delayed to
a higher order. This is why the perturbative result (\ref{pertexp})
consists of an oscillating series of ever higher divergences. What
is needed is an approximation technique in which the gravitational
response is able to keep pace with what is going on in other
sectors.

Note that any finite bare mass drops out of the exact result (\ref{exact})
in the limit $R \rightarrow 0$. This makes for an interesting contrast with
the usual program of renormalization. Without gravity one would adjust the
bare mass $m_0$ to be whatever divergent quantity is necessary to produce
the measured mass $m_{\rm meas}$,
\begin{equation}
m_0 c^2 = m_{\rm meas} c^2 - \frac{e^2}{8 \pi \epsilon_0 R} \; .
\end{equation}
Of course the same procedure would work with gravity as well,
\begin{equation}
m_0 c^2 = m_{\rm meas} c^2 - \frac{e^2}{8 \pi \epsilon_0 R} +
\frac{G m_{\rm meas}^2}{2 R} \; .
\end{equation}
The difference with gravity is that we have an alternative: keep $m_0$ finite
and let the dynamical cancellation of divergences produce a unique result for
the measured mass. This would fulfill the old dream of deriving particle
masses from their self-interactions. It would also mean that fundamental
particle masses represent disguised quantum gravitational effects which
would provide sensitive tests of the theory of quantum gravity if only we
possessed the analytical tools to predict them.

\section{Current Approaches to Quantum Gravity}

Many fine physicists have burned away their lives grappling with the
problem of quantum gravity. There is not space in this article to discuss
all their efforts, nor do I possess the expertise for it. I will here review
five of the most popular approaches that are still being pursued. Three of
them derive from the particle theory belief that general relativity must be
changed; this common origin and methodology dictates that they should be 
presented consecutively. The remaining two approaches derive from the 
relativist's belief that quantum general relativity might be alright if 
studied nonperturbatively. I will cite review articles and books specific 
to each approach but I would also like to recommend the general review 
article by Carlip \cite{Carlip1}.

\subsection{Superstring Theory}

A central point to understanding string theory is that it cannot be
formulated the way all other fundamental theories are, by giving the 
dynamical variables and the equations they obey. We do not know what the 
fundamental dynamical variables of string theory are, nor the equations 
they obey. What we have instead is a formalism for perturbatively 
computing what is the usual observable of a quantum field theory, the 
S-matrix.\footnote{Maldacena's AdS/CFT correspondence \cite{Maldacena}
is now widely accepted as providing a nonperturbative formulation of string 
theory (in terms of an ordinary quantum field theory) for the boundary 
conditions associated with the most symmetric solution of general 
relativity with a negative cosmological constant.} The reasons for this 
are historical so I will summarize how string theory was developed.

String theory began in the late 1960's as an attempt to understand the
strong interactions. Experiment had shown a series of resonances whose
mass-squared $m^2$ increases approximately linearly as a function of angular
momentum $J$, starting from a positive intercept,
\begin{equation}
m^2 = +m_0^2 + \Delta m^2 J \; . \label{linris}
\end{equation}
It was obvious to everyone that the strong interactions could not be
treated perturbatively (no one then suspected that the strong interaction
would become weaker at high energies) so, instead of proposing quantum 
field theories, physicists tried to guess scattering amplitudes which 
incorporated such resonances. The first to succeed was Gabriele Veneziano 
\cite{Veneziano}, who proposed what would become known as the 4-particle,
open string tree amplitude. Miguel Virasoro found the analogous amplitude
for a closed string \cite{Virasoro}. These 4-particle amplitudes were 
quickly generalized to give $N$-particle scattering for open \cite{openN} 
and for closed strings \cite{closedN}.

The early string amplitudes had linearly rising resonances (\ref{linris})
but they suffered from three problems:
\begin{itemize}
\item{They did not include fermions;}
\item{They contained resonances with the wrong sign to come from physical
particles; and}
\item{They started from a negative intercept, rather than a positive one.}
\end{itemize}
Particles with the wrong sign are called ``ghosts'' and they can be viewed 
as making the theory decay instantly through a kinetic instability like the 
Ostrogradskian instability I discussed in section 2.5. Particles with a 
negative mass-squared are called ``tachyons.'' People sometimes make the
mistake of thinking of them as particles that move faster than the speed
of light but what they really are is instabilities of the potential energy. 
In a field theory such an instability would not be serious, it just 
means the field decays to some value with a lower potential energy, as 
happens when electroweak symmetry breaking takes place in the Standard
Model. But string theory is not based upon a field theory so there is no
principle to tell us how the perturbative background shifts --- or even 
what the perturbative background is on the fundamental level. All we have 
is the S-matrix about a handful of backgrounds, and the appearance of a 
tachyon in the spectrum implies that this S-matrix cannot be correct.

It was my University of Florida colleague, Pierre Ramond who worked out 
how to incorporate free fermions in 1971 \cite{Ramond}. By this time it had 
been recognized that string theory scattering amplitudes can be written
as integrals with respect to the coordinates of a string $X^{\mu}(\sigma)$,
similar to the way that ordinary quantum field theory scattering amplitudes
can be written as integrals with respect to the spacetime coordinates of
a point particle $x^{\mu}$. In neither case are the true dynamical degrees
of freedom these coordinates; for quantum field theory the true dynamical
degrees of freedom are the various fields, no one knows what they are for 
string theory.

Shortly after Ramond's work, Neveu and Schwarz added a new kind of string 
fermions to produce the amplitudes for interacting bosons \cite{ANJHS}.
Although these models had supersymmetry in the string coordinate space,
the amplitudes themselves were not supersymmetric. In 1976 Gliozzi, Scherk 
and Olive showed how to get supersymmetric amplitudes by combining Ramond's
formalism with that of Neveu and Schwarz, and then projecting out a
certain sector which includes the problematic tachyon \cite{GSO}. This was 
the birth of superstring theory.

Ramond's work is tremendously significant for quantum gravity because it
represents the first appearance of a supersymmetry which connects 
fermions and bosons. There are many interesting things about supersymmetry
but its importance for quantum gravity derives from a fact I mentioned
in section 2.4: the 0-point energies of bosons and fermions with the same 
mass $m$ and wave number $k = \Vert \vec{k}\Vert$ cancel,
\begin{eqnarray}
{\rm Bosons} & \Longrightarrow & +\frac12 \sqrt{m^2 c^4 + (\hbar c k)^2} 
\; , \\
{\rm Fermions} & \Longrightarrow & -\frac12 \sqrt{m^2 c^4 + (\hbar c k)^2} 
\; .
\end{eqnarray}
Supersymmetry involves a tight relation between fermions and bosons,
which is necessary if this cancellation is to do any good for quantum 
gravity. With this tight relation,  every correction from the 0-point 
motion of bosons in the theory tends to be canceled by an opposite 
correction from the 0-point motion of fermions.

At this point I must digress to discuss the implications supersymmetry
has for the cosmological constant $\Lambda$. Unbroken supersymmetry
is only consistent with $\Lambda$ being zero or negative, not positive.
If it exists in nature then supersymmetry must be badly broken at low 
energies because unbroken supersymmetry predicts that every known 
particle has a ``super-partner'' (bosonic super-partners for known 
fermions and fermionic super-partners for known bosons) with the same 
mass, and not a single one of these super-partners has been observed. 

One does not need superstring theory in order to have supersymmetry. In 
fact, supersymmetric algebras and/or quantum field theories were developed 
before superstring theory by Golfand and Likhtman \cite{GL}, by Volkov and 
Akulov \cite{VA}, by Volkov and Soroka \cite{VS} and by Wess and Zumino 
\cite{JWBZ}. Supersymmetry had also been extended to gravity to produce a 
class of models known as {\it supergravity} by Freedman, van Nieuwenhuizen and 
Ferrara \cite{FvNF} and by Deser and Zumino \cite{DZ}. (I shall have much 
more to say about supergravity in the next subsection.) A fact of great 
importance for this exposition is that {\it the cosmological constant can 
have any sign in generic theories of gravity but it can only be negative or 
zero in supergravity and superstring theory.} So the only way supergravity
or superstring theory can be consistent with the observed acceleration of 
the current universe \cite{SNIA,Yun} is to suppose that we are now in a 
metastable state which will eventually decay to a universe which is either 
decelerating or actually contracting.

Recall that the initial string amplitudes had three problems: no fermions,
ghosts and a tachyon. We have just seen that superstring theory resolves
the first problem and the third one. My University of Florida colleague, 
Charles Thorn was among those who proved that the ghosts drop out in 
certain key dimensions: $D=26$ dimensions for bosonic string theory and 
$D=10$ dimensions for superstrings \cite{RCB,PGCBT}. 

The ``no-ghost'' theorems actually came in 1972, before the invention of 
superstring theory, so string theory still had the tachyon problem. It
also had massless particles: a spin one particle in the open string
amplitudes and a spin two particle in the closed string amplitudes. 
Although these particles pose no problem for stability, they are no more
part of the observed spectrum of strongly interacting particles than 
tachyons. The funny dimensions dictated by the no-ghost theorems also 
seemed to preclude using string theory to describe the strong interactions. 
And recall from section 2.4 that Politzer \cite{Pol} and Wilzcek and Gross 
\cite{WG} showed in 1973 that the currently accepted theory of the strong 
interactions gets weaker at high energies, which meant perturbation theory 
can be used to make predictions. These predictions were confirmed, by which 
point few people were interested in string theory. 

In 1974 Joel Scherk and John Schwarz made a virtue of the massless 
particles by showing that they could be the photon and the graviton, 
respectively \cite{JSJHS1}. A year later Scherk and Schwarz proposed that 
the six extra dimensions of (what would become) superstring theory were 
``compactified'' \cite{JSJHS2}. A dimension which is compactified does 
not extend to macroscopic distances the way the three known spatial 
dimensions do; instead it is rolled up into a circle (or more complicated 
shape) of radius so small that we cannot observe motion in this direction. 
With the advent of superstring theory the next year all the ingredients 
were in place for a potential theory of everything.

At first very few people were interested. The majority of particle 
physicists were working out the consequences of what would become known
as the {\it Standard Model}. More and more people also began studying
the {\it Grand Unified Theories} which were suggested by the fact that 
the three Standard Model coupling constants become comparable at very 
high energies. Proper, mainstream particle theorists did {\it not} worry 
about quantizing gravity! I recall one of my graduate professors being 
asked for a reference on supersymmetry and supergravity and replying, 
with lofty disdain:
\begin{quote}
{\it Supersymmetry is one of those subjects which we here at Harvard try to 
discourage students from studying.}
\end{quote}
Students who wished to study quantum gravity at Harvard did so through a
sort of ``underground railroad'' in which Sidney Coleman signed the
forms but Stanley Deser was our true adviser.

Supersymmetry was eventually accepted by particle theorists as a way of
solving a hierarchy problem somewhat less severe than the one I discussed
in section 3.2 with regard to the cosmological constant. Supersymmetry
also allowed particle theorists to continue using perturbation theory,
which is our chief analytical tool, to very high energies. Quantum gravity
got accepted shortly thereafter. Part of the reason for this is that the
scale then envisaged for Grand Unification ($10^{16}~{\rm GeV})$ is only
three orders of magnitude below the Planck scale ($10^{19}~{\rm GeV}$).
Another reason is that the extra components of a higher dimensional
metric (it has $\frac12 D (D+1)$ components in $D$ spacetime dimensions)
could be regarded as the other kinds of bosonic particles, thereby 
unifying gravity and the other forces. For example, the 15 components 
of a 5-dimensional metric could be regarded as the ten components of
our 4-dimensional metric, plus a 4-component vector potential and a 
scalar particle.

Superstring theory was rehabilitated because higher dimensional general
relativity isn't a very good way of unifying all the forces. For one
thing, it wasn't clear how to make it incorporate fermions correctly.
For another, increasing the number of dimensions makes the divergence
problem worse. It is easy to understand why from the discussion of section
2.4. Recall that the first order gravitational response to 0-point motion
in a static gravitational field of wave number $\vec{p}$ involves a
divergent mode sum of the form,
\begin{equation}
\int^{K} \!\!\! \frac{d^3k}{(2\pi)^3} \frac{\mathcal{E}^4}{[E(\vec{k}) \!+\!
E(\vec{p} \!-\! \vec{k})]^3} = A K^4 + B K^2 p^2 + C \ln(K^2) p^4 +
{\rm Finite} \; , \label{4prob}
\end{equation}
where $\mathcal{E}^4$ consists of a variety of terms quartic in $\vec{p}$ 
and $\vec{k}$, $K$ is the cutoff and $A$, $B$ and $C$ are numbers of order 
one. The details of how extra dimensions are compactified don't matter in 
the regime of large wave numbers which gives rise to the problem. So going 
to six spacetime dimensions amounts to changing the wave vector in 
(\ref{4prob}) from $\vec{k} = (k_1,k_2,k_3)$ to $\vec{k} = 
(k_1,k_2,k_3,k_4,k_5)$ to give,
\begin{equation}
\int^{K} \!\!\! \frac{d^5k}{(2\pi)^5} \frac{\mathcal{E}^4}{[E(\vec{k}) \!+\!
E(\vec{p} \!-\! \vec{k})]^3} = A K^6 + B K^4 p^2 + C K^2 p^4 + D \ln(K^2) p^6 
+ {\rm Finite} \; \; . \label{6prob}
\end{equation}
This requires even more unacceptable counterterms than the 4-dimensional
theory!

Interest in superstring theory surged in 1984 when Green and Schwarz 
showed that the theory naturally incorporates the right kinds of fermions
and is also likely to be finite \cite{MBGJHS}. The statement about finiteness
might seem surprising because the no-ghost theorems require superstring
theory to exist in $D=10$ spacetime dimensions and I have just explained
that increasing the dimension makes the divergences of quantum general
relativity worse, not better. Of course the explanation is that superstring
theory is {\it not} general relativity; it doesn't even have the metric as
one of its fundamental degrees of freedom. Although superstrings do 
incorporate gravity, stress-energy is not the source of gravity at high 
wave number, so superstrings violate the 3rd of the four propositions listed
at the beginning of section 3.

All of this raises the question of what superstring theory is on the
fundamental level. The unsatisfactory answer is that no one knows! Based 
on the way we have of expressing the perturbative S-matrix one might think 
the fundamental variable should be a {\it string field}, that is, a field
whose argument is a string's position in spacetime $X^{\mu}(\sigma)$,
just like a normal field could be regarded as depending upon a particle's
position $x^{\mu}$. String field theories can indeed be constructed which
reproduce the perturbative string S-matrix \cite{Giddings}. The 
earliest ones were in a noncovariant formalism due to Kaku and Kikkawa 
\cite{MKKK}, but the 1980's witnessed the development of lovely invariant 
Lagrangians by Witten \cite{Witten1} and Horowitz, Lykken, Rohm and 
Strominger \cite{HLRS}. The trouble with these string field theories is
that they must be {\it nonlocal}, that is, they cannot be expressed in
terms of just the dynamical variable and a finite number of its derivatives
\cite{DJGAJ}. This is not some defect of how a string field theory was 
constructed from the perturbative S-matrix, it {\it must} be true in order for 
superstrings to avoid divergences. Instead of the coupling to the energy
of a mode, superstring field theories couple to the energy times exponentials 
of the wave number,
\begin{eqnarray}
{\rm General\ Relativity} & \Longrightarrow & E(\vec{k}) = \hbar c 
\Vert \vec{k}\Vert \; , \\
{\rm String\ Field\ Theory} & \Longrightarrow &
\hbar c \Vert \vec{k}\Vert \times e^{-\alpha' \Vert \vec{k} \Vert^2} \; ,
\label{stringcoup}
\end{eqnarray}
where $\alpha'$ is the Regge slope parameter. Of course these exponentials
make the integrals converge at high $\Vert \vec{k}\Vert$, but one has to
wonder if they engender new problems.

Anyone who has studied quantum mechanics knows that exponentiating the 
derivative operator effects a spatial translation,
\begin{equation}
e^{\Delta x \, d/dx} f(x) = \int_{-\infty}^{\infty} \! \frac{dk}{2\pi}
\, e^{ik x} \times \Bigl[e^{i k \Delta x} \widetilde{f}(k) \Bigr] = f(x \!+\!
\Delta x) \; . 
\end{equation}
Exponentiating the square of a derivative, as in (\ref{stringcoup}), 
involves a superposition over translations of all distances out 
to infinity. And one must of course include {\it temporal} as well as 
spatial derivatives or else there would be a massive violation of 
Lorentz invariance. Recall from section 2.5 that adding even a single
higher time derivative results in new degrees of freedom which have
the opposite kinetic energy to the lower derivative degrees of freedom.
It turns out that the problem grows worse the more derivatives you add:
there is an extra degree of freedom for each new time derivative and
essentially half of these new degrees of freedom carry negative kinetic
energy. That is why we cannot modify the gravitational equations of
motion to include the $C^2$ counterterm, or the higher derivative
counterterm whose necessity for pure gravity proven by Goroff and Sagnotti 
\cite{GS,Ven}! So is it alright to have exponentials of the derivative
operator in string field theory?

Not every nonlocal field theory succumbs to the Ostrogradskian instability
\cite{Holger}. But the nonlocality of string field theory must be 
restricted to entire functions of the derivative operator or else the
conservation of probability would not work out correctly. And the 
definition of an entire function is that it converges to its Taylor series
expansion. This means that string field theory can be viewed as the 
limit of a sequence of ever-higher derivative models, which grow more and
more unstable. One sometimes hears the statement that the extra degrees
of freedom might decouple in the limit because they get driven to 
infinite frequency, but recall the fallacy of this argument from section 
2.5: it plays on one's intuition, from lower derivative theories, that 
high frequency modes cannot be excited because there is only a limited 
amount of free energy. That is only true when all degrees of freedom have 
positive energy; when there are negative energy degrees of freedom, the 
high frequency modes can be excited by also exciting modes with the 
opposite energy. Far from dropping out, high frequency modes of
negative energy {\it dominate} because there are so many more of 
them!\footnote{In view of the recurring efforts to legitimize entire 
functions of the derivative operator I should observe that, were they 
allowed, we could solve the divergence problems of quantum gravity without 
superstrings \cite{Moffat}.}

All of this led David Eliezer and me to conclude in 1989 that string field 
theory must suffer from the Ostrogradskian instability \cite{EW}. It doesn't 
show up in perturbation theory simply because the nonlocality is restricted 
to interactions, but it is present in the full theory. Of course the only
reason for studying string field theory is to get a nonperturbative
definition of string theory, so our result means either that string theory
is wrong or else that some other formalism defines string theory on the
nonperturbative level. Most string theorists take the latter view, but
there has so far been no other nonperturbative way of defining the 
formalism in general.\footnote{Note again that Maldacena's AdS/CFT
correspondence \cite{Maldacena} is now widely accepted as providing a 
nonperturbative formulation of string theory (in terms of an ordinary 
quantum field theory) for the boundary conditions associated with the 
most symmetric solution of general relativity with a negative cosmological 
constant.}

It has been notoriously difficult to derive testable predictions from 
superstring theory. A notable exception to this is the 1988 observation by
Antoniadis, Bachas, Lewellen and Tomaras that, if string supersymmetry, 
which must be broken at the low energy scales we can access in experiments, 
should happen to be broken {\it perturbatively}, then there has to be at 
least one ``large'' extra dimension within experimental reach \cite{ABLT}.
This was elaborated further \cite{Ignatios} and has led to a huge amount of 
work which I won't attempt to review, but I do wish to relate a funny story 
about it. The story concerns a ``phenomenologist'' --- the kind of particle 
theorist who works closely with experimentalists to explain data --- who 
was frustrated by the sometimes incomprehensible mathematics of string 
theory. After a seminar by Antoniadis she exclaimed:
\begin{quote}
{\it I can't usually understand much of what the string theorists tell me.
But} {\bf this,} {\it this looks like a prediction!}
\end{quote}

That concludes the portion of string history which has the greatest
relevance to this article. Of course there have been many more developments
in superstring theory. Some of the principal ones are:
\begin{itemize}
\item{In 1995 Witten suggested that relations between the various
10-di\-men\-sion\-al superstring theories point to an 11-dimensional parent
that was dubbed {\it M-Theory} \cite{Witten}.}
\item{Also in 1995, Polchinski showed that extended objects called 
{\it D-branes}, on which open strings can end, are necessary to realize 
one of the relations between superstrings \cite{Polchinski}.}
\item{The first microphysical derivation, in 1996, of the Bekenstein-Hawking
entropy of a black hole by Strominger and Vafa \cite{ASCV}. This is the
most successful application of superstring theory, although one should 
note that subsequent derivations were made (for example, from loop quantum
gravity \cite{Abhay}) and Carlip has suggested that the result follows from
conformal symmetry near the horizon rather than from the details of
superstring theory \cite{Carlip2}.}
\item{Maldacena's 1997 conjecture that string theory on one spacetime 
manifold is equivalent to a certain quantum field theory without gravity 
on a lower-dimensional manifold \cite{Maldacena}. This is a terrifically
important insight because it might serve as the long-sought, 
nonperturbative definition of superstring theory. It has also been 
applied to nuclear physics and even condensed matter physics with
interesting results.}
\item{The 2000 discovery by Busso and Polchinski that vast numbers of 
discrete choices, called {\it flux vacua}, can be made in compactifying 
superstring theory \cite{RBJP}. The number of these choices is estimated 
to run from between $10^{100}$ and $10^{500}$ \cite{SAMD}, and no principle 
is yet known to fix which one is taken.}
\item{The 2003 suggestion by Susskind that the {\it Anthropic Principle}
might usefully constrain where we are in the vast {\it string landscape} 
of flux vacua \cite{Susskind}. The Anthropic Principle states that physical 
laws must be such that humans of some other form of intelligent life can
exist to observe them.}
\end{itemize}

The last two developments on my list, and the fact that string theory 
predicts the wrong sign for the cosmological constant, have led to a 
slackening of interest in superstrings as a fundamental theory of 
everything, although there is still much work on applications. (I will 
discuss one in the next subsection.) In the face of numbers like $10^{500}$, 
some physicists now speak of a ``theory of anything'' rather than a theory 
of everything. And even veterans of decades of string research find it 
difficult to accept the Anthropic Principle. A personal anecdote might best 
convey the current state of affairs. Early in the spring 2007 semester my 
University of Florida colleague, Charles Thorn, began a seminar by announcing 
his belief that:
\begin{quote}
{\it String theory is just a technique for summing the leading terms in 
the 1/N expansion of QCD.}
\end{quote}
After years of hearing more ambitious assessments this was so shocking
that I checked to be sure I had understood correctly. Charles confirmed 
that I had; in his current view, the effort to regard superstrings as a 
fundamental theory of everything was a blind alley. Later that year I
related Charles' pronouncement to string theory colleagues on three
continents and solicited their own opinions. About half of them agreed 
with him, more often the younger people.

String theory has occupied some of the best minds of particle theory for 
many decades, and it would require more than the space allotted for this
article to do justice to their efforts. Nor do I possess the expertise for
it. Let me instead refer the interested reader to some standard books by 
the leaders of this field. I recommend the texts by Green, Schwarz and 
Witten \cite{GSW}, by Polchinski \cite{JP}, by Zwiebach \cite{BZ}, by 
Kiritsis \cite{EK} and by Becker, Becker and Schwarz \cite{BBS}. In fairness 
I should also mention popular science books which are critical of string 
theory by Smolin \cite{Smolin1} and by Woit \cite{Woit}.

\subsection{On-Shell Finiteness}

Although superstring theory may be short on successful predictions, it
does provide a very efficient way of organizing perturbative calculations
which involve corrections from many components of tensor fields such as
the metric. In fact the simplest method for computing the lowest order 
scattering amplitude for 4-gravitons in general relativity is to use 
superstring theory and then take a certain limit \cite{Sannan}. A few
more tricks from superstring theory, and a {\it lot} of hard work by
many physicists over the course of two decades, has led Bern, Dixon and 
Roiban to conjecture that a version of supergravity might actually be 
{\it On-Shell Finite} to all orders in perturbation theory \cite{BDR}. 
I shall first explain what this means, what the theory is which may have 
this property, and recent results concerning it. Then I will review the 
developments that led to this possibility.

Recall that renormalizable quantum field theories have divergences which 
can be absorbed into redefinitions of free parameters. Once this is done
one can compute not only scattering amplitudes but also the expectation
values of products of the field operators at different points, which are
known as {\it correlation functions}. In contrast, an on-shell finite
theory has scattering amplitudes which are finite, without the need for 
renormalization, but its correlation functions are not necessarily finite 
or even renormalizable.\footnote{However, one can construct redefinitions 
of the original fields which give finite correlation functions.} 
On-shell finiteness has particular importance for theories of gravity
and supergravity because they cannot be perturbatively renormalizable, but
divergences tend to cancel out of their scattering amplitudes. For example,
pure gravity, without any matter, is on-shell finite at first order in
perturbation theory \cite{HV}. One has to go to the second order before 
divergences occur \cite{GS,Ven}. 

Recall that a supersymmetry relates fermions to bosons. This is a very 
good thing for divergences because bosons contribute a positive 0-point
energy whereas fermions contribute a negative 0-point energy, so if
one contrives a tight relation between the two kinds of particles then
it is conceivable that quantum corrections from fermions will cancel
the divergences in quantum corrections from bosons, order-by-order in
perturbation theory. Just that seems to be happening in a model called,
$N=4$ {\it Super-Yang-Mills} theory. I should explain that ``Yang-Mills''
theories are nonlinear generalizations of electrodynamics which can 
describe forces such as the weak and strong interactions. They do not 
have to be supersymmetric, but the term ``$N=4$'' means this particular
model has the maximum amount of supersymmetry possible for a Yang-Mills 
theory. Although this supersymmetry almost certainly makes the model 
finite, it comes at a terrible price: $N=4$ Super-Yang-Mills theory cannot 
incorporate the known particles of the Standard Model. Further, the 
exact cancellation of divergences is only valid as long as the particles 
are all massless, which is very far from the observed universe in which
only the photon, the ``gluon'' which carries the strong force, and the 
graviton appear to be massless.

The gravity model which might be finite is called, ``$N=8$ Supergravity.'' 
Just like $N=4$ Super-Yang-Mills theory, its cannot be a realistic model 
of the universe because its particles are all massless, and they lack other 
essential properties of the known particles in the Standard Model. The 
spectrum of $N=8$ Supergravity consists of: a graviton, 8 spin $\frac32$
particles called {\it gravitinos}, 28 vectors bosons, 56 spin $\frac12$
fermions and 70 scalar bosons. The term $N=8$ means this model has the 
maximum number of independent supersymmetries for a theory of gravity.

It might seem strange that people are so fascinated with a theory which they
know cannot describe the universe. The reason is that supergravity models
with less supersymmetry do have a chance to describe physics. Although
these models are finite at first order in perturbation theory, and even 
at second order, it had been believed that they must suffer uncontrollable
divergences at third order \cite{doubters}. Those expectations are based on 
demonstrating that counterterms exist at higher orders which are not
forced to vanish by any known symmetry or dynamical relation. There is a
widespread belief in particle physics that any divergence which {\it can} 
happen, {\it must} happen. Weinberg refers to such beliefs as 
``folk-theorems'', and this one has never been checked because it was just 
too difficult to make the required computations. The interest in $N=8$ 
Supergravity derives from the fact that it has recently become possible
to check for divergences in this model {\it and they aren't present} at 
third order \cite{3loop}. 

Of course the third order result prompted supergravity experts to 
re-examine the old arguments \cite{doubters} and they soon explained why
the divergence fails to occur \cite{BHS}. They also predicted a new 
divergence at higher order \cite{BHS}. There was even a famous bet about 
this between Zvi Bern and Kelly Stelle. The funny thing is, when the 
prediction was checked, the theory was again seen to be finite \cite{4loop}. 
As a good scientist, Stelle has admitted he lost the bet and is hard at
work trying to understand what happened. 

The computational technology which made this possible began to
be developed in the late 1980's by Zvi Bern and David Kosower. They 
noticed how much more efficient string theory is than the existing
applications of Yang-Mills theory at computing scattering amplitudes.
Recall that Yang-Mills theories can describe the strong interactions,
which are an important background for very high energy experiments.
People needed an efficient way of computing Yang-Mills amplitudes with
lots of external particles and Bern and Kosower developed one by taking 
suitable limits in string theory \cite{ZBDAK1}. By 1990-91 they had 
extracted the essential simplifications of string theory and worked out 
how to do efficient computations directly in Yang-Mills theory \cite{ZBDAK2}. 
In 1992 they were joined by Lance Dixon and the trio derived a number of 
useful results about the strong interactions during the mid 1990's 
\cite{BDK1}. 

By the turn of the century Bern, Kosower and Dixon had developed a 
technique they called the ``unitarity method'' in which perturbative 
corrections could be generated from the lowest order results \cite{BDK2}. 
The idea is a more sophisticated version of the old bootstrap program
from the 1960's. If we express the S-matrix as $S=I + iT$, then unitarity
implies,
\begin{equation}
S^{\dagger} \times S \qquad \Longrightarrow \qquad -i (T \!-\! T^{\dagger})
= T^{\dagger} \times T \; . \label{unitarity}
\end{equation}
So if the $T$ matrix consists of a perturbative series expansion, one can
determine the $N$th order contribution to $(T - T^{\dagger})$ from sums of
products of lower order terms on the right hand side of (\ref{unitarity}).
Then the $N$th order contribution to $(T + T^{\dagger})$ can be inferred by 
a a procedure known as ``bootstrapping'', and the stage is set to push
one order higher \cite{boot1,boot2,boot3}. This procedure works best
for supersymmetric Yang-Mills theory because then the bootstrap is
simplest, and the bootstrap works best of all for $N=4$ Super-Yang-Mills
\cite{BDDK}.

The reader might wonder why an efficient technique for computing 
Super-Yang-Mills amplitudes has any relevance for supergravity. The 
connection is provided by the Kawai-Lewellen-Tye (KLT) relations between 
open and closed string amplitudes \cite{KLT}. The KLT relations say that 
a lowest order gravitational scattering amplitude can be factorized into 
sums of products of Yang-Mills amplitudes. These relations are relatively 
simple to see from string theory but so difficult to recognize for 
ordinary field theory that they were only inferred using string theory. Of 
course superstring theory requires ten dimensions, which we don't want,
and it also contains infinite towers of super-massive particles that are 
present in neither $N=4$ Super-Yang-Mills nor $N=8$ Supergravity. Both 
problems are avoided by taking the low energy limit \cite{BDDPR}. 
So the key to computing high order perturbative corrections to the
sattering amplitudes of $N=8$ Supergravity is first to decompose these 
amplitudes into products of zeroth order amplitudes using the unitarity 
method, then evaluate the later in terms of $N=4$ Super-Yang-Mills 
amplitudes \cite{nots}. 

The most recent results about $N=8$ Supergravity \cite{4loop} mean 
something is wrong with our thinking about what sorts of divergences 
can happen. No one knows how high the cancellations extend in $N=8$ 
Supergravity, or if they might apply to more realistic models. We 
could be witnessing the start of a revolution. The latest progress
on this subject is so new that there are no books or long review 
articles. I highly recommend the short article by Bern, Carrasco and 
Johansson \cite{Erice}.

\subsection{Asymptotic Safety}

Recall from section 2.6 that, if one doesn't consider the $R^2$ and $C^2$
counterterms to be part of the gravitational field equations then the
divergences of quantum gravity get worse at each order in perturbation 
theory. The same thing happens if you include the 4th derivative
counterterms but regard them as perturbations, rather than as part of
the 0th order field equations. In that case each new order in perturbation
theory requires the introduction of counterterms which contain two more
derivatives of the metric. Accepting this escalating series of counterterms
might seem crazy, in view of the fact it could be avoided by simply
considering the $R^2$ and $C^2$ counterterms to be part of the 0th order 
equations. However, the procedure has a saving virtue: {\it if we regard 
the higher derivative counterterms as perturbations then they do not add 
new degrees of freedom to the theory, and the Ostrogradskian instability of 
section 2.5 is avoided.} 

The program of {\it Asymptotic Safety} is to accept the escalating series 
of perturbative counterterms and attempt to show that interesting predictions 
can be derived from the resulting formalism \cite{SWasymp}. To make this
approach intelligible I will first explain why regarding the counterterms 
as perturbative avoids new degrees of freedom. Then I discuss how the 
counterterms affect the theory's ability to make predictions. Finally, I
will review progress on the subject.

The discussion of degrees of freedom is simplest in the context of a
one dimensional, point particle whose position as a function of time
is $q(t)$. Suppose that the Lagrangian is that of a harmonic oscillator
with a higher derivative,
\begin{equation}
L = -\frac{g m}{2 \omega^2} \, \ddot{q}^2 + \frac{m}2 \, \dot{q}^2 
-\frac{m \omega^2}2 \, q^2 \; .
\end{equation}
Here $m$ is the particle's mass, $\omega$ is a frequency and $g$ is a
small, positive number I wish to think of as a coupling constant. 
The Euler-Lagrange equation,
\begin{equation}
\frac{g}{\omega^2} \frac{d^4 q}{dt^4} + \ddot{q} + \omega^2 q = 0 \; ,
\label{qELE}
\end{equation}
has the general initial value solution,
\begin{equation}
q(t) = A_+ \cos(k_+ t) + B_+ \sin(k_+ t) + A_- \cos(k_- t) + B_- \sin(k_- t)
\; ,
\end{equation}
where the two frequencies and the four combination coefficients are,
\begin{equation}
k_{\pm} = \frac{\omega}{\sqrt{2g}} \Bigl[1 \mp \sqrt{1\!-\!4g}\Bigr]^{\frac12}
\quad , \quad A_{\pm} = \frac{k_{\mp}^2 q_0 \!+\! \ddot{q}_0}{k_{\mp}^2 
\!-\! k_{\pm}^2} \quad , \quad 
B_{\pm} = \frac{k_{\mp}^2 \dot{q}_0 \!+\! {d^3 q_0}/{dt^3} }{k_{\pm} 
(k_{\mp}^2 \!-\! k_{\pm}^2)} \; .
\end{equation}
The $+$ mode carries positive energy; it is the mode that would appear
even for $g=0$. The $-$ mode carries negative energy; it is the new,
higher derivative degree of freedom.

Suppose we regard the higher derivative term as a perturbation. That amounts 
to making the substitution,
\begin{equation}
q_{\rm pert} = \sum_{n=0}^{\infty} g^n x_n(t) \; ,
\end{equation}
in the Euler-Lagrange equation (\ref{qELE}) and then segregating terms with
the same powers of $g$. The resulting series of equations is,
\begin{eqnarray}
\ddot{x}_0 + \omega^2 x_0 & = & 0 \; , \\
\ddot{x}_1 + \omega^2 x_1 & = & -\frac1{\omega^2} \frac{d^4 x_0}{dt^4} \; , \\
\ddot{x}_2 + \omega^2 x_2 & = & -\frac1{\omega^2} \frac{d^4 x_1}{dt^4} \; ,
\end{eqnarray}
and so on. Note that the higher derivative terms always enter as sources,
evaluated at the lower order solution. They don't appear at 0th order, so
one only needs $q_0$ and $\dot{q}_0$ to get a unique solution $x_0(t)$.
The higher derivative term does appear in the 1st order equation, but
only in the form of $d^4x_0(t)/dt^4$, which was already fixed at 0th order,
so one again needs just $q_0$ and $\dot{q}_0$ to determine $x_1(t)$. 
Because the equation is linear the resulting series can be summed up and
gives,
\begin{equation}
q_{\rm pert} = q_0 \cos(k_+ t) + \frac{\dot{q}_0}{k_+} \, \sin(k_+ t) \; .
\end{equation}
Note that the higher derivative term did have an effect, it shifted the
frequency of oscillation from the 0th order result of $\omega$ to $k_+$.
However, we have avoided the higher derivative degrees of freedom which
give rise to the Ostrogradskian instability.

Although explicit results are difficult to obtain when the equations
become nonlinear, one can prove that this procedure works generally
\cite{EW,JLM}, so it should apply to the higher derivative counterterms 
of quantum gravity. Of course it cannot serve as a nonperturbative 
definition of quantum gravity (in this I dispute the official position 
of Asymptotic Safety) but I have already explained in section 2.3 that 
even the asymptotic series solutions one derives from perturbation 
theory should be wonderfully accurate at low energies. The problem is
the escalating series of counterterms, each with its own completely arbitrary, 
finite part. Those finite parts can only be fixed by requiring some
prediction of the theory to agree with measurement. {\it But each time
this is done one loses a potential prediction, and it must be done an
infinite number of times!} In such a case one wonders what the resulting
formalism can be used to predict?

One answer to this question was provided by John F. Donoghue 
\cite{Donoghue}. He pointed out that all counterterms must take the form 
of polynomials of the Fourier wave vector $\vec{p}$ such as we found
in expression (\ref{locdiv}). The finite contributions (\ref{nonloc})
contain some terms of this form --- and they will be rendered ambiguous
by the arbitrary finite parts of the counterterms. However, there are
also terms in (\ref{nonloc}) of the form $p^4 \ln(L^2 p^2)$ which could
never have been part of a counterterm and are {\it logarithmically enhanced}
with respect to the $p^4$ counterterms in the small $p$ regime which is
most accessible to experiment. That is how a unique result was obtained 
for the first quantum gravitational correction (\ref{dEarth}) to the 
Earth's potential \cite{gravpot}. This way of using a nonrenormalizable 
quantum field theory to make predictions is known as {\it low energy 
effective field theory}.

The program of Asymptotic Safety aims at the even better possibility
that we might be able to {\it predict the finite parts of all the 
counterterms in such a way that quantum general relativity could be used
at all energies.} Explaining how this works in any detail would go far 
beyond the scope of this article but I will try to put the idea across 
in simple terms. Recall from our discussion of renormalization in section 
2.4 that the strengths of interactions change as one varies the energy scale. 
This {\it running of coupling constants} has many important consequences
for particle physics. For example, forcing the strong interactions into
a regime for which perturbation theory is reliable provides our principal
check on the theory, and the fact that the various interactions become
comparable at a very high energy is the strongest evidence for Grand
Unification. The new counterterms one finds in quantum general relativity
vary with the energy scale as well. The hope of Asymptotic Safety is that
this variation might carry each of them to a unique value called a {\it
nontrivial ultraviolet fixed point}. If this proves to be the case then
we would not have to use up predictions in order to determine the finite
parts of counterterms. Instead we would just set them to their values at
the ultraviolet fixed point and this should be very near to the correct
result at sufficiently high energies.

The possibility of Asymptotic Safety was recognized by Steven Weinberg 
in 1979 \cite{SWasymp}. The equation which governs how coupling constants 
change is known as the {\it Exact Renormalization Group Equation} and
was derived by Christof Wetterich in 1992 \cite{Wetterich}. One can never
solve this for the infinity of necessary counterterms so what is done 
instead is to study the way selected combinations of counterterms flow. 
Exploring these truncations cannot prove Asymptotic Safety, but it could 
{\it disprove} the conjecture, and the results so far are consistent with 
the existence of an ultraviolet fixed point. However, it has been 
computationally impossible to include the first really problematic 
counterterm --- the one whose divergent coefficient was computed by Goroff 
and Sagnotti \cite{GS,Ven}. Much recent work on Asymptotic Safety has been 
done by Oliver Lauscher and Martin Reuter \cite{OLMR1}, Roberto Percacci 
\cite{Percacci} and by Percacci and Daniele Perini \cite{RPDP}. Interested 
readers should consult the review by Lauscher and Reuter \cite{OLMR2}.

\subsection{Loop Quantum Gravity}

Recall that relativists suspect the problems of quantum general
relativity derive from using perturbation theory. They therefore
insist on a very careful formulation which is not based upon
perturbing around any background geometry. That has far-reaching
consequences. It means that relativists can not accept, as a complete
description of quantum gravity, the asymptotic scattering states 
employed by particle theorists, nor can they accept the decay rates 
and cross sections we use as observables, nor even our inner product! 
So the full apparatus of quantum mechanics (states, inner product, 
observables) must be constructed from the beginning and with an 
unprecedented level of rigor and generality. It should be noted that 
we do not possess such a general formulation even for quantum 
electrodynamics, which is the best understood and most thoroughly 
tested quantum field theory.

Loop quantum gravity is based upon a Hamiltonian for general
relativity which was developed by Abhay Ashtekar in 1986
\cite{Ashtekar}. Recall that any Hamiltonian formalism has
``coordinates'' (the $Q$'s) and conjugate momenta (the $P$'s), and
it is usual to consider quantum mechanical states to be functions of
the coordinates. The term ``loop'' in the name comes from the way 
Ashtekar's coordinate variables are organized into ``Wilson loops'',
a sort of line integral (which also involves continuous multiplication 
of matrices) around closed loops. Under certain assumptions one can
show that this is the unique way of organizing the quantum theory 
\cite{LOST}.

Any Hamiltonian formalism which involves one or more of the
fundamental forces will possess {\it constraint equations}. Solving
these at the required level of rigor and generality is a major
problem for loop quantum gravity \cite{Nicolai}. To understand what
a constraint equation is, recall the discussion in section 2.1
concerning the triune nature of fundamental force fields:
\begin{itemize}
\item{One part of the field can be changed arbitrarily by a symmetry
transformation  which is associated with the conservation law;}
\item{Another part of the field is completely fixed by its sources; and}
\item{The final part consists of independent degrees of freedom.}
\end{itemize}
Solving the constraint equations is what determines the part of the
force field which is completely fixed by its sources. The constraint
equation of electrodynamics is Gauss's law ($\epsilon_0 \vec{\nabla}
\cdot \vec{E} = \rho$), another equation is fixed by current conservation
and the remaining two are dynamical. In general relativity four of the ten
Einstein equations (\ref{Einstein}) are constraints, four are fixed by
stress-energy conservation and the remaining two are dynamical.

In the Hamiltonian formalism of loop quantum gravity one doesn't think
of the constraint equations as determining field operators, the way I did 
in section 2, but rather as restricting how state wave functions can
depend upon the coordinates and also what operators can be observed. It 
is simple enough to work this out perturbatively to a given order, but that 
sort of perturbative solution is exactly what relativists wish to avoid! 
This makes things tougher but perhaps not impossible. The constraint 
equations of electrodynamics and the other fundamental forces have been 
solved nonperturbatively, as have all but one of the quantum gravitational 
constraints. Unfortunately, there is not yet a general solution to the 
final constraint.

The absence of explicit states or observables makes it difficult to 
compute much in full loop quantum gravity. In a number of symmetry
reduced models (see, e.g. \cite{lqc}) the program has been completed and
the resulting quantum theory sheds light on important conceptual issues 
such as the meaning of time and dynamics in background independent physics
and the likely fate of the most interesting classical singularities in
quantum gravity. Work has also been done on developing approximation 
procedures to construct explicit observables \cite{Dittrich}.

A measure of the difficulty involved is the effort that went into deriving 
the free graviton propagator \cite{gravprop}. This is a completely trivial 
exercise for particle theorists but it required several years of hard work 
using loop quantum gravity precisely because the concepts involved in its
construction are highly nontrivial in a background independent approach.
At the current stage, it also seems to involve a number of choices which 
indicates that the formalism is not yet complete. 

The field of loop quantum gravity has grown to include work on quantum 
cosmology, black hole entropy, the issue of information loss, coupling 
to matter, path integral quantization and the breaking or deformation of 
Poincar\'e invariance. Because this approach is far removed from my own 
area of expertise it is best that interested readers consult recent 
reviews by the leaders of the field. I recommend the articles by Ashtekar 
and Lewandowski \cite{AAJL}, by Smolin \cite{Smolin}, and the books by 
Rovelli \cite{Rovelli} and Thiemann \cite{Thiemann}.

\subsection{Causal Dynamical Triangulations}

I have admitted my bias as a particle theorist who loves being able to
get approximate results using perturbation theory, and I have tried to 
be honest about the possibility that these results might be erroneous, as 
so many of my relativist friends believe. However, exact calculations 
are unlikely to be attainable for quantum gravity, so the most fruitful 
way of questioning perturbation theory is to develop better 
approximation techniques. An example of work along these lines is the 
program of {\it Causal Dynamical Triangulations} by Jan Ambjorn, Jerzy 
Jurkiewicz and Renate Loll.

The idea behind Causal Dynamical Triangulations is to numerically
simulate a formal expression of quantum general relativity in terms
of an infinite number of integrations, one for each of the ten
components of the metric at each point in spacetime. Most everyone
is familiar with the use of computers to numerically evaluate a
finite number of integrals. Of course there are an infinite number
of points in continuum spacetime, and no computer can simulate more
than a finite number of integrations. So one first discretizes the
formal representation using a technique devised in 1961 by Tullio
Regge for approximating classical general relativity without the use 
of coordinates \cite{Regge}. Then the issue becomes how to take the 
continuum limit so that one plausibly recovers the correct theory.

{\it Recovering the correct continuum theory is highly nontrivial!}
In much simpler models for which the answer was known, physicists
discovered that one typically has to make parameters change with the
level of discretization in sensitive ways. And sometimes, one must
introduce new parameters which don't seem to be present in the
continuum theory. Of course we don't know exactly what the result
should be for quantum gravity but we do know it should have four
spacetime dimensions on macroscopic scales. With the most straightforward
quantum and computer adaptations of Regge's approach \cite{MEAAAM,Hamber} 
there does not seem to be any way of taking a continuum limit for pure
gravity which will have this property \cite{BBKP,BVB}, although the
issue is still open when matter is added \cite{EHY}.

To surmount the problem Ambjorn and Loll developed a two dimensional 
variation of Regge's discretization in which time is given a preferred 
status \cite{JARL}. Subsequent work with Jurkiewicz produced a four 
dimensional model \cite{AJL1} which seems to remain four dimensional as
the number of elements in the triangulation increases \cite{AJL2}. On
large scales the resulting spacetime looks like the most symmetric
solution of the classical Einstein equations with a positive cosmological 
constant \cite{AJL2}. It is still not clear how to take the continuum
limit, or if that limit even exists. And it is of course unknown if
the continuum limit will reproduce Newtonian gravitation at large 
distances, or contain gravitational radiation. One should not expect
progress too soon; it required decades of labor to attain better than 10\% 
accuracy with numerical simulations of strong interactions in the low 
energy regime for which perturbation theory fails \cite{LQCD}. A very 
recent review of Causal Dynamical Triangulations is \cite{AJL3}.

\section{Cosmology}

For years quantum gravity was a realm of speculation in which it was
even respectable to deny the need for a quantum theory of gravity. That
has changed recently with the advent of the first recognizable quantum
gravitational data and it has revolutionized the field. These data come
from cosmology, and many more are likely on the way, so any discussion of
quantum gravity must include cosmology. I shall begin by introducing the
classical metric which describes cosmology, and explaining how the
Einstein equations constrain this metric. Then I discuss why primordial
inflation is necessary and the simplest class of models which provide
the stress-energy to support it. The key point of this section is why
quantum gravitational effects are enhanced during inflation and why the
results on cosmological perturbations \cite{WMAP,SDSS} represent the
first ever recognized quantum gravitational data.

\subsection{FRW Geometry}

On the largest scales the universe seems to have no special origin or
special directions \cite{KT,Steve}. These properties are known are {\it
homogeneity} and {\it isotropy}, respectively. The universe also seems
to be devoid of spatial curvature \cite{WMAP}. The spacetime geometry
consistent with these three features is characterized by the following
invariant element,
\begin{equation}
ds^2 = -c^2 dt^2 + a^2(t) d\vec{x} \cdot d\vec{x} \; . \label{ds^2}
\end{equation}
(The acronym ``FRW'' is formed from the names of three cosmologists:
Alex\-ander Friedman, Howard Percy Robertson and Arthur
Geoffrey Walker.) The coordinate $t$ represents physical time, the
same as it does in flat space. However, the physical distance
between $\vec{x}$ and $\vec{y}$ is not given by their Euclidean
norm, $\Vert \vec{x} - \vec{y} \Vert$, but rather by $a(t) \Vert
\vec{x} - \vec{y} \Vert$. Because it converts coordinate distance
into physical distance $a(t)$ is known as the {\it scale factor}.

Although the scale factor is not directly measurable, three simple observable
quantities can be constructed from it,
\begin{equation}
z \equiv \frac{a_0}{a(t)} - 1 \quad , \quad
H(t) \equiv \frac{\dot{a}}{a} \quad , \quad q(t) \equiv -
\frac{a \ddot{a}}{\dot{a}^2} = -1 - \frac{\dot{H}}{H^2} \; . \label{H&q}
\end{equation}
The {\it redshift} $z$ gives the proportional increase in the wavelength of
light emitted at time $t$ and received at the current time, $t_0$. (I am 
ignoring the special relativistic Doppler shift.) Redshift is often used 
to measure cosmological time, even for epochs from which we detect no 
radiation. The {\it Hubble parameter} $H(t)$ gives the rate at which the 
universe is expanding. It's current value is, $H_0 = (70.5 \pm 1.3) \; 
\frac{\rm km}{\rm s\ Mpc} \simeq 2.3 \times 10^{-18}~{\rm Hz}$ \cite{WMAP}. 
The {\it deceleration parameter} $q(t)$ is less well measured. Observations 
of Type Ia supernovae are consistent with a current value of $q_0 \simeq -.6$
\cite{SNIA,Yun}.

Astronomers infer $H_0$ and $q_0$ by constructing {\it Hubble Plots}.
Suppose the light from a distant star contains a distinctive absorption line
measured at the wave length $\lambda$. If the same line occurs at wave length
$\lambda_E$ on Earth, we say the star's redshift is $z = \lambda/\lambda_E
-1$. One can also measure the flux of energy ${\cal F}$ from the star. If we
understand the star well enough to know it should emit radiation at
luminosity ${\cal L}$ (that is why Type Ia supernovae are important) we can
infer its luminosity distance $d_L$, which is the distance the star would be
at if the scale factor was one,
\begin{equation}
{\cal F} = \frac{\cal L}{4\pi d_L^2} \quad \Longrightarrow \quad d_L =
\sqrt{\frac{\cal L}{4 \pi {\cal F}}} \,.
\end{equation}
A Hubble Plot is a graph of $z$ versus $d_L$ for many distant stars.

Stars throughout the universe move with respect to their local
environments at typical velocities of about $10^{-3}$ the speed of
light $c$. This motion gives rise to a special relativistic Doppler
shift of $\Delta z \sim \pm 10^{-3}$. If spacetime was not
expanding, this shift would be the only source of nonzero $z$, and
averaging over many stars at the same luminosity distance would give
zero redshift. That is just what happens for stars within our
galaxy. However, the luminosity distances of stars in distant
galaxies are observed to grow approximately linearly with their
redshifts,
\begin{equation}
c^{-1} H_0 d_L = z + \frac12 (1 - q_0) z^2 + O(z^3) \; . \label{Hplot}
\end{equation}
One really does get $H_0$ from the slope, although inferring $q_0$
requires extending the plot to $z \sim 1$, at which point the expansion
breaks down and one must use the Einstein equations.

I cannot forbear to comment on the inconvenient minus sign in the definition
(\ref{H&q}) of $q(t)$. It was placed there because almost all theoretical
physicists were certain the current universe must be decelerating before the
measurement was done in 1998. (Checking this belief was not a priority; the
head of one of the two teams who measured $q_0$ told me he wasn't even
funded at the time!) Physicists like to define the parameters of equations
to be positive so that one can infer general trends at a glance. Generations
of physicists have cursed Benjamin Franklin for proposing the arbitrary sign
convention that lead to electrons --- which are the principal charge carrier
of our electrical industry --- having negative charge! But the universe
played a trick on us and it is actually accelerating, rather than
decelerating, so future generations will curse the minus sign we clever
theorists inserted in the definition (\ref{H&q}) of $q(t)$. Aside from
general amusement, this tale should serve to caution readers about placing
too much confidence (which means, any confidence at all) in the pronouncements
of the scientific establishment on issues that have not yet been subjected
to experimental and observational scrutiny.

\subsection{Einstein's Equations for FRW}

Homogeneity and isotropy restrict the stress-energy tensor to only an
energy density $\rho(t)$ and a pressure $p(t)$,
\begin{equation}
T_{00} = \rho(t) \qquad , \qquad T_{0i} = 0 \qquad , \qquad T_{ij} = p(t)
g_{ij},
\end{equation}
where $i$ and $j$ are spatial indices. In this geometry Einstein's
equations take the form,
\begin{eqnarray}
3 H^2 - c^2 \Lambda & = & \frac{8 \pi G}{c^2} \, \rho \; , \label{rhoeqn} \\
-2 \dot{H} -3 H^2 + c^2 \Lambda & = & \frac{8 \pi G}{c^2} \, p \label{peqn}\; .
\end{eqnarray}
It is usual to redefine the energy density and pressure so as to absorb the
cosmological constant,
\begin{equation}
\rho \longrightarrow \rho + \frac{c^4 \Lambda}{8 \pi G} \qquad {\rm and}
\qquad p \longrightarrow p - \frac{c^4 \Lambda}{8 \pi G} \; .
\end{equation}
When this is done, the current energy density is,
\begin{equation}
\rho_0 = \frac{3 c^2 H_0^2}{8 \pi G} \simeq 8.5 \times 10^{-10}~{\rm J/m}^3
\; .
\end{equation}
This is the rest mass energy of about 5.7 Hydrogen atoms per cubic meter.

By differentiating (\ref{rhoeqn}) and then adding $3 H$ times (\ref{rhoeqn})
plus (\ref{peqn}), we derive a relation between the energy density and
pressure known as stress-energy conservation,
\begin{equation}
\dot{\rho} = - 3 H (\rho + p) \; . \label{Econs}
\end{equation}
If we also assume a constant equation of state, $w \equiv
p(t)/\rho(t)$, then relation (\ref{Econs}) can be used to express the energy
density in terms of the scale factor,
\begin{equation}
\rho(t) = \rho_1 \Bigl(\frac{a(t)}{a_1}\Bigr)^{-3(1+w)} \; .
\label{rhow}
\end{equation}
The substitution of (\ref{rhow}) in (\ref{rhoeqn}) gives,
\begin{equation}
a(t) = a_1 \Bigl[1 + \frac32 (1 + w) H_1 (t - t_1)\Bigr]^{\frac{2}{3
(1 + w)}} \; . \label{aw}
\end{equation}

The cases of $w = +\frac13$, 0, $-\frac13$ and $-1$ correspond to
radiation, non-relativistic matter, spatial curvature, and vacuum
energy (which includes the cosmological constant), respectively,
\begin{eqnarray}
\mbox{Radiation} & \Longrightarrow & \rho \propto a^{-4} \; ,
\quad a(t) \propto (H_1 t)^{\frac12} \; , \label{raddom} \\
\mbox{Non-Relativistic Matter} & \Longrightarrow & \rho \propto a^{-3} \; ,
\quad a(t) \propto (H_1 t)^{\frac23} \; , \label{matdom} \\
\mbox{Curvature} & \Longrightarrow & \rho \propto a^{-2} \; , \quad
a(t) \propto H_1 t \; , \label{curdom} \\
\mbox{Vacuum Energy} & \Longrightarrow & \rho \propto 1 \; , \qquad a(t)
\propto e^{H_1 t} \; . \label{dS}
\end{eqnarray}
The actual universe seems to be composed of at least three of the pure types,
so the scale factor does not have a simple time dependence. However, as long as
each type is separately conserved, we can use (\ref{rhow}) to conclude that,
\begin{equation}
\rho(t) = \frac{\rho_{\rm rad}}{a^4(t)} + \frac{\rho_{\rm mat}}{a^3(t)} +
\frac{\rho_{\rm cur}}{a^2(t)} + \rho_{\rm vac} \; . \label{rhotot}
\end{equation}
As the universe expands, the relative importance of the four types
changes. Whenever a single type predominates, we can infer $a(t)$
from (\ref{aw}). This different dependence is one reason it makes
sense to think of an early universe dominated by radiation
(\ref{raddom}), evolving to a universe dominated by
non-relativistic matter (\ref{matdom}). It is also how one can
understand that the current universe seems to be making the
transition to domination by vacuum energy (\ref{dS}).

Under certain conditions there can be significant energy flows
between three of the pure types of stress-energy. For example, as
the early universe cooled, massive particles changed from behaving
like radiation to behaving like non-relativistic matter. This
change would increase $\rho_{\rm mat}$ and decrease
$\rho_{\rm rad}$ in Eq.~(\ref{rhotot}). The parameter that cannot
change is that of the spatial curvature, $\rho_{\rm cur}$. I should
not actually have regarded spatial curvature as a type of
stress-energy, but rather as an additional parameter in the
homogeneous and isotropic metric (\ref{ds^2}). I avoided this
complication because the extra terms it gives in the Einstein
equations (\ref{rhoeqn}-\ref{peqn}) can be subsumed into the
energy density and pressure, and because the measured value of
$\rho_{\rm cur}/a_0^2$ is consistent with zero \cite{WMAP}.

\subsection{Primordial Inflation}

The cosmology in which a radiation dominated universe evolves to matter
domination is a feature of what is known as the Big Bang scenario.
Although strongly supported by observation \cite{KT,Steve}, the composition of
$\rho$ at the start of radiation domination (at which the scale factor is
$a_{\rm rad}$) does not seem natural,
\begin{equation}
\frac{\rho_{\rm rad}}{a_{\rm rad}^4} \gg \rho_{\rm vac} \qquad , \qquad
\frac{\rho_{\rm rad}}{a_{\rm rad}^4} \gg \frac{\rho_{\rm cur}}{a_{\rm rad}^2} 
\; . \label{hier}
\end{equation}
It might be expected instead that each of the three terms was
comparable, in which case the universe would quickly become
dominated by vacuum energy. There is no accepted explanation for
the first inequality of (\ref{hier}), or for the seeming
coincidence that $\rho_{\rm mat}/a_0^3 \sim \rho_{\rm vac}$.
However, the second inequality of (\ref{hier}) finds a natural
explanation in the context of {\it Primordial Inflation}.

Inflation is defined as a phase of accelerated expansion, that is,
$q(t) < 0$ with $H(t) > 0$. From the current values of the cosmological
parameters one can see that the universe is in such a phase now. Recall
from section 3.2 that explaining why this is happening is one thing a
theory of quantum gravity might do. However, for now I wish to discuss
primordial inflation, which is conjectured to have occurred at something
like $10^{-37}$ seconds after the beginning of the universe with a
Hubble parameter 55 orders of magnitude larger than it is today. I will
be more specific later about what might cause inflation but, for now,
let us assume it is driven by a vacuum energy $\rho_{\rm vac}' \gg 
\rho_{\rm vac}$ (remember it is only $\rho_{\rm cur}$ which cannot 
change) and that it begins at scale factor $a_{\rm inf}$. If the energy 
densities of curvature and vacuum energy are comparable at the beginning 
of inflation then their ratio by the onset of radiation domination is,
\begin{equation}
\frac{\rho_{\rm cur}}{a_{\rm inf}^2 \, \rho_{\rm vac}'} \sim 1 \qquad
\Longrightarrow \qquad \frac{\rho_{\rm cur}}{a_{\rm rad}^2 \, \rho_{\rm vac}'}
\sim \Bigl(\frac{a_{\rm inf}}{a_{\rm rad}}\Bigr)^2 \; . \label{curve}
\end{equation}
In the next subsection I will explain why this number is smaller than
about $10^{-51}$. Inflation makes the other types of stress-energy even
smaller, but there are mechanisms through which the primordial vacuum energy 
$\rho_{\rm vac}'$ can be converted into matter and radiation. This process, 
which I will not discuss, is known as {\it reheating}.

Inflation also explains how the large scale universe became so nearly
homogeneous and isotropic. This explanation is crucial because
gravity makes even tiny inhomogeneities grow, and the process has
had 13.7 billion years to operate. It is believed that the galaxies
of today's universe had their origins in quantum fluctuations
of magnitude ${\Delta \rho}/\rho \simeq 10^{-5}$, which occurred
during the last 60 e-foldings of inflation. The imprint of these
fluctuations in the cosmic microwave background has recently been
imaged with unprecedented accuracy by the WMAP satellite \cite{WMAP}.
They also show up in large scale structure surveys \cite{SDSS}. If
this view is correct, these observations represent the first quantum
gravitational data.

\subsection{The Smoothness Problem}

There are many reasons for believing that the very early universe
underwent a phase of primordial inflation \cite{Linde}. I will confine
myself to reviewing how inflation resolves the {\it smoothness problem}.
This can be summed up in the question, why does the large scale universe
possess such a simple geometry (\ref{ds^2})?

To understand the problem we need to compare the distance light can
travel from the beginning of the universe to the time of some
observable event, with the distance it can travel from then to the
present. Recall from special relativity that light rays travel along
paths of zero invariant length, $ds^2 = 0$. From the invariant
element (\ref{ds^2}) we see that the radial position of a light ray
obeys, $dr = \pm c dt/a(t)$. The minus sign gives the {\it past
light-cone} of the point $x^{\mu} = (c t_0,\vec{0})$, whereas the
plus sign gives the {\it future light-cone} of a point $x^{\mu} = (c
t_{beg},\vec{0})$ at the beginning of the universe,
\begin{equation}
R_{\rm past} = \int_{t_{obs}}^{t_0} \!\!\! dt \, \frac{c}{a(t)}
\qquad , \qquad R_{\rm future} = \int_{t_{beg}}^{t_{obs}} \!\!\! dt \,
\frac{c}{a(t)} \; . \label{lcones}
\end{equation}
We can observe thermal radiation from the time of decoupling ($z_{dec}
\simeq 1089$) whose temperature is isotropic to one part in $10^5$. This
is a much higher degree of thermal equilibrium than exists in the air
in any office! Unless the universe simply began this way --- which seems
unlikely --- this equilibrium must have been established by processes
acting at or below the speed of light. In other words, we must have
$R_{\rm future} > R_{\rm past}$.

Suppose that, during the period $t_1 \le t \le t_2$, the deceleration
parameter is constant $q(t) = q_1$. In that case we can obtain explicit
expressions for the Hubble parameter and the scale factor in terms of their
values at $t=t_1$,
\begin{equation}
H(t) = \frac{H_1}{1 + (1 \!+\! q_1) H_1 (t \!-\! t_1)} \quad {\rm and }
\quad a(t) = a_1 \Bigl[1 + (1 \!+\! q_1) H_1 (t \!-\! t_1)\Bigr]^{\frac1{1
\!+\!q_1}} \; .
\end{equation}
These expressions permit us to evaluate the fundamental integral involved
in the past and future light-cones (\ref{lcones}),
\begin{equation}
\int_{t_1}^{t_2} \!\!\! dt \, \frac{c}{a(t)} = \frac{c}{a_1 H_1 q_1}
\Bigl[1 \!+\! (1\!+\! q_1) H_1 (t
\!-\!t_1)\Bigr]^{\frac{q_1}{1\!+\!q_1}} \Bigl\vert_{t_1}^{t_2} =
\frac{c}{q_1} \left\{ \frac1{a_2 H_2} - \frac1{a_1 H_1} \right\} .
\end{equation}

Although $q_0$ is negative, this is a recent event ($z \simeq 1$)
which followed a long period of nearly perfect matter domination
with $q = +\frac12$. Much before the time of matter-radiation
equality ($z_{eq} \simeq 3300$) the universe was almost perfectly
radiation-dominated, which corresponds to $q = +1$. To simplify the
computation we will ignore the recent phase of acceleration and also
the transition periods,
\begin{equation}
a(t) H(t) = a_0 H_0 \cases{ \sqrt{1 + z} & $\forall \, z \le z_{eq}$ \cr
\frac{1 + z}{\sqrt{1 + z_{eq}}} & $\forall \, z \ge z_{eq}$} \; .
\label{ass}
\end{equation}
The cosmic microwave radiation was emitted within about a hundred redshifts
of $z_{dec} < z_{eq}$, so the past light-cone is,
\begin{equation}
R_{\rm past} = \frac{2c}{a_0 H_0} \left\{ \frac1{\sqrt{1 \!+\! 0}} -
\frac1{ \sqrt{1 \!+\! z_{dec}}} \right\} \simeq \frac{2c}{a_0 H_0}
\; . \label{past}
\end{equation}
The future light-cone derives from both epochs and it depends slightly upon
the beginning redshift, $z_{beg}$,
\begin{equation}
R_{\rm future} = \frac{2c}{a_0 H_0} \left\{ \frac1{\sqrt{1 \!+\!
z_{dec}}} - \frac1{\sqrt{1 \!+\! z_{eq}}} \right\} + \frac{c}{a_0
H_0} \left\{\frac{\sqrt{1 \!+\! z_{eq}}}{1 \!+\! z_{eq}} -
\frac{\sqrt{1 \!+\! z_{eq}}}{1 \!+\! z_{beg}} \right\} \; .
\end{equation}
One maximizes $R_{\rm future}$ by taking $z_{beg} \rightarrow
\infty$, but it isn't enough,
\begin{equation}
\lim_{z_{beg} \rightarrow \infty} R_{\rm future} \simeq \frac{c}{a_0
H_0} \Biggl\{ \frac2{\sqrt{z_{dec}}} - \frac1{\sqrt{z_{eq}}}
\Biggr\} \; .
\end{equation}
Under the assumption of $q = +1$ before $z_{eq}$ we are forced to
conclude that the 2-dimensional surface we can see from the time of
decoupling consists of about,
\begin{equation}
\Bigl(\frac{R_{\rm past}}{R_{\rm future}}\Big)^2 \simeq
\frac{z_{dec}}{[1 \!-\! \frac12
(\frac{z_{dec}}{z_{eq}})^{\frac12}]^2} \simeq 2200 \; ,
\end{equation}
regions which cannot have exchanged even a photon since the
beginning of time! So how did these 2200 different regions reach
equilibrium to one part in $10^5$?

The problem grows worse the earlier one believes the universe was
homogeneous. For example, the seven lightest nuclear species were
almost all produced during Nucleosynthesis at about $z_{nuc} \simeq
10^9$ and their isotopic abundances seem to be uniform over the
observed universe. For Nucleosynthesis the radii of the past and
future light-cones are,
\begin{eqnarray}
\lefteqn{R_{\rm past} = \frac{2c}{a_0 H_0} \left\{ \frac1{\sqrt{1
\!+\! 0}} - \frac1{\sqrt{1 \!+\! z_{dec}}} \right\} } \nonumber \\
& & \hspace{4.8cm} + \frac{c}{a_0 H_0} \Biggl\{ \frac1{\sqrt{1 \!+\!
z_{eq}}} - \frac{\sqrt{1 \!+\! z_{eq}}}{1 \!+\! z_{nuc}} \Biggr\}
\simeq \frac{2c}{a_0 H_0} \; , \qquad \\
\lefteqn{R_{\rm future} = \frac{c}{a_0 H_0} \left\{ \frac{\sqrt{1
\!+\! z_{eq}}}{1 \!+\! z_{nuc}} - \frac{\sqrt{1 \!+\! z_{eq}}}{1 
\!+\! z_{beg}} \right\} \simeq \frac{c}{a_0 H_0}
\, \frac{\sqrt{z_{eq}}}{z_{nuc}} \; .}
\end{eqnarray}
So the assumption of $q=+1$ for all time before $t_{eq}$ implies
that the number of causally disconnected regions at the time of
Nucleosynthesis which comprise the current universe is about,
\begin{equation}
\Bigl(\frac{R_{\rm past}}{R_{\rm future}}\Big)^2 \simeq \frac{4
z_{nuc}^2}{z_{eq}} \simeq 10^{15} \; ,
\end{equation}
The corresponding numbers for the phase of quark-gluon plasma ($z
\sim 10^{12}$) and the electroweak phase transition ($z \sim
10^{15}$) are $10^{21}$ and $10^{27}$, respectively. One could
quibble about the extent to which we know the universe was
homogeneous at these times but it seems obvious something is very
wrong with the assumption of positive deceleration throughout cosmic
history.

This embarrassment resulted from the fact that the upper limit of
integration dominates $R_{\rm future}$ for positive deceleration.
Inflation solves the problem by positing a very early epoch of
negative deceleration. Let us suppose the deceleration is $q =
-1$ for $z > z_{rad}$. In that case our simplified model of cosmic
history (\ref{ass}) generalizes to,
\begin{equation}
a(t) H(t) \simeq a_0 H_0 \cases{ \sqrt{1 + z} & $0 < z < z_{eq}$ \cr
\frac{z}{\sqrt{z_{eq}}} & $z_{eq} < z < z_{rad}$ \cr
\frac{z_{rad}^2}{\sqrt{z_{eq}} \, z} & $z_{rad} < z < z_{inf}$} \; .
\label{newass}
\end{equation}
Now let us compute the past and future light-cones of some event at
redshift $z$ in the radiation dominated period. Of course our
previous result of $R_{\rm past} \simeq 2c/a_0 H_0$ is still valid.
However, the initial phase of acceleration makes a profound change
in the future light-cone. During acceleration it is the lower limit of
integration which dominates the future light-cone, and the result can
be made as large as desired simply by increasing the redshift $z_{inf}$
at the beginning of inflation. Under the assumption that $z_{inf} \gg
z_{rad} \gg z$ we get,
\begin{eqnarray}
R_{\rm future} & \simeq & \frac{c}{a_0 H_0} \Biggl\{ \frac{\sqrt{z_{eq}}}{z}
- \frac{\sqrt{z_{eq}}}{z_{rad}}\Biggr\} -\frac{c}{a_0 H_0} \Biggl\{
\frac{\sqrt{z_{eq}}}{z_{rad}} - \frac{\sqrt{z_{eq}} \, z_{inf}}{z_{rad}^2}
\Biggr\} \; , \qquad \\
& \simeq & \frac{c}{a_0 H_0} \, \frac{\sqrt{z_{eq}} \, z_{inf}}{z_{rad}^2} \; .
\end{eqnarray}

Inflation explains the smoothness of the large scale universe by supposing
that everything we now see derived from a region which was small enough
for causal processes to make it homogeneous and isotropic. Then inflation
stretched it out and the various pieces slowly came back into contact,
after inflation, still looking very much alike. In fact the inhomogeneities
we now see on less than cosmic scales derived from almost 14 billion years
of gravitational collapse operating on a universe that was smooth to one
part in $10^5$ just after inflation.

The usual assumption that $z_{rad} \simeq 10^{26}$ derives from supposing
that radiation domination commences at a scale of about $10^{13}~{\rm GeV}$.
If we require $R_{\rm future} \gtwid R_{\rm past}$ then $z_{inf} \gtwid
10^{51}$. (Many models of inflation vastly exceed this minimum.) Note that
primordial inflation not only solves the smoothness problem, it also explains
why the spatial curvature (\ref{curve}) is small, which is an observed fact
\cite{WMAP}.

\subsection{Slow Roll Scalar-Driven Inflation}

The case for an early period of accelerated expansion is very strong,
and the idea was suggested even before the advent of inflation 
\cite{precursors}. However, a completely satisfactory mechanism for 
causing accelerated expansion has yet to be identified, either for
primordial inflation or for the current phase. Guth's proposal \cite{Guth} 
failed to have a satisfactory ending but this was quickly corrected by the
{\it slow roll scalar} models proposed by Linde \cite{chaotic} and by Albrecht 
and Steinhardt \cite{newinf}. Although many other classes of models have 
since then been devised, and none are without problems, these are the 
simplest and I will describe them. 

Slow roll scalar models are based upon a hypothetical spin zero (scalar) 
field called {\it the inflaton} $\varphi(t,\vec{x})$. (Not having a good
candidate for this field from fundamental theory is one problem with these
models.) The Lagrangian for $\varphi(t,\vec{x})$ also involves the
metric $g_{\mu\nu}(t,\vec{x})$,
\begin{equation}
L = \int \!\! d^3x \, \sqrt{-{\rm det}(g_{\alpha\beta})} \, \Biggl\{-\frac12
\sum_{\mu=0}^3 \sum_{\nu=0}^3 \frac{\partial \varphi}{\partial x^{\mu}}
\frac{\partial \varphi}{\partial x^{\nu}} \, \Bigl(g^{-1}\Bigr)^{\mu\nu} -
V(\varphi)\Biggr\} , \label{inflaton}
\end{equation}
where $\det(g_{\alpha\beta})$ is the determinant of the metric and $g^{-1}$
is its matrix inverse. Note how the metric, which is the true measure of 
lengths and times, modifies the infinitesimal coordinate volume $d^3x$ and
the derivatives. This is typical of the way it couples to matter in general 
relativity. 

The stress-energy tensor of the inflaton is,
\begin{equation}
T_{\mu\nu} = \frac{\partial \varphi}{\partial x^{\mu}}
\frac{\partial \varphi}{\partial x^{\nu}} - \frac12 g_{\mu\nu} 
\sum_{\rho =0}^3 \sum_{\sigma =0}^3 \frac{\partial \varphi}{\partial x^{\rho}}
\frac{\partial \varphi}{\partial x^{\sigma}} \Bigl(g^{-1}\Bigr)^{\rho\sigma}
- g_{\mu\nu} V(\varphi) \; .
\end{equation}
One can see that the term involving the scalar potential has the same form 
as the cosmological constant $\Lambda$ in the Einstein equations 
(\ref{Einstein}). By itself, this contribution would tend to make the
universe accelerate, but one must also reckon with the kinetic terms which
involve derivatives. In order for inflation to start, the scalar potential
energy must dominate over its kinetic energy throughout a region somewhat
larger than light can cross from the beginning of the universe. This is
not anywhere near as bad as the terrific mismatches we found in the 
previous subsection but it does concern inflationary cosmologists because
such an initial condition can only have been an accident. Estimating how 
unlikely this accident is depends upon what one believes about how the 
universe began, and also involves tricky questions about how to compute 
probabilities. However, all estimates give very small numbers (I have heard
$10^{-120}$) for the chances of it happening.

Rather than start from an inhomogeneous configuration I will simply assume
the initial condition was homogeneous and isotropic. I will also assume
spatial flatness, which would in any case result, approximately, from a
long phase of inflation. This means the metric takes the FRW form (\ref{ds^2}) 
and that the scalar depends only upon time, $\varphi(t,\vec{x}) \rightarrow 
\varphi_0(t)$. The nontrivial Einstein equations are,
\begin{eqnarray}
3 H^2 - c^2 \Lambda & = & \frac{8\pi G}{c^2} \Bigl[ \frac{\dot{\varphi}_0^2}{
2 c^2} + V(\varphi_0)\Bigr] \label{Ephi1} \; , \\
-2\dot{H} - 3 H^2 + c^2 \Lambda & = & \frac{8\pi G}{c^2} \Bigl[ 
\frac{\dot{\varphi}_0^2}{2 c^2} - V(\varphi_0)\Bigr] \; . \label{Ephi2}
\end{eqnarray}
The deceleration parameter (times $H^2$) can be computed from a linear
combination of these equations,
\begin{equation}
q H^2 = -\dot{H} - H^2 = \frac{8 \pi G}{3 c^2} \Bigl[ \frac{\dot{\varphi}_0^2}{
c^2} - V(\varphi_0) - \frac{c^4 \Lambda}{8\pi G}\Bigr] \; .
\end{equation}
So the condition for accelerated expansion is,
\begin{equation}
V(\varphi_0) + \frac{c^4 \Lambda}{8\pi G} > \frac{\dot{\varphi}_0^2}{c^2} \; .
\label{accon}
\end{equation}

The equation for the homogeneous scalar $\varphi_0(t)$ is,
\begin{equation}
\ddot{\varphi}_0 + 3 H \dot{\varphi}_0 + c^2 V'(\varphi_0) = 0 \; , 
\label{phiev}
\end{equation}
where $V'(\varphi) \equiv \partial V/\partial \varphi$. The simplest
model of scalar-driven inflation, and one which is still consistent with
all data \cite{WMAP}, consists of a constant plus a quadratic term,
\begin{equation}
V(\varphi) = V_0 + \frac12 \Bigl(\frac{mc}{\hbar}\Bigr)^2 \, \varphi^2
\qquad \Longrightarrow \qquad V'(\varphi) = \Bigl(\frac{mc}{\hbar}\Bigr)^2
\; \varphi \; . \label{m2phi2}
\end{equation}
Substituting (\ref{m2phi2}) into (\ref{phiev}) gives the equation for a
damped harmonic oscillator,
\begin{equation}
\ddot{\varphi}_0 + 3 H \dot{\varphi}_0 + \Bigl(\frac{mc^2}{\hbar}\Bigr)^2
\, \varphi_0 = 0 \; , \label{damped}
\end{equation}
The term $3 H \dot{\varphi}_0$ in equations (\ref{phiev}) and (\ref{damped}) 
is known as {\it Hubble friction}. For sufficiently large $H(t)$ one can 
see that it makes the scalar over-damped so that it slowly rolls down the 
quadratic potential (\ref{m2phi2}). That makes the scalar kinetic energy
small, which will enforce the condition (\ref{accon}) for accelerated
expansion provided two more conditions hold.
 
The first of these extra conditions is that the constant $V_0$ must be
chosen to almost cancel the cosmological constant,
\begin{equation}
V_0 = -\frac{c^4 \Lambda}{8\pi G} + \rho_{\rm vac} \; , \label{fine}
\end{equation}
where $\rho_{\rm vac} \sim 6 \times 10^{-10}~{\rm J/m}^3$ is the currently 
observed value of the vacuum energy \cite{SNIA,Yun}. Of course the absence
of any explanation for this choice is the problem of the cosmological 
constant that I discussed in section 3.2. The second extra condition is that 
the scalar must start with a large enough initial value, and a small enough 
initial time derivative, so that (\ref{accon}) holds initially. The absence 
of any explanation for this is part of what is known as the {\it initial
condition problem.}

If one assumes these two conditions then Hubble friction causes the scalar 
to slowly roll down its potential. As it rolls, the Hubble parameter grows
smaller, which makes Hubble friction less effective. Eventually the system
becomes under-damped and begins oscillating. If the scalar is coupled to
other fields (the quantum corrections from which must be prevented from
distorting its potential too much --- another problem with this class of 
models!) then this phase of oscillations can result in hot, 
radiation-dominated universe \cite{Slava}. By choosing the initial value
of the scalar sufficiently large one can make the phase of inflation
last arbitrarily long, although making it too large can force the system
to a regime known as {\it eternal inflation} in which quantum fluctuations 
actually push the scalar up its potential.

I have described scalar slow-roll models to show that primordial inflation 
can be supported in a relatively simple way, even if this requires a number 
of arbitrary assumptions. There are other models, none of which is without 
problems. Although I very much doubt any of these models is correct, I 
consider the case for an early phase of accelerated expansion to be
to be overwhelming. Finding a realistic model which causes this phase is
one of the things I hope quantum gravity can do.

\subsection{The Strength of Quantum Effects during Inflation}

Leonard Parker was the first to give a quantitative assessment of how
spacetime expansion affects quantum processes \cite{Parker} but one can
understand some of the things he found in a qualitative way. In particular,
I will try to explain three crucial facts:
\begin{itemize}
\item{The expansion of spacetime strengthens quantum effects;}
\item{This strengthening is greatest during accelerated expansion;}
\item{Just as in flat space, the largest quantum effects derive from
the lightest particles {\it provided} they can avoid being driven to
zero amplitude by the expansion of the universe.}
\end{itemize}

Recall from section 2 that quantum fields obey exactly same equations as 
their classical counterparts, so one can understand quantum effects as the 
classical response to the 0-point motion which is required by the uncertainty 
principle. For example, we saw that vacuum polarization works the same way 
as classical polarization in a medium if one simply accepts that each mode
of the electron field has 0-point motion. The amount of 0-point motion a
field experiences is controlled by its free field mode functions, and
whatever increases the amount of 0-point motion will strengthen quantum 
effects. For example, we observed from the oscillatory factors of
$e^{i E t/\hbar}$ in the electron mode functions (\ref{Psimodefunc}) that
the combination of an electron of wave vector $\vec{k}$ and a positron
of wave vector $\vec{p} - \vec{k}$ can only remain coherent for a time
$\Delta t$ of about,
\begin{equation}
\Delta t \sim \frac{\hbar}{E(\vec{k}) \!+\! E(\vec{p} \!-\! \vec{k})} \qquad
{\rm where} \qquad E(\vec{k}) \equiv \sqrt{m^2 c^4 \!+\! \hbar^2 c^2
\Vert \vec{k}\Vert^2} \; . \label{ETunc}
\end{equation}
That is the only quantum mechanics one needs. The remainder of the
computation consists of using the Lorentz force law to find the classical
polarization induced by such a pair being acted upon for time $\Delta t$
by a static electric field $\widetilde{\vec{E}}(\vec{p})$,
\begin{equation}
e \Delta \vec{x} \sim \frac{e^2 c^2 \Delta t^2 \widetilde{\vec{E}}(\vec{p})}{
E(\vec{k}) \!+\! E(\vec{p} \!-\! \vec{k})} \sim \frac{e^2 c^2 \hbar^2
\widetilde{\vec{E}}(\vec{p})}{[E(\vec{k}) \!+\! E(\vec{p} \!-\! \vec{k})]^3}
\; .
\end{equation}
Summing over all modes $\vec{k}$ gives (up to a factor of $\frac83$) the
actual first order result (\ref{trueP}-\ref{kint}) for the vacuum polarization
due to a static field,
\begin{equation}
\widetilde{\vec{P}}(\vec{p}) \sim \int \!\! \frac{d^3k}{(2\pi)^3} \,
\frac{e^2 \hbar^2 c^2 \widetilde{\vec{E}}(\vec{p})}{[E(\vec{k}) \!+\!
E(\vec{p} \!-\! \vec{k})]^3} \; .
\end{equation}

The reason spacetime expansion strengthens quantum effects is that physical
wave numbers redshift. Because the FRW geometry is invariant under 
spatial translations, particles are still labeled by constant wave 
vectors $\vec{k}$. However, because $\Vert \vec{k}\Vert = 2\pi/\lambda$
is the inverse of a coordinate wavelength, not the physical wavelength 
$a(t) \lambda$, it is really the combination $\vec{k}/a(t)$ which tends
to enter physical expressions. The actual mode functions of a massive
particle are complicated but it is not a bad approximation to think
of the term $E(\vec{k}) \Delta t/\hbar$ in the phase of a mode function
generalizing to,
\begin{equation}
\frac{E(\vec{k}) \Delta t}{\hbar} \Longrightarrow \int_{t}^{t + \Delta t}
\!\!\!\!\!\!\!\!\!\! dt' \, \frac{E(t',\vec{k})}{\hbar} \qquad {\rm where}
\qquad E(t,\vec{k}) = \sqrt{m^2 c^4 \!+\! \hbar^2 c^2 \Vert \vec{k} \Vert^2/
a^2(t)} \; . \label{phase}
\end{equation}
The expansion of $a(t')$ always makes the accumulated phase smaller than it 
would have been for constant scale factor, so the mode can persist longer
and we see why the expansion of spacetime strengthens quantum effects.

Just as in flat space (that is, $a(t) = 1$) particles with the smallest
masses remain coherent longest, which is why almost all vacuum polarization
comes from electrons and positrons, even though there are many other
charged particle fields. Setting $m=0$ in expression (\ref{phase}) gives
the same integral (\ref{lcones}) we saw in section 5.4,
\begin{equation}
\lim_{m \rightarrow 0} \int_{t}^{t + \Delta t} \!\!\!\!\!\!\!\!\!\! dt' \, 
\frac{E(t',\vec{k})}{\hbar} = c \Vert \vec{k}\Vert \times
\int_{t}^{t + \Delta t} \!\!\!\!\!\!\!\!\!\! dt' \, \frac1{a(t')} \; .
\label{a(t)phase}
\end{equation}
Now recall the key distinction between accelerated expansion and deceleration:
\begin{itemize}
\item{For deceleration ($q(t) > 0$) the integral (\ref{a(t)phase}) grows
without bound as $\Delta t$ increases; but}
\item{For acceleration ($q(t) < 0$) the integral (\ref{a(t)phase}) approaches
a constant as $\Delta t$ goes to infinity.}
\end{itemize}
The first fact means that modes in a decelerating universe must eventually
become decoherent, even though they persist longer than for a static
universe. The second fact means that the 0-point motion of a sufficiently 
long wavelength mode can persist forever during inflation.

This is not quite the end of the story because almost all particles 
develop a symmetry known as {\it conformal invariance} when they become
massless. This symmetry causes their mode functions to fall off like
powers of the scale factor $a(t)$. For example, if we set the electron
mass to zero in (\ref{Psimodefunc}) and then account for the FRW geometry,
the electron mode function becomes,
\begin{eqnarray}
\varepsilon_i(t,\vec{x};\vec{k},s) & = & \sqrt{\frac{\hbar}{2 k}} \,
e^{-i k ct + i \vec{k} \cdot \vec{x}} \, u_i(\vec{k},s) \; , \\
& \longrightarrow & \Bigl[a(t)\Bigr]^{-\frac32} \sqrt{\frac{\hbar}{2 k}} 
e^{-i k c \! \int_0^t \! dt'\!/a(t') + i \vec{k} \cdot \vec{x}} \, 
u_i(\vec{k},s) \; , \label{amploss}
\end{eqnarray}
where the spinor wave function $u_i(\vec{k},s)$ is unchanged from its 
flat space value (\ref{uspin}) with $m=0$. Of course this decreasing 
amplitude tends to suppress 0-point motion, which weakens quantum effects.

To complete the third point on my list I need to show that there are 
massless particles whose mode functions avoid being driven to zero. It 
turns out there are two such particles:
\begin{itemize}
\item{Scalars like the inflaton (\ref{inflaton}) but with zero 
potential $V(\varphi)$; and}
\item{Gravitons.}
\end{itemize}
It also turns out that the mode functions of gravitons obey the same
equations as those of the massless inflaton \cite{Grishchuk}, so I will
specialize to the latter. 

We can understand the physics of massless inflatons by specializing the 
Lagrangian (\ref{inflaton}) to the FRW geometry (\ref{ds^2}) but still 
considering $\varphi(t,\vec{x})$ to be an arbitrary function of space 
and time, and then using Parseval's theorem,
\begin{eqnarray}
L & = & \int \!\! d^3x \, a^3(t) \Biggl\{ \frac12 \Bigl(\frac{\dot{\varphi}(t,
\vec{x})}{c}\Bigr)^2 - \frac12 \, \Bigl\Vert \frac{\vec{\nabla} 
\varphi(t,\vec{x})}{a(t)} \Bigr\Vert^2 \Biggr\} \; , \\
& = & \int \!\! \frac{d^3k}{(2\pi)^3} \, a^3(t) \Biggl\{ \frac1{2 c^2} \, 
\vert \dot{\widetilde{\varphi}}(t,\vec{k})\vert^2 - \frac{\Vert \vec{k} 
\Vert^2}{2 a^2(t)} \, \vert \widetilde{\varphi}(t,\vec{k})\vert^2 
\Biggr\} \; . \label{Linfl}
\end{eqnarray}
It follows that the field at each Fourier wave vector $\vec{k}$ behaves
as an independent harmonic oscillator with time-dependent mass and
frequency,
\begin{equation}
m(t) = m_0 a^3(t) \qquad {\rm and} \qquad \omega(t) = \frac{kc}{a(t)} \; .
\end{equation}

If we use $q(t)$ to represent the position of such an oscillator its 
Lagrangian and energy are,
\begin{equation}
L(t) = \frac{m(t)}2 \Bigl[ \dot{q}^2(t) - \omega^2(t) q^2(t)\Bigr] \quad 
\Longrightarrow \quad E(t) = \frac{m(t)}2 \Bigl[ \dot{q}^2(t) + \omega^2(t)
q^2(t)\Bigr] \; .
\end{equation}
All the usual theorems of quantum mechanics apply to this system. In
particular, the state with minimum energy at any fixed time $t$ must be,
\begin{equation}
E_{\rm min}(t) = \frac{\hbar k c}{2 a(t)} \; .
\end{equation}
However, the time dependence of the scale factor means that the minimum
energy state at one time is not the same at other times! The usual choice
for the ground state under these circumstances is called {\it Bunch-Davies 
vacuum} \cite{BD} and it corresponds to the state which was minimum energy 
in the distant past.

Let us find an expression for $q(t)$, analogous to (\ref{HOq}), in terms 
of the raising an lowering operators $\alpha^{\dagger}$ and $\alpha$ for 
Bunch-Davies vacuum $\vert \Omega \rangle$. It takes the form,
\begin{equation}
q(t) = \alpha \varepsilon(t) + \alpha^{\dagger} \varepsilon^*(t) \qquad
{\rm where} \qquad \Bigl[\alpha,\alpha^{\dagger}\Bigr] = 1 \qquad {\rm and}
\quad \alpha \Bigl\vert \Omega \Bigr\rangle = 0 \; .
\end{equation}
The mode functions obey,
\begin{equation}
\ddot{\varepsilon}(t) + 3 H(t) \dot{\varepsilon}(t) + \Bigl(\frac{kc}{a(t)}
\Bigr)^2 \varepsilon(t) = 0 \quad {\rm and} \quad \varepsilon(t) 
\dot{\varepsilon}^*(t) - \dot{\varepsilon}(t) \varepsilon^*(t) = 
\frac{i \hbar}{m_0 a^3(t)} \; . \label{MMCE}
\end{equation}
These equations are too difficult to solve for general scale factor $a(t)$
but the solution for $H(t) = H_I$ (which corresponds to the $q(t) = -1$
paradigm we employed for primordial inflation in section 5.4) is,
\begin{equation}
a(t) \propto e^{H_I t} \qquad \Longrightarrow \qquad \varepsilon(t) =
\sqrt{\frac{\hbar H_I^2}{2 m_0 c^3 k^3}} \Biggl[1 - \frac{i c k}{H_I a(t)} 
\Biggr] \exp\Bigl[\frac{ick}{H_I a(t)}\Bigr] \; . \label{deS}
\end{equation}
Now evaluate the expectation value of the energy operator in Bunch-Davies 
vacuum, first for arbitrary $a(t)$ and then for (\ref{deS}),
\begin{eqnarray}
\Bigl\langle \Omega \Bigl\vert E(t) \Bigr\vert \Omega \Bigr\rangle
& = & \frac12 m_0 a^3(t) \Biggl[ \dot{\varepsilon}(t) \dot{\varepsilon}^*(t)
+ \Bigl(\frac{k c}{a(t)}\Bigr)^2 \varepsilon(t) \varepsilon^*(t) \Biggr]
\; , \\
& \longrightarrow & \frac{\hbar k c}{2 a(t)} + \frac{\hbar H_I^2 a(t)}{4 c k} 
= \frac{\hbar k c}{a(t)} \Biggl[\frac12 + \Bigl(\frac{H_I a(t)}{2 k c}\Bigr)^2
\Biggr] \; . \qquad \label{N+half}
\end{eqnarray}

Recall that this system has the energy eigenstates of a harmonic oscillator
with energies $(N + \frac12) \hbar \omega(t)$ at any fixed instant in time.
So it is completely valid to regard the final expression in (\ref{N+half}) 
as giving the occupation number,
\begin{equation}
N(t) = \Bigl[\frac{H_I a(t)}{2 k c} \Bigr]^2 \; . \label{N(t)}
\end{equation}
Hence we see that Bunch-Davies vacuum starts out empty. Expression (\ref{deS}) 
reveals that, for $k c \gg H_I a(t)$ the mode function oscillates and falls 
off like $1/a(t)$, which is how a massless, conformally invariant scalar
behaves. As the scale factor grows, the occupation number (\ref{N(t)})
increases, and both the oscillations in the mode function and the decrease 
of its amplitude slow down. A key event is {\it first horizon crossing}
when $k c = H(t) a(t)$, at which point the occupation number becomes order 
one and the mode function begins approaching a constant. After that the 
occupation number becomes exponentially large, as does the amount of energy 
in this single mode.

One might worry that the behavior I have just sketched is very special to
the case of $q(t) = -1$ but that is not true. As long as the universe is
accelerating the product $H(t) a(t)$ grows, and one can see from equation
(\ref{MMCE}) that modes which start with $k c \gg H(t) a(t)$ oscillate and 
fall off,
\begin{equation}
kc \gg H(t) \, a(t) \quad \Longrightarrow \quad \varepsilon(t) \sim
\sqrt{\frac{\hbar}{2 m_0 k c}} \, \frac1{a(t)} \exp\Bigl[-i k c \!\! 
\int_0^t \!\! dt' \, \frac1{a(t')}\Bigr] \; .
\end{equation}
They are also drawn towards first horizon crossing. If inflation persists
long enough for them to reach it, the $(kc/a)^2 $ term of (\ref{MMCE}) drops
out and one can see that the mode function approaches a constant. So the
behavior I found for (\ref{deS}) is actually generic to any inflating 
geometry.

Of course this is why massless inflatons and gravitons show enhanced 
quantum effects during inflation. Before closing I should comment on
the sheer wonder of what we found in expression (\ref{N+half}). {\it 
This represents the 0-point energy of one single mode!} Of course one
must divide it over the vast 3-volume of the inflating universe, but
summing over the modes which have experienced first horizon crossing
gives a small, macroscopic effect. Another important comment is that 
{\it this result derives from the long wavelength regime in which
perturbative quantum general relativity should be valid}, even if 
quantum gravity is described, on the fundamental level, by some other
theory, or if perturbative methods do not give the correct asymptotic
series. My final comment is that there is no natural mechanism for
keeping the inflaton massless, and if it develops a large
mass that will suppress quantum effects during inflation the same way
it would in flat space. By contrast, nothing needs to be done to keep
the graviton massless, so this long wavelength regime during inflation
is a natural place to look for quantum gravitational effects.

\subsection{Quantum Gravitational Data}

I have already discussed scalar-driven inflation and why one expects 
enhanced quantum effects during this epoch. In this subsection I will
sketch the theory behind the first quantum gravitational observables ever 
measured, which are cosmological perturbations. The original work on 
tensor perturbations was done by Starobinsky \cite{AAS}; Mukhanov and 
Chibisov did the first calculation of scalar perturbations \cite{MC}. For 
more details I recommend the excellent recent text by Mukhanov \cite{Slava}.

Cosmological perturbations are spacetime dependent fluctuations of the
full scalar and metric fields around the spatially homogeneous background 
values described in section 5.5,
\begin{eqnarray}
\varphi(t,\vec{x}) & = & \varphi_0(t) + \delta \varphi(t,\vec{x}) \; , \\
g_{\mu\nu}(t,\vec{x}) & = & \overline{g}_{\mu\nu}(t) + h_{\mu\nu}(t,\vec{x})
\; .
\end{eqnarray}
Here $\overline{g}_{\mu\nu}(t)$ stands for the FRW metric (\ref{ds^2}),
\begin{equation}
\overline{g}_{00}(t) = -1 \qquad , \qquad \overline{g}_{0i}(t) = 0 \qquad , 
\qquad \overline{g}_{ij}(t) = a^2(t) \delta_{ij} \; .
\end{equation}
It is typical to express the graviton using just scalar and tensor fields
because the linearized Einstein equations cause the vector fields to
vanish,
\begin{eqnarray}
h_{00}(t,\vec{x}) & = & -2 \phi(t,\vec{x}) \; , \\
h_{0i}(t,\vec{x}) & = & - c \frac{\partial}{\partial x^i} \, 
B(t,\vec{x}) \; , \\
h_{ij}(t,\vec{x}) & = & -2 a^2(t) \psi(t,\vec{x}) \delta_{ij} \!-\! 2 c^2
\frac{\partial}{\partial x^i} \frac{\partial}{\partial x^j} \, E(t,\vec{x})
\!+\! a^2(t) h^{TT}_{ij}(t,\vec{x}) \; . \qquad
\end{eqnarray}
Here the dynamical graviton field $h_{ij}^{TT}$ is both transverse and
traceless,
\begin{equation}
\sum_{i=1}^3 \frac{\partial}{\partial x^i} \, h^{TT}_{ij}(t,\vec{x}) = 0 = 
\sum_{i=1}^3 h_{ii}^{TT}(t,\vec{x}) \; .
\end{equation}

Dynamical gravitons are invariant under linearized coordinate 
transformations and their spatial Fourier transforms obey the same 
equation (\ref{MMCE}) as a massless inflaton,
\begin{equation}
\Biggl[ \Bigl(\frac{\partial}{\partial t}\Bigr)^2 + 3 H(t) 
\frac{\partial}{\partial t} + \frac{k^2 c^2}{a^2(t)} \Biggr] \,
\widetilde{h}^{TT}_{ij}(t,\vec{k}) = 0 \; .
\end{equation}
We have already seen that for $k c \gg H(t) a(t)$ the solutions are
oscillatory with amplitudes that fall off like $1/a(t)$. Long after 
first horizon crossing, in the regime for which $k c \ll H(t) a(t)$ the
term proportional to $(kc/a)^2$ drops out and the solution goes to a 
linear combination of a constant and the falling indefinite integral,
\begin{equation}
\int^t \!\! dt' \frac1{a^3(t')} \; .
\end{equation}

The following combinations of the scalar fields are invariant under
linearized coordinate transformations,
\begin{eqnarray}
\Phi(t,\vec{x}) & \equiv & \phi(t,\vec{x}) - \frac{\partial}{\partial t}
\Bigl[B(t,\vec{x}) \!-\! \dot{E}(t,\vec{x}) \!+\! 2 H(t) E(t,\vec{x})\Bigr]
\; , \\
\Psi(t,\vec{x}) & \equiv & \psi(t,\vec{x}) + H(t) \Bigl[B(t,\vec{x}) \!-\! 
\dot{E}(t,\vec{x}) \!+\! 2 H(t) E(t,\vec{x})\Bigr] \; , \\
\Xi(t,\vec{x}) & \equiv & \delta \varphi(t,\vec{x}) - \dot{\varphi}_0(t)
\Bigl[B(t,\vec{x}) \!-\! \dot{E}(t,\vec{x}) \!+\! 2 H(t) E(t,\vec{x})\Bigr]\; .
\end{eqnarray}
The linearized Einstein equations for indices $\mu=i$ and $\nu = j$ imply 
$\Psi = \Phi$, whereupon the linearized $\mu = 0$, $\nu =0$ and $\mu = 0$, 
$\nu = i$ equations become,
\begin{eqnarray}
\Biggl[6 H \frac{\partial}{\partial t} \!+\! 6 H^2 \!-\!
\frac{2 c^2 \nabla^2}{a^2}\Biggr] \Phi & = & \frac{8\pi G}{c^4} \Biggl[
\dot{\varphi}_0^2 \Phi - \dot{\varphi}_0 \dot{\Xi} - c^2 V'(\varphi_0) \Xi
\Biggr] \; , \label{E00} \\
-\frac2{a} \frac{\partial}{\partial x^i} \Bigl[ \dot{\Phi} \!+\! H \Phi \Bigr]
& = & -\frac1{a} \frac{\partial}{\partial x^i} \Bigl[ \frac{8 \pi G}{c^4}
\, \dot{\varphi}_0 \Xi\Bigr] \; . \label{E0i}
\end{eqnarray}
Of course equation (\ref{E0i}) relates the scalar perturbation to the scalar
part of the metric perturbation,
\begin{equation}
\Xi(t,\vec{x}) = \frac{c^4}{4\pi G} \Biggl[ \frac{\dot{\Phi}(t,\vec{x})
\!+\! H(t) \Phi(t,\vec{x})}{\dot{\varphi}_0(t)} \Biggr] \; .
\end{equation}
Substituting this in (\ref{E00}), taking the spatial Fourier transform, 
and making some simplifications eventually results in what has been termed
the {\it Mukhanov Equation} for $\widetilde{\Phi}(t,\vec{k})/\dot{\varphi
}_0(t)$,
\begin{equation}
\Biggl[ \Bigl(\frac{\partial}{\partial t}\Bigr)^2 + 
H \frac{\partial}{\partial t} + \frac{k^2 c^2}{a^2(t)} -
\Bigl(\frac{\ddot{\theta}(t) \!+\! H(t) \dot{\theta}(t)}{\theta(t)} \Bigr)
\Biggr] \Biggl( \frac{\widetilde{\Phi}(t,\vec{k})}{\dot{\varphi}_0(t)} 
\Biggr) = 0 \; , \label{Meqn}
\end{equation}
where the function $\theta(t)$ is,
\begin{equation}
\theta(t) \equiv \frac{H(t)}{a(t) \dot{\varphi}_0(t)} \; .
\end{equation}

The solution of (\ref{Meqn}) is qualitatively similar to that of the
massless inflaton. In the regime $k c \gg H(t) a(t)$ the field oscillates
and its amplitude falls off. Long after first horizon crossing we can again
drop the term proportional to $(kc/a)^2$ and the solution becomes a linear
combination of $\theta(t)$ and the indefinite integral,
\begin{equation}
\theta(t) \int^t dt' \frac1{a(t') \theta^2(t')} \; .
\end{equation}
It is this second solution which dominates long after first horizon
crossing.

For both the scalar perturbations of $\widetilde{\Phi}(t,\vec{k})$ and the
tensor perturbations of $\widetilde{h}^{TT}_{ij}(t,\vec{k})$ we can get
good approximate solutions for the mode functions in the regime 
$kc \gg H(t) a(t)$. Canonical quantization fixes their normalization.
Long after first horizon crossing we can again get good solutions, one of
which approaches a constant with the other falling. The strength of the
each perturbation is quantified by its {\it power spectrum}. If the spatial
Fourier transform of some quantum field $F(t,\vec{x})$ approaches a constant
we define its power spectrum $\mathcal{P}_{F}(k)$ as,
\begin{equation}
\Bigl\langle \Omega \Big]\vert \widetilde{F}(t,\vec{k}) 
\widetilde{F}^{\dagger}\!(t,\vec{k}') \Bigr\vert \Omega \rangle \equiv
\frac{\mathcal{P}_{\!F}(k)}{4 \pi k^3} \times (2\pi)^3 \delta^3(\vec{k} \!-\!
\vec{k}') \; . \label{powerdef}
\end{equation}

It is quite a challenge to connect the early, oscillatory regime --- for
which the normalization is known --- to the late time regime, long after
first horizon crossing \cite{WMS}. Approximate results can be obtained
by crudely matching the early and late time solutions at the time $t_k$
for which wave number $k$ experiences first horizon crossing,
\begin{equation}
k c \equiv H(t_k) a(t_k) \; .
\end{equation}
These mode-matching results are,
\begin{eqnarray}
\mathcal{P}_{\Phi}(k) & \sim & \frac{G H^2(t_k)}{1 \!+\! q(t_k)} \; , 
\label{PPhi} \\
\mathcal{P}_{h}(k) & \sim & G H^2(t_k) \; . \label{Ph}
\end{eqnarray}

It should be noted that (\ref{PPhi}-\ref{Ph}) are the {\it primordial} 
power spectra. What is actually measured is the result of complicated but 
quite well understood physics that occurs after inflation, during the 
subsequent epochs of radiation domination and matter domination. This is
included through a known {\it transfer function}, so the challenge for 
fundamental theory is to compute the primordial power spectra. It should also 
be noted that the primordial power spectra are almost independent of $k$ 
because $H(t)$ is nearly constant during inflation. The data are therefore 
typically organized into scalar and tensor {\it spectral indices},
\begin{eqnarray}
n_s(k) & \equiv & 1 + \frac{\partial \ln[\mathcal{P}_{\Phi}(k)]}{\partial
\ln(k)} \; , \\
n_t(k) & \equiv & \frac{\partial \ln[\mathcal{P}_{h}(k)]}{\partial \ln(k)} 
\; ,
\end{eqnarray}
reported at a fiducial wave number. Lately there has been an effort to
report the first derivative of the scalar spectral index as well \cite{WMAP}.

One can see from expressions (\ref{PPhi}-\ref{Ph}) that the scalar power
spectrum is enhanced by a factor of $1/(1 +q)$, relative to the tensor 
one. Because $q$ is very near $-1$ during inflation this is a large
enhancement, and that is why the tensor signal has so far not been detected. 
Getting a nonzero measurement for $\mathcal{P}_h(k)$ is very important 
because it would tell us the scale of primordial inflation. It would also
be the first proof that gravitons exist and are quantized.

I want to emphasize that primordial perturbations are a quantum gravitational
effect. Many quantum gravity experts dismiss them because we have so far not 
detected the tensor signal from gravitons, but this is wrong-headed. There is 
no problem with pure gravitons at the lowest order in perturbation theory; 
the problem at lowest order comes from matter, and the scalar perturbation 
signal probes precisely this effect at an energy scale potentially as high 
as $10^{13}~{\rm GeV}$. If I had been told only one of the two perturbation 
spectra could be measured and asked to choose which one, I would have picked 
the scalar. {\it This is priceless information,} although it would be better 
if we has a unique theory of inflation which made specific predictions. As it 
is, one can accommodate almost any data by changing the model of inflation, 
and none of the models proposed so far is very compelling.

I should like to close with two more comments, one about experiment and the
other about theory. First, there are a lot more data on the way. The European
Space Agency has launched the Planck satellite which will try to resolve the
primordial tensor spectrum. And this is just one of many experiments. The 
most exciting idea to me personally is the proposal to use the 21 centimeter 
line out to huge redshifts such as $z \sim 50$ \cite{21cm}. Foregrounds will 
be an enormous problem, and it will require years of labor, but there is 
potentially enough data present to determine the tensor-to-scalar ratio $r$ 
to one part in $10^8$, compared with the current bound of $r < .22$
\cite{WMAP}!

My comment about theory is that we are just beginning to understand
how to compute quantum corrections during inflation. Nothing like the
asymptotic scattering theory of flat space quantum field theory yet
exists for this environment. Results such as (\ref{PPhi}-\ref{Ph}) 
receive quantum corrections which Weinberg has studied \cite{SWpower}. 
He found a peculiar thing: although these corrections are suppressed by 
a very small factor of $G H^2 \ltwid 10^{-12}$, in addition to the 
factor of $G H^2$ already present in the lowest order result (\ref{PPhi}), 
they contain time dependent enhancement factors that grow like 
$\ln[a(t)/a(t_k)]$, where $t_k$ is the time at which the mode experienced 
first horizon crossing. Nick Tsamis and I have found similar factors in 
quantum gravitational corrections to the metric and to the graviton 
self-energy \cite{NCT}, and Shun-Pei Miao and I have found them in quantum 
gravitational corrections to fermion mode functions \cite{SPM}. For the 
modes whose spatial variation we can resolve today there have been at most 
about 120 e-foldings, so these {\it infrared logarithms} represent a 
terrific enhancement, but not enough to make corrections observable. (But 
they {\it might} become observable if the 21-centimeter observations fulfill 
their promise!) However, this is just because one insists on being able 
to resolve the spatial variation. If one studies something which is 
spatially constant, like quantum corrections to the vacuum energy, then 
there can be infrared logarithms from modes which experienced first horizon 
crossing early during a very long period of inflation, and there seems no 
limit on the size of the effect they might give. Which is interest, because
it just so happens that there is this little problem understanding the 
vacuum energy . . .

\section {Conclusions}

This article began with a list of seven questions. I will devote a
paragraph to each question and the answers that have been developed.
Then I will make a few additional points and comment on the future.

{\it What is the distinction between classical physics and quantum
physics that makes general relativity give such a wonderful classical
theory of gravity and such a problematic quantum one?} There is no
difference between the equations of motion for Heisenberg field operators
and those of the corresponding classical theory. Nor is there any distinction
in what it means to solve those equations; in both cases the general solution
consists of expressing the dynamical variable at any time in terms of its
initial values. These initial values are the fundamental degrees of freedom
of physics, and it is usual to label them by their Fourier wave number
$\vec{k}$. The key distinction between classical and quantum is what the
initial values are: in classical physics they are numbers and each of
them can be set to any value; in quantum physics each mode $\vec{k}$ contains
one or more pairs of non-commuting operators which obey the uncertainty
principle. One consequence of this is that no mode can have less than a
minimum $0$-point energy. When we say that general relativity gives a
superb classical theory of gravity, we mean that its results are
indistinguishable from nature {\it when we set all the large $\Vert \vec{k}
\Vert$ modes to zero.} There is no problem doing that classically, but it is
not permitted in quantum mechanics. It is the influence of these high
$\Vert \vec{k}\Vert$ modes which makes general relativity so problematic as
a quantum theory.

{\it Why do we have to quantize gravity?} Because part of any force
field is entirely determined by its sources, and the matter fields which
source gravity are indisputably quantum mechanical, whether or not gravitons
are quantized or even exist. It used to be argued that we could take the
metric field to be sourced by the expectation value of the matter stress-energy
tensor but that view is not tenable under the simplest interpretation of the
observed anisotropies in the cosmic ray microwave background. If the predicted
tensor component of these anisotropies can be imaged there will be direct
evidence for the existence and quantum nature of gravitons as well.

{\it Why do quantum field theories have divergences?} Because continuum
field theories have an infinite number of modes, and the 0-point motion of
each one of them contributes a little to typical quantum effects. Spacetime
may well be discrete on some level but the expansion of the universe by the
staggering factor of $e^{120} \sim 10^{52}$ (if we accept the reality of
primordial inflation) means that this discreteness cannot be responsible for
keeping quantum gravitational effects so small.

{\it Why are the divergences of quantum general relativity worse than those
of the other forces?} Because the other forces couple to charges which are
the same for all modes, whereas gravity couples to stress-energy, which
grows with the wave number. This means that first order quantum corrections
to the gravitational field equations produce divergences not just on terms
which have two derivatives of the metric or zero derivatives of it, but also
on terms which contain four derivatives of the metric. Two terms of this
type are possible, the ``$R^2$ counterterm'' and the ``$C^2$ counterterm.''

{\it How bad is the problem?} Adding the $R^2$ and $C^2$
counterterms to the gravitational field equations would allow the
divergences of quantum gravity to be renormalized to all orders, the
same way as with other forces \cite{KS}. However, increasing the
number of derivatives in a field equation introduces new degrees of
freedom. The new degree of freedom associated with the $R^2$
counterterm is a positive energy particle with spin zero, which
poses no essential problem. Unfortunately, the new degrees of
freedom associated with the $C^2$ counterterm comprise a negative
energy particle with spin two, which would make the universe blow up
instantly. The $C^2$ counterterm is necessary at first order in
gravity plus scalar particles such as the Higgs \cite{HV}. It is
also needed for gravity plus electromagnetism \cite{DvN} and for
gravity plus the weak or strong interactions \cite{DTvN}. It is not
necessary at first order for pure gravity in four spacetime
dimensions \cite{HV}, but this theory requires an equally
unacceptable counterterm at second order \cite{GS,Ven}. Hence we
must either add unacceptable counterterms, which gives a finite
theory that is virulently unstable, or else low order perturbation
theory makes divergent predictions.

{\it What are the main schools of thought about quantizing gravity and
why do they disagree?} The problem with perturbative quantum general
relativity arises from a conflict between four things: continuum field
theory, quantum mechanics, general relativity and perturbation theory.
Because the first two items on this list are not likely to be at fault the
schools of thought on quantum gravity differ on which of the last two they
suspect. Particle theorists come from a long tradition of quickly exploiting
perturbation theory to get results and then rejecting models which fail to
measure up. This brought great success with the Standard Model, so it seems
reasonable to particle theorists that they should trust perturbation theory
and reject general relativity. Relativists come from a long tradition in which
general relativity was many times alleged to have problems that always
disappeared when a sufficiently careful analysis was made. So it makes sense
to relativists that they should reject perturbation theory and instead focus
on a painstakingly rigorous formulation of quantum general relativity.

{\it What would we do with the theory of quantum gravity if we had it?}
The strength of quantum gravity corrections from a mode of energy $E$ seems
to be roughly $G E^2/\hbar c^5 \sim (E/10^{19}~{\rm GeV})^2$. Although we
cannot access interesting energies in the laboratory, nature reaches these
scales in four cases: the initial singularity, the final stages of black
hole collapse, the final stages of black hole evaporation, and during
primordial inflation. Ideas about the first three are still speculative but
the simplest interpretation of current data is that the primordial
perturbations in the gravitational potential of the Universe derive from
quantum matter fluctuations during primordial inflation.

Establishing phenomenological contact with a weak interaction typically
involves exploiting its special properties. The unique properties of gravity
are:
\begin{enumerate}
\item{One of the gravitational parameters is the cosmological constant
$\Lambda$;}
\item{It determines the maximum speed at which signals can propagate;}
\item{Gravitons have zero mass without being driven to zero amplitude by
the expansion of the universe; and}
\item{The gravitational interaction energy is negative.}
\end{enumerate}
Physicists suspect that a successful theory of quantum gravity will
explain the enormous disparity between the measured value of the cosmological
constant (\ref{Lamval}) and the natural scales of fundamental theory. The use
of first order perturbation theory in the context of point 2 leads to finite
predictions for the blurring of images, fluctuations in luminosity and a
broadening of spectral lines. Some of these effects may be observable in the
not-too-distant future. Point 3 suggests that inflationary cosmology is a
natural venue to search for quantum gravitational effects. And point 4 is the
basis for the old dream of being able to compute the masses of fundamental
particle from the interactions of their own force fields.

An interesting and possibly significant fact strikes one about all
four of the current approaches to quantizing gravity reviewed in
section 4. Each of them involves negative energy in some form:
\begin{itemize}
\item{The particle theorists' dreams concerning superstrings and on-shell
finiteness exploit the negative $0$-point energy of fermionic
superpartners;}
\item{The higher counterterms allowed in asymptotic safety would induce
negative energy degrees of freedom if they were treated
nonperturbatively; and}
\item{The relativists' dream that quantum general relativity will regulate
its own ultraviolet divergences relies upon negative gravitational
interaction energies.}
\end{itemize}
Perhaps this is more than a curiosity.

My answer to the question posed by the title of this article is that we
are very far from having a complete quantum theory of gravity. A measure
of the reliability of current thought (including my own) is this quotation
from a renowned string theorist in reaction to the initial observations which
indicated the universe is accelerating \cite{SNIA}:
\begin{quote}
{\it I'm sure the data is wrong because string theory predicts a negative
cosmological constant.}
\end{quote}
I recount these words not because string theorists are bad
physicists but rather because they are among the best our species
has ever produced. That even they failed points up the folly of
trying to guess natural law with nothing more to go on than
mathematics and aesthetics. If that was not obvious two and a half
decades ago it is surely beyond dispute now.

Neither the nature of the problem, nor the prescription for its inevitable
solution are unique in man's long struggle to understand the universe. The
historian Colin McEvedy had this to say about the intellectual stagnation of
late Roman civilization \cite{CME}:
\begin{quote}
\item{\it Speculation ran way beyond the testable and dwindled into
metaphysics; technology remained tradition-bound and sluggish. Only the
evolution of a scientific stance --- one foot inside the boundary of the
known, the other just outside --- could have guaranteed the superiority,
and consequently the integrity, of Mediterranean society, and the world
was still too young for that.}
\end{quote}
The good news for quantum gravity is that we shall not have to
endure centuries of darkness until a more powerful mode of thought
emerges from the ashes of our failures. The process of achieving a
scientific stance is underway; the first data have been taken, and
many more are coming. Understanding what they have to teach us will
likely be a long and painful process, and I expect that most of what
we currently believe will need to be abandoned. But the outcome
cannot be in doubt and those who finally win it for us will write in
the book of human history.

\begin{center}
{\bf Acknowledgements}
\end{center}

Quantum gravity is a strange land whose features have only been
glimpsed by the hardiest of theoretical physicists. One doesn't
attempt the journey without guides, and I was fortunate to have the
best: Stanley Deser and Bryce DeWitt. One also isn't advised to go
alone, and I have enjoyed more than three decades of friendship and
collaboration with Nikolaos Tsamis. Among many other things, I am
grateful to him for hosting me at the lovely University of Crete
during the composition of this article. And I must thank the following
colleagues for advice on the manuscript: Jan Ambjorn, Abhay Ashtekar, 
Cecile DeWitt, Gary Horowitz, Elias Kiritsis, Costas Kounnas, Renate 
Loll, Shun-Pei Miao, Sohyun Park, Tomislav Prokopec, Martin Reuter, 
Jan Smit, Lee Smolin, Theodore Tomaras. This work was partially 
supported by NSF grants PHY-0653085 and PHY-0855021, by FQXi grant 
RFP2-08-31 and by the Institute for Fundamental Theory at the University 
of Florida.


\begin{thebibliography}{99}

\bibitem{Bryce} B. S. DeWitt, Phys. Rev. {\bf 160} (1967) 1113-1148;
1195-1239; 1239-1256.

\bibitem{HV} G. 't Hooft and M. Veltman, Ann. Inst. Henri Poincar\'e
{\bf XX} (1974) 69-94.

\bibitem{DvN} S. Deser and P. van Nieuwenhuizen, Phys. Rev. Lett. {\bf 32}
245-247; {\it Phys. Rev.} {\bf D10}, (1974) 401-410; Phys. Rev. {\bf D10}
(1974) 411-420.

\bibitem{DTvN} S. Deser, H.-S. Tsao and P. van Nieuwenhuizen, {\it Phys.
Lett.} {\bf B50} (1974) 491-493; {\it Phys.  Rev.} {\bf D10} (1974) 3337-3342.

\bibitem{KS} K. S. Stelle, Phys. Rev. {\bf D16} (1977) 953-969.

\bibitem{GS} M. H. Goroff and A. Sagnotti, Phys. Lett. {\bf B160} (1985)
81-86; Nucl. Phys. {\bf B266} (1986) 709-736.

\bibitem{Ven} A. E. M. van de Ven, Nucl. Phys. {\bf B378} (1992) 309-366.

\bibitem{SNIA} A. G. Riess {\it et al.}, Astron. J. {\bf 116} (1998)
1009-1038, astro-ph/9805201;
S. Perlmutter {\it et al.}, Astrophys. J. {\bf 517} (1999) 565-586,
astro-ph/9812133.

\bibitem{Yun} Y. Wang and P. Mukherjee, Astrophys. J. {\bf 650} (2006) 1-6,
astro-ph/0604051; U. Alam, V. Sahni and A. A. Starobinsky, JCAP {\bf 0702}
011, 2007, astro-ph/0612381.

\bibitem{Lamb} W. E. Lamb and R. C. Retherford, Phys. Rev. {\bf 72} (1947)
241-243.

\bibitem{Will} C. M. Will, {\it Theory and experiment in gravitational physics}
(Cambridge University Press, 1981).

\bibitem{AS} M. Abramowitz and I. Stegun, {\it Handbook of Mathematical
Functions} (Dover, New York, 1964).

\bibitem{ADM} R. Arnowitt, S. Deser and C. W. Misner, Phys. Rev. Lett.
{\bf 4} (1960) 375-377; Phys. Rev. {\bf 120} (1960) 313-320; Phys. Rev.
{\bf 120} (1960) 321-324; Ann. Phys. {\bf 33} (1965) 88-107.

\bibitem{Pol} D. H. Politzer, Phys. Rev. Lett. {\bf 30} (1973) 1346-1349.

\bibitem{WG} F. Wilczek and D. H. Gross, Phys. Rev. Lett. {\bf 30} (1973)
1343-1346.

\bibitem{JS} J. Schwinger, J. Math. Phys. {\bf 2} (1961) 407-432.

\bibitem{MB} K. T. Mahanthappa, Phys. Rev. {\bf 126} (1962) 329-340;
P. M. Bakshi and K. T. Mahanthappa, J. Math. Phys. {\bf 4} (1963) 1-11;
J.  Math. Phys. {\bf 4} (1963) 12-16.

\bibitem{LK} L. V. Keldysh, Zh. Eksp. Teor. Fiz. {\bf 47} (1964) 1515-1527.

\bibitem{SKrev} K. C. Chou, Z. B. Su, B. L. Hao and L. Yu, Phys. Rept.
{\bf 118} (1985) 1-131; R. D. Jordan, Phys. Rev. {\bf D33} (1986) 444-454;
E. Calzetta and B. L. Hu, Phys. Rev. {\bf D35} (1987) 495-509.

\bibitem{MO} M. Ostrogradsky, Mem. Ac. St. Petersbourg \textbf{VI 4} (1850)
385.

\bibitem{Guth} A. H. Guth, Phys. Rev. {\bf D23} (1981) 347-356.

\bibitem{WMAP} WMAP Collaboration (E. Komatsu et al.), Astrophys. J. Suppl.
{\bf 180} (2009) 330-376.

\bibitem{SDSS} W. J. Percival et al., Mon. Not. Roy. Astron. Soc. {\bf 381}
(2007) 1053-1066, arXiv:0705.3323.

\bibitem{Hubble} E. Hubble, Proc. Nat. Acad. Sci. USA {\bf 15} (1929)
168-173.

\bibitem{gravpot} N. E. J. Bjerrum-Bohr, J. F. Donoghue and B. R. Holstein,
Phys. Rev. {\bf D67} (2003) 084033, Erratum-ibid. {\bf D71} (2005) 069903,
hep-th/0211072.

\bibitem{HE} S. W. Hawking and G. F. R. Ellis, {\it The large scale
structure of space-time} (Cambridge University Press, 1973).

\bibitem{sing} A. Borde, A. H. Guth and A. Vilenkin, Phys. Rev. Lett. {\bf 90}
(2003) 151301, gr-qc/0110012.

\bibitem{Weinberg} S. Weinberg, {\it Quantum Theory of Fields, Volume 1:
Foundations} (Cambridge University Press, 1995).

\bibitem{Carroll} S. M. Carroll, Living Rev. Rel. {\bf 4} (2001) 1
astro-ph/0004075.

\bibitem{DETF} A. J. Albrecht et al., ``Report of the Dark Energy Task Force,''
astro-ph/0609591.

\bibitem{Jackson} J. D. Jackson, {\it Classical Electrodynamics
Third Edition} (Wiley, New York, 1999).

\bibitem{Ford1} R. T. Thompson and L. H. Ford, Phys. Rev. {\bf D74} (2006)
024012, gr-qc/0601137; J. Borgman and L. H. Ford, Phys. Rev. {\bf D70} (2004)
064032, gr-qc/0307043.

\bibitem{smear} W. Pauli, Helv. Phys. Acta Suppl. {\bf 4} (1956) 69;
S. Deser, Rev. Mod. Phys. {\bf 29} (1957) 417-423.

\bibitem{Ford2} L. H. Ford and R. P. Woodard, Class. Quant. Grav. {\bf 22}
(2005) 1637-1647, gr-qc/0411003.

\bibitem{Ford3} C. H. Wu and L. H. Ford, Phys. Rev. {\bf D64} (2001)
045010, quant-ph/0012144.

\bibitem{JS2} J. Schwinger, Phys. Rev. {\bf 73} (1948) 416.

\bibitem{Weisskopf} V. S. Weisskopf, Phys. Rev. {\bf 56} (1939) 72-85.

\bibitem{Carlip1} S. Carlip, Rept. Prog. Phys. {\bf 64} (2001) 885-942,
gr-qc/0108040.

\bibitem{Maldacena} J. M. Maldacena, Adv. Theor. Math. Phys. {\bf 2} (1998)
231-252, hep-th/9711200.

\bibitem{Veneziano} G. Veneziano, Nuovo Cim. {\bf A57} (1968) 190-197.

\bibitem{Virasoro} M. A. Virasoro, Phys. Rev. {\bf 177} (1969) 2309-2311.

\bibitem{openN} C. J. Goebel and B. Sakita, Phys. Rev. Lett. {\bf 22}
(1969) 257-260; H. M. Chan, Phys. Lett. {\bf B28} (1969) 425-428.

\bibitem{closedN} J. A. Shapiro, Phys. Lett. {\bf B33} 361-362.

\bibitem{Ramond} P. Ramond, Phys. Rev. {\bf D3} (1971) 2415-2418.

\bibitem{ANJHS} A. Neveu and J. H. Schwarz, Nucl. Phys. {\bf B31} (1971)
86-112.

\bibitem{GSO} F. Gliozzi, J. Scherk and D. Olive, Phys. Lett. {\bf B65}
282-286; Nucl. Phys. {\bf B122} (1977) 253-290.

\bibitem{GL} Yu. A. Golfand and E. P. Likhtman, JETP Lett. {\bf 13} (1971)
323-326.

\bibitem{VA} D. V. Volkov and V. P. Akulov, JETP Lett. {\bf 16} (1972)
438-440; Phys. Lett. {\bf B46} (1973) 109-110.

\bibitem{VS} D. V. Volkov and V. A. Soroka, JETP Lett. {\bf 18} (1973) 312-314;
Theor. Math. Phys. {\bf 20} (1974) 829-834.

\bibitem{JWBZ} J. Wess and B. Zumino, Nucl. Phys. {\bf B70} (1974) 39-50;
Phys. Lett. {\bf B49} (1974) 52-54.

\bibitem{FvNF} D. Z. Freedman, P. van Nieuwenhuizen and S. Ferrara,
Phys. Rev. {\bf D13} (1976) 3214-3218.

\bibitem{DZ} S. Deser and B. Zumino, Phys. Lett. {\bf B62} (1976) 335-337.

\bibitem{RCB} R. C. Brower, Phys. Rev. {\bf D6} (1972) 1655-1662.

\bibitem{PGCBT} P. Goddard and C. B. Thorn, Phys. Lett. {\bf B40} (1972) 
235-238.

\bibitem{JSJHS1} J. Scherk and J. H. Schwarz, Nucl. Phys. {\bf B81} (1974)
118-144.

\bibitem{JSJHS2} J. Scherk and J. H. Schwarz, Phys. Lett. {\bf B57} (1975)
463-466.

\bibitem{MBGJHS} M. B. Green and J. H. Schwarz, Phys. Lett. {\bf B149}
(1984) 117-122.

\bibitem{Giddings} S. B. Giddings and S. A. Wolpert, Commun. Math. Phys. 
{\bf 109} (1987) 177-190.

\bibitem{MKKK} M. Kaku and K. Kikkawa, Phys. Rev. {\bf D10} (1974) 1110-1133;
Phys. Rev. {\bf D10} (1974) 1823-1843.

\bibitem{Witten1} E. Witten, Nucl. Phys. {\bf B268} (1986) 253-294.

\bibitem{HLRS} G. T. Horowitz, J. Lykken, R. Rohm and A. Strominger,
Phys. Rev. Lett. {\bf 57} (1986) 283-286.

\bibitem{DJGAJ} D. J. Gross and A. Jevicki, Nucl. Phys. {\bf B283}
(1987) 1-49; Nucl. Phys. {\bf B287} (1987) 225-250.

\bibitem{Holger} D. L. Bennett, H. B. Nielsen and R. P. Woodard,
Phys. Rev. {\bf D57} (1998) 1167-1170, hep-th/9707088.

\bibitem{Moffat} D. Evens, J. W. Moffat, G. Kleppe and R. P. Woodard,
Phys. Rev. {\bf D43} (1991) 499-519; G. Kleppe and R. P. Woodard, Nucl. 
Phys. {\bf B388} (1992) 81-112, hep-th/9203016.

\bibitem{EW} D. A. Eliezer and R. P. Woodard, Nucl. Phys. {\bf B325} (1989)
389-469.

\bibitem{ABLT} I. Antoniadis, C. Bachas, D. C. Lewellen and T. N. Tomaras,
Phys. Lett. {\bf B207} (1988) 441-446.

\bibitem{Ignatios} I. Antoniadis, Phys. Lett. {\bf B246} (1990) 377-384.

\bibitem{Witten} E. Witten, ``Some comments on string dynamics,'' in
{\it STRINGS 95: Future Perspective in String Theory} (World Scientific,
Ridge Edge, NJ, 1996) Eds. I. Bars, P. Bouwknegt, J. Minahan, D.
Nemeschansky, K. Pilch, H. Saleur and N. Warner, pp. 501-523, hep-th/9507121.

\bibitem{Polchinski} J. Polchinski, Phys. Rev. Lett. {\bf 75} (1995) 
4724-4727, hep-th/9510017.

\bibitem{ASCV} A. Strominger and C. Vafa, Phys. Lett. {B379} (1996) 99-104,
hep-th/9601029.

\bibitem{Abhay} A. Ashtekar, J. Baez, A. Corichi and K. Krasnov, Phys. 
Rev. Lett. {\bf 80} 1998 904-907, gr-qc/9710007.

\bibitem{Carlip2} S. Carlip, Phys. Rev. Lett. {\bf 88} (2002) 241301,
gr-qc/0203001.

\bibitem{RBJP} R. Bousso and J. Polchinski, JHEP {\bf 06} (2000) 006,
hep-th/0004134.

\bibitem{SAMD} S. Ashok and M. R. Douglas, JHEP {\bf 0410} (2004) 060,
hep-th/0307049.

\bibitem{Susskind} L. Susskind, ``The Anthropic Landscape of String Theory,''
in {\it Universe or Multiverse?} (Cambridge University Press, 2007)
Ed. B. Carr, pp. 247-266, hep-th/0302219.

\bibitem{GSW} M. Green, J. H. Schwarz and E. Witten, {\it Superstring theory,
Vol. 1: Introduction} (Cambridge University Press, 1987);
{\it Superstring theory, Vol. 2: Loop amplitudes, anomalies and phenomenology}
(Cambridge University Press, 1987).

\bibitem{JP} J. Polchinski, {\it String Theory, Vol. 1: An introduction
to the bosonic string} (Cambridge University Press, 1998);
{\it String Theory, Vol. 2: Superstring theory and beyond} (Cambridge
University Press, 1998).

\bibitem{BZ} B. Zwiebach, {\it A First Course in String Theory} (Cambridge
University Press, 2004).

\bibitem{EK} E. Kiritsis, {\it String Theory in a Nutshell} (Princeton
University Press, 2007).

\bibitem{BBS} K. Becker, M. Becker and J. H. Schwarz, {\it String Theory and
M-Theory: A Modern Introduction} (Cambridge University Press, 2007).

\bibitem{Smolin1} L. Smolin, {\it The Trouble with Physics: The Rise of String
Theory, the Fall of Science, and What Comes Next} (Houghton Mifflin, New York,
2006).

\bibitem{Woit} P. Woit, {\it Not Even Wrong --- The Failure of String Theory
and the Search for Unity in Physical Law} (Basic Books, New York, 2006).

\bibitem{Sannan} S. Sannan, Phys. Rev. {bf D34} (1986) 1749-1758.

\bibitem{BDR} Z. Bern, L. J. Dixon and R. Roiban, Phys. Lett. {\bf B644}
(2007) 265-271, hep-th/0611086.

\bibitem{doubters} 
S. Deser, J. H. Kay and K. S. Stelle, Phys. Rev. Lett. {\bf 38} (1977) 527-530;
S. Ferrara and B. Zumino, Nucl. Phys. {\bf B134} (1978) 301-326;
N. Marcus and A. Sagnotti, Nucl. Phys. {\bf B256} (1985) 77-108;
P. S. Howe and K. S. Stelle, Int. J. Mod. Phys. {\bf A4} (1989) 1871-1912.

\bibitem{3loop} Z. Bern, J. J. Carrasco, L. J. Dixon, H. Joansson, D. A.
Kosower and R. Roiban, Phys. Rev. {\bf D7} (2008) 105019, arXiv:0808.4112;
Phys. Rev. Lett. {\bf 98} (2007) 161303, hep-th/0702112.

\bibitem{BHS} G. Bossard, P. S. Howe and K. S. Stelle, Gen. Rel. Grav.
{\bf 41} (2009) arXiv:0901.4661.

\bibitem{4loop} Z. Bern, J. J. Carrasco, L. J. Dixon, H. Johansson and R.
Roiban, ``The Ultraviolet Behavior of N=8 Supergravity at Four Loops,''
\hfil\break arXiv:0905.2326.

\bibitem{ZBDAK1} Z. Bern and D. A. Kosower, 
Phys. Rev. {\bf D38} (1988) 1888-1892.

\bibitem{ZBDAK2} Z. Bern and D. A. Kosower, Nucl. Phys. {\bf B379} (1992)
451-561; Nucl. Phys. {\bf B362} (1991) 389-448; Phys. Rev. Lett. {\bf 66}
(1991) 1669-1672.

\bibitem{BDK1} Z. Bern, L. J. Dixon and D. A. Kosower, 
Ann. Rev. Nucl. Part. Sci. {\bf 46} (1996) 109-148, hep-ph/9602280; 
Nucl. Phys. {\bf B437} (1995) 259-304, hep-ph/9409393; 
Nucl. Phys. {\bf B412} (1994) 751-816, hep-ph/9306240; 
Phys. Rev. Lett. {\bf 70} (1993) 2677-2680, hep-ph/9302280;
Phys. Lett. {\bf B302} (1993) 299-308, Erratum-ibid. {\bf B318} (1993) 648,
hep-ph/9212308.

\bibitem{BDK2} Z. Bern, L. J. Dixon and D. A. Kosower, JHEP {\bf 0001} (2000)
027, hep-ph/0001001;
Nucl. Phys. {\bf B513} (1998) 3-86, hep-ph/9708239.

\bibitem{boot1} Z. Bern, L. J. Dixon and D. A. Kosower, Phys. Rev. {\bf D73}
(2006) 065013, hep-ph/0507005.

\bibitem{boot2} Z. Bern, N. E. J. Bjerrum-Bohr, D. C. Dunbar and H. Ita,
JHEP {\bf 0511} (2005) 027, hep-ph/0507019.

\bibitem{boot3} C. F. Berger, Z. Bern, L. J. Dixon, D. Forde and D. A. 
Kosower, Phys. Rev. {\bf D75} (2007) 016006 hep-ph/0607014;
Phys. Rev. {\bf D74} (2006) 036009, hep-ph/0604195.

\bibitem{BDDK} Z. Bern, L. J. Dixon, D. C. Dunbar and D. A. Kosower,
Phys. Lett. {\bf B394} (1997) 107-115, hep-ph/9611127;
Nucl. Phys. {\bf B435} (1995) 59-101, hep-ph/9409265;
Nucl. Phys. {\bf B425} (1994) 217-260, hep-ph/9403226.

\bibitem{KLT} H. Kawai, D. C. Lewellen and S. H. H. Tye, Nucl. Phys.
{\bf B269} (1986) 1-23.

\bibitem{BDDPR} Z. Bern, L. J. Dixon, D. C. Dunbar, M. Perelstein and J. S. 
Rozowsky, Nucl. Phys. {\bf B530} (1998) 401-456, hep-th/9802162.

\bibitem{nots} Z. Bern, N. E. J. Bjerrum-Bohr and D. C. Dunbar, JEHP 
{\bf 0505} (2005) 056, hep-th/0501137.

\bibitem{Erice} Z. Bern, J. J. M. Carrasco and H. Johansson, ``Progress on
Ultraviolet Finiteness of Supergravity,'' arXiv:0902.3765.

\bibitem{SWasymp} S. Weinberg, ``Ultraviolet Divergences in Quantum Theories
of Gravitation,'' in {\it General Relativity: An Einstein Centenary Survey}
(Cambridge University Press, 1979) Eds. S. W. Hawking and W. Israel, 
pp. 790-831.

\bibitem{JLM} X. Ja\'en, J. Llosa and A. Molina, Phys. Rev. {\bf D34}
(1986) 2302-2311.

\bibitem{Donoghue} J. F. Donoghue, Phys. Rev. Lett. {\bf 72} (1994) 
2996-2999, gr-qc/9310024;
Phys. Rev. {\bf D50} (1994) 3874-3888, gr-qc/9405057.

\bibitem{Wetterich} C. Wetterich, Phys. Lett. {\bf B301} (1993) 90.

\bibitem{OLMR1} O. Lauscher and M. Reuter, JHEP {\bf 0510} (2005) 050,
hep-th/0508202;
Phys. Rev. {\bf D66} (2002) 025026, hep-th/0205062;
Int. J. Mod. Phys. {\bf A17} (2002) 993-1002, hep-th/0112089;
Class. Quant. Grav. {\bf 19} (2002) 483-492, hep-th/0110021;
Phys. Rev. {\bf D65} (2002) 025013, hep-th/0108040.

\bibitem{Percacci} R. Percacci, ``Asymptotic Safety,'' arXiv:0709.3851;
Phys. Rev. {\bf D73} (2006) 041501, hep-th/0511177;

\bibitem{RPDP} R. Percacci and D. Perini, Phys. Rev. {\bf D68} (2003) 
044018, hep-th/0304222;
Phys. Rev. {\bf D67} (2003) 081503, hep-th/0207033.

\bibitem{OLMR2} O. Lauscher and M. Reuter, Lect. Notes Phys. {\bf 721}
(2007) 265-285.

\bibitem{Ashtekar} A. Ashtekar, Phys. Rev. Lett. {\bf 57} (1986)
2244-2247; Phys. Rev. {\bf D36} (1987) 1587-1602.

\bibitem{LOST} J. Lewandowski, A. Okolow, H. Sahlmann and T. Thiemann,
Commun. Math. Phys. {\bf 267} (2006) 703-733, gr-qc/0504147.

\bibitem{Nicolai} H. Nicolai, K. Peeters and M. Zamaklar, Class.
Quant. Grav. {\bf 22} (2005) R193-R247, hep-th/0501114.

\bibitem{lqc} A. Ashtekar, Gen. Rel. Grav. {\bf 41} (2009) 707-741,
arXiv:0812.0177.

\bibitem{Dittrich} B. Dittrich, Class. Quant. Grav. {\bf 23} (2006)
6156-6184, gr-qc/0507106; 
Gen. Rel. Grav. {\bf 39} (2007) 1891-1927, gr-qc/0411013.

\bibitem{gravprop} C. Rovelli, Phys. Rev. Lett. {\bf 97} (2006)
151301, gr-qc/0508124; E. Bianchi, L. Modesto, C. Rovelli and S.
Speziale, Class. Quant. Grav. {\bf 23} (2006) 6989-7028,
gr-qc/0604044.

\bibitem{AAJL} A. Ashtekar and J. Lewandowski, Class. Quant. Grav. 
{\bf 21} (2004) R53-R152, gr-qc/0404018.

\bibitem{Smolin} L. Smolin, ``The main postulates and results of
loop quantum gravity,'' in {\it Deserfest: A Celebration of the Life
and Works of Stanley Deser} (World Scientific, Hackensack, 2006),
Eds. J. T. Liu, M. J. Duff, K. S. Stelle and R. P. Woodard, pp.
266-302.

\bibitem{Rovelli} C. Rovelli, {\it Quantum Gravity} (Cambridge University
Press, 2004).

\bibitem{Thiemann} T. Thiemann, {\it Introduction to Modern Canonical
Quantum General Relativity} (Cambrdige University Press, 2007).

\bibitem{Regge} T. Regge, Nuovo Cim. {\bf 19} (1961) 558-571.

\bibitem{MEAAAM} M. E. Agishtein and A. A. Migdal, Nucl. Phys. {\bf B385}
(1992) 395-412, hep-lat/9204004.

\bibitem{Hamber} H. W. Hamber, Nucl. Phys. {\bf B400} (1993) 347-389.

\bibitem{BBKP} P. Bialas, Z. Burda, A. Krzywicki and B. Petersson,
Nucl. Phys. {\bf B472} (1996) 293-308, hep-lat/9601024.

\bibitem{BVB} B. V. de Baker, Phys. Lett. {\bf B389} (1996) 238-242,
hep-lat/9603024.

\bibitem{EHY} H. S. Egawa, S. Horata and T. Yukawa, Nucl. Phys. Proc.
Suppl. {\bf 106} (2002) 971-973, hep-lat/0110042.

\bibitem{JARL} J. Ambjorn and R. Loll, Nucl. Phys. {\bf B536} (1998)
407-434, hep-th/9805108.

\bibitem{AJL1} J. Ambjorn, J. Jurkiewicz and R. Loll, Phys. Rev.
Lett. {\bf 85} (2000) 924-927, hep-th/0002050.

\bibitem{AJL2} J. Ambjorn, J. Jurkiewicz and R. Loll, Phys. Rev.
Lett. {\bf 93} (2004) 131301, hep-th/0404156.

\bibitem{LQCD} S. Durr {\it et al.}, Science {\bf 322} (2008) 1224-1227,
arXiv:0906.3599.

\bibitem{AJL3} J. Ambjorn, J. Jurkiewicz and R. Loll, ``Quantum gravity as
sum over spacetimes,'' arXiv:0906.3947.

\bibitem{KT} E. W. Kolb and M. S. Turner, {\it The Early Universe}
(Addison-Wesley, Redwood City, CA, 1990).

\bibitem{Steve} S. Weinberg, {\it Cosmology} (Oxford University Press, 2008).

\bibitem{Linde} A. D. Linde, {\it Particle Physics and Inflationary
Cosmology} (Harwood, Chur, Switzerland, 1990).

\bibitem{precursors}
R. Brout, F. Englert and E. Gunzig, Ann. Phys. {\bf 115} (1978) 78-106;
R. Brout, F. Englert and P. Spindel, Phys. Rev. Lett. {\bf 43} (1979)
417-420;
A. A. Starobinsky, Phys. Lett. {\bf B91} (1980) 99-102;
D. Kazanas, astrophys. J. {\bf 241} (1980) L59-L63;
K. Stao, Phys. Lett. {\bf B99} (1981) 66-70.

\bibitem{chaotic} A. Linde, Phys. Lett. {\bf B108} (1982) 389-393.

\bibitem{newinf} A. Albrecht and P. J. Steinhardt, Phys. Rev. Lett. {\bf 48}
(1982) 1220-1223.

\bibitem{Slava} V. Mukhanov, {\it Physical Foundations of Cosmology}
(Cambridge University Press, 2005).

\bibitem{Parker} L. Parker, Phys. Rev. {\bf 183} (1969) 1057-1068.

\bibitem{Grishchuk} L. P. Grishchuk, Sov. Phys. JETP {\bf 40} (1975) 409-415.

\bibitem{BD} N. D. Birrell and P. C. W. Davies, {\it Quantum Fields in
Curved Space} (Cambridge Univ. Press, 1982).

\bibitem{AAS} A. A. Starobinsky, JETP Lett. {\bf 30} (1979) 682-685.

\bibitem{MC} V. F. Mukhanov and G. V. Chibisov, JETP Lett. {\bf 33} (1981)
532-535.

\bibitem{WMS} L. M. Wang, V. F. Mukhanov and P. J. Steinhardt, Phys. Lett.
{\bf B414} (1997) 18-27, astro-ph/9709032.

\bibitem{21cm} S. R. Furlanetto, S. P. Oh and F. H. Briggs, Phys. Rept.
{\bf 433} (2006) 181-301, astro-ph/0608032.

\bibitem{SWpower} S. Weinberg, Phys. Rev. {\bf D74} (2006) 023508,
hep-th/0605244;
Phys. Rev. {\bf D72} (2005) 043514, hep-th/0506236.

\bibitem{NCT} N. C. Tsamis and R. P. Woodard, Nucl. Phys. {\bf B474}
(1996) 235-248, hep-ph/9602315;
Ann. Phys. {\bf 253} (1997) 1-54;
Phys. Rev. {\bf D54} (1996) 2621-2639.

\bibitem{SPM} S. P. Miao and R. P. Woodard, Phys. Rev. {\bf D74} (2006)
024021, gr-qc/0603135; Class. Quant. Grav. {\bf 23} (2006) 1721-1762,
gr-qc/0511140.

\bibitem{CME} C. McEvedy, {\it The Penguin Atlas of Ancient History} (Penguin
Books, Harmondsworth, 1967).

\end{thebibliography}
\end{document}